\documentclass[12pt, reqno]{amsart}
\usepackage[utf8]{inputenc}
\usepackage[a4paper, margin=1.0in]{geometry}
\usepackage{setspace}
\usepackage{footmisc}

\usepackage{amsfonts, amssymb, amsthm}
\usepackage{mathtools} 
\numberwithin{equation}{section}
\usepackage{mathpazo}
\newtheoremstyle{sltheorem}
  {2em}{2em}{\upshape}{}{\bfseries}{.}{ }{}
\theoremstyle{sltheorem}
\newtheorem{theorem}{Theorem}
\newtheorem{assumption}{Assumption}
\newtheorem{definition}{Definition}

\newtheorem{lemma}{Lemma}
\newtheorem{remark}{Remark}
\newtheorem{corollary}{Corollary}
\newtheorem{step}{Step}
\usepackage[hypertexnames=false,hidelinks]{hyperref}
\newtheorem*{supplementlemma}{Lemma \supplementresultnumber}
\newtheorem*{supplementcorollary}{Corollary \supplementresultnumber}
\newenvironment{restatedlemma}[1]
  {\def\supplementresultnumber{\ref*{#1}}\begin{supplementlemma}}
  {\end{supplementlemma}}
\newenvironment{restatedcorollary}[1]
  {\def\supplementresultnumber{\ref*{#1}}\begin{supplementcorollary}}
  {\end{supplementcorollary}}
\usepackage{enumitem}
\setlist{itemsep=0.2em, topsep=0.2em}
\usepackage{graphicx}
\usepackage{booktabs}
\usepackage{array}
\usepackage{threeparttable}
\usepackage{dcolumn}
\usepackage{adjustbox}
\usepackage{pdflscape}
\usepackage[figuresright]{rotating}
\usepackage{caption}
\usepackage{subcaption}
\usepackage[english]{babel}
\usepackage{tikz}
\usepackage[round]{natbib}
\usepackage{bibunits}
\usepackage{float}
\def\*#1{\mathbf{#1}}
\def\+#1{\boldsymbol{#1}}
\renewcommand{\*}[1]{\mathbf{#1}}
\renewcommand{\+}[1]{\boldsymbol{#1}}
\newcommand{\rank}{\operatorname{rank}}

\allowdisplaybreaks
\newcommand{\no}{\noindent}
\newcommand{\br}{\mathbb{R}}
\newcommand{\be}{\mathbb{E}}
\newcommand{\cg}{\mathcal{G}}
\newcommand{\crc}{\mathcal{R}}
\newcommand{\lmax}{\lambda_{\max}}
\newcommand{\lmin}{\lambda_{\min}}
\newcommand{\WP}{\widehat{\mathbf{P}}}
\newcommand{\TP}{\widetilde{\mathbf{P}}}

\makeatletter
\def\section{\@startsection{section}{1}%
  \z@{-.5\baselineskip\@plus-.7\baselineskip}{.5\baselineskip}%
  {\normalfont\large\bfseries\centering}}

\def\subsection{\@startsection{subsection}{2}%
  \z@{-.5\linespacing\@plus-.7\linespacing}{.5\linespacing}%
  {\normalfont\normalsize\bfseries\centering}}

\newcommand{\appendixtableofcontents}{%
  \par
  \noindent\begin{minipage}{\textwidth}%
    \centering\normalfont\bfseries\setstretch{1.0}%
    {\large Contents\par}%
  \end{minipage}\par
  \nobreak\vspace{.5\baselineskip}%
  \begingroup
    \hypersetup{linkcolor=black}%
    \setTrue{atoc}%
    \@input{\jobname.atoc}%
    \if@filesw
      \@xp\newwrite\csname tf@atoc\endcsname
      \immediate\@xp\openout\csname tf@atoc\endcsname \jobname.atoc\relax
    \fi
    \global\@nobreakfalse
  \endgroup
  \addvspace{\baselineskip}%
}
\newcommand{\appendixsection}[1]{%
  \section{#1}%
  \addcontentsline{atoc}{section}{\protect\tocappendix{Appendix}{\thesection}{#1}}%
}
\newcommand{\appendixsubsection}[1]{%
  \subsection{#1}%
  \addcontentsline{atoc}{subsection}{\protect\tocsubsection{}{\thesubsection}{#1}}%
}
\makeatother

\makeatletter
\def\@settitle{%
  \begin{center}%
    \normalfont\Large\bfseries
    \begin{minipage}{0.9\textwidth}
    \centering
    \setstretch{1.0}
    \@title
    \end{minipage}
  \end{center}%
  \vspace{1.0em}%
}

\def\@setauthors{%
  \begin{center}%
    \normalfont\normalsize \authors
  \end{center}%
}
\makeatother

\makeatletter
\renewenvironment{abstract}{%
  \ifx\maketitle\relax
    \ClassWarning{\@classname}{Abstract should precede \protect\maketitle\space in AMS classes}%
  \fi
  \global\setbox\abstractbox=\vtop\bgroup
    \normalfont\small
    \list{}{\labelwidth\z@
      \leftmargin3pc \rightmargin\leftmargin
      \listparindent\normalparindent \itemindent\z@
      \parsep\z@ \@plus\p@}%
    \item[\hskip\labelsep\bfseries Abstract.]%
}{%
  \endlist\egroup
}
\makeatother

\begin{document}

\title[Estimation of Panel Data Models with Nonlinear Factor Structure]{Estimation of Panel Data Models with Nonlinear Factor Structure$^\star$}%

\dedicatory{\normalfont \footnotesize \today}

\author[C. Maschmann and J. Westerlund]{
  Christina Maschmann$^\blacktriangle$ \and
  Joakim Westerlund$^\vartriangle$}

 \thanks{$^\star$Previous versions of this paper have been presented at the 8th Swedish Conference in Economics in Lund, the New York Camp Econometrics XIX, the 11th Italian Congress of Econometrics and Empirical Economics in Palermo, the 2025 Workshop in Econometrics and Statistics in Gothenburg, the Annual Conference of the International Association for Applied Econometrics in Torino, the 30th International Panel Data Conference in Montpellier, and the 13th World Congress of the Econometric Society in Seoul. The authors would like to thank workshop and conference participants, and in particular Amedeo Andriollo, Badi Baltagi, Jad Beyhum, Kirill Borusyak, Jan Ditzen, Grigory Franguridi, Hugo Freeman, Simon Freyaldenhoven, Xiyu Jiao, Adam Lee, Luca Margaritella, Simon Reese, Robin Sickles, Mikkel Sølvsten, Ovidijus Stauskas, Taining Wang and Martin Weidner for many valuable comments and suggestions. Thank you also to the Knut and Alice Wallenberg Foundation for financial support through a Wallenberg Academy Fellowship as well as the Stiftelsen f\"or fr\"amjande av ekonomisk forskning vid Lunds universitet for conference participation funding.}
 \thanks{$^\blacktriangle$Maschmann: Department of Economics, Lund University. Email: \texttt{christina.maschmann@nek.lu.se.} \\ $^\vartriangle$Westerlund: Department of Economics, Lund University, and Department of Economics, Deakin University. Email: \texttt{joakim.westerlund@nek.lu.se}.}

\begin{abstract}
\no Panel data models with unobserved heterogeneity in the form of interactive effects standardly assume that the time effects -- or ``common factors'' -- enter linearly. This assumption is restrictive because it concerns an unobserved component of the model, for which a particular functional form is rarely justified. By contrast, linearity in the observable regressors can often be motivated by economic theory or empirical convention. Linearity in the factors has persisted mainly because it is convenient and improves on standard fixed effects. This paper relaxes that assumption by combining the common correlated effects (CCE) approach with sieve methods. The resulting estimator -- abbreviated ``SCCE'' -- preserves key advantages of CCE, including computational simplicity and good small-sample and asymptotic properties, while allowing for a broader class of factor structures that nests the linear case. This makes it suitable for a wide range of empirical applications.
\end{abstract}
\par

\maketitle

\no \textbf{JEL Classification:} C13; C14; C33;

\vspace{1cm}
%%%%%%%%%%%%%%%%%%%%%%%%%%%%%%%%%%%%%%%%%%%%%%%
		%%%%%%% INTRODUCTION %%%%%%%
%%%%%%%%%%%%%%%%%%%%%%%%%%%%%%%%%%%%%%%%%%%%%%%
\section{Introduction} \label{intro}
Consider the panel variable $y_{i,t} \in \br$, observable for $i=1, \dots, N$ cross-sectional units and $t=1, \dots, T$ time periods. Following the bulk of the existing literature on interactive effects models for such variables (see \citealp{chudik_pesaran_2015}, and \citealp{karabiyik2019}, for surveys), we assume that $y_{i,t}$ depends linearly on a vector of regressors $\*x_{i,t} \in \mathbb{R}^{d}$. The main difference compared to said literature is that in the present paper the interactive effects are not necessarily linear. The particular model that we will be considering is given by
\begin{align}
y_{i,t} &= \*x_{i,t}'\+\beta + g_i(\*f_t) + \varepsilon_{i,t}, \label{yi}\\
\*x_{i,t} &= \*G_i(\*f_t) + \*v_{i,t},\label{xi}
\end{align}
where $\varepsilon_{i,t} \in \mathbb{R}$ and $\*v_{i,t} \in \mathbb{R}^{d}$ are idiosyncratic errors that are assumed to be  independent of one another. However, this does not mean that $\*x_{i,t}$ is exogenous. The reason is the presence of unobserved heterogeneity in the form of interactive effects, as captured by $g_i(\*f_t)$ and $\*G_i(\*f_t)$. Here $\*f_t\in \mathbb{R}^{m}$ is a vector of unobserved common factors, whose dimension, $m$, is fixed and does not need to be known. The way that $\*f_t$ enters \eqref{yi} and \eqref{xi} is determined by $g_i(\cdot) \in \mathbb{R}$ and $\*G_i(\cdot) = [G_{1,i}(\cdot),\ldots, G_{d,i}(\cdot)]'\in \mathbb{R}^{d}$, where $g_i(\cdot),G_{1,i}(\cdot),\ldots, G_{d,i}(\cdot) : \mathbb{R}^m \to \mathbb{R}$ are some unknown, real-valued functions. For reasons that will soon become clear, it is convenient to write $g_i(\cdot)$ and $\*G_i(\cdot)$ as functions of $\*f_t$, and to let dependence on $i$ be absorbed by these functions. This is not necessary, though, but we can just as well write $g_i(\*f_t) = g_t(\+\gamma_i)$, where $\+\gamma_i\in \mathbb{R}^{m}$ is a vector of factor loadings, or even $g_i(\*f_t) = g(\*f_t,\+\gamma_i)$. The same is true for $\*G_i(\cdot)$. Either way, because the unobserved interactive effects enter both \eqref{yi} and \eqref{xi}, $\*x_{i,t}$ is potentially endogenous. Our goal is to consistently estimate and conduct inference on $\+\beta$ in this case. 

The existing interactive effects literature can be divided into two main strands (see, for example, \citealp{chudik_pesaran_2015}, and \citealp{karabiyik2019}). One is based on the common correlated effects (CCE) approach of \citet{pesaran2006}. This approach is extremely simple to implement as it has a closed form and it does not require accurate estimation of the number of factors, $m$; however, this simplicity has a price in that $g_i(\cdot)$ and $\*G_i(\cdot)$ must be linear, such that $g_i(\*f_t) =\+\gamma_i' \*f_t $ and $\*G_i(\*f_t) = \+\Gamma_i' \*f_t$, where $\+\Gamma_i\in \mathbb{R}^{m\times d}$.\footnote{As \citet{DeVos2019} point out, requiring $\*G_i(\cdot)$ to be linear is restrictive not only in itself but also because it rules out models in which there are regressors that enter \eqref{yi} nonlinearly.} The other strand is based on the quasi maximum likelihood (QML) approach of \citet{bai2009}. This approach is more general than CCE in that it does not require $\*G_i(\cdot)$ to be linear but then it is also computationally much more involved.\footnote{There are some studies based on generalized methods of moments (see, for example, \citealp{Ahn2013}, and \citealp{Robertson2015}). These are similar to those based on QML in that $\*G_i(\cdot)$ is not required to be linear.} Both approaches require that $g_i(\cdot)$ is linear. 

Requiring $g_i(\cdot)$ to be linear is very convenient because it means that given a consistent estimate of the space spanned by $\*f_t$, the source of the endogeneity of $\*x_{i,t}$ can be netted out using simple least squares (LS) orthogonal projections.\footnote{As is well known from the classical common-factor literature, factors and loadings cannot be identified separately; at best, one can consistently estimate only the spaces they span.} Of course, just because linearity is simple, does not mean it is correct, a fact that is now being increasingly recognized in the literature. For example, in climate economics, unobserved factors such as common technological adaptation or weather are likely to exert nonlinear effects. Temperature's effect on productivity and agricultural yield, for instance, often exhibits threshold behavior; small variations may have little impact, but larger variations can after a certain point become extremely damaging (see \citealp{hsiang16}). In empirical asset pricing, it is well known that conventional linear factor models cannot price securities whose payoffs are nonlinear functions of the factors, especially in the presence of derivative securities (see \citealp{bansal}). Therefore, nonlinear factor models offer more appropriate and flexible pricing kernels. In macroeconomics, evidence indicates that inflation responds disproportionally to large, common economic shocks, while small shocks have little or no effect (see \citealp{bobeica}). Another example, which we will come back to in our empirical illustration, is from labor economics. \citet{voig} investigates the effects of intersectoral technology-skill complementarity on skill demand. He notes that technological changes in one sector of the economy are likely to have nonlinear effects on the skill demand in other sectors. These are a few examples but there are many more. In fact, it is difficult to imagine a scenario of empirical relevance in which it is known that $g_i(\cdot)$ and $\*G_i(\cdot)$ are linear. This scenario is ruled out almost by construction, as $g_i(\cdot)$ and $\*G_i(\cdot)$ are a part of the unobserved part of the model.

Although the functional forms of $g_i(\cdot)$ and $\*G_i(\cdot)$ are generally unknown, the relationship between $y_{i,t}$ and $\*x_{i,t}$ is often better understood. Economic theory frequently implies linearity, and empirical work commonly supports it. A vast majority of studies therefore assumes that the relationship is linear, and so do we. What we have in mind is an empirical researcher interested in estimating the average marginal effect of $\*x_{i,t}$ on $y_{i,t}$. Consistent with previous research in the area, this part of the model is taken to be linear, which means that the average marginal effect is given by $\+\beta$. The model is also parsimonious, which is again consistent with previous research, and this raises concerns about unobserved heterogeneity due to omitted variables. Little is known about the nature of this heterogeneity, except for the fact that it is likely present. The researcher therefore wants to infer $\+\beta$ while controlling for as much unobserved heterogeneity as possible.\footnote{This scenario is inspired in part by cross-sectional studies such as \citet{Cattaneo03072018}, and \citet{GALBRAITH2020609}. Here there are ``core'' regressors that are known to enter the model for the dependent variable linearly, and there are ``cofounders'' that may enter nonlinearly. In this part of the literature, the cofounders are assumed to be observed and the main challenge is how to control for their -- potentially nonlinear -- effect.} The approach in this paper does exactly that.

Our main contribution is that we allow $g_i(\cdot)$ and $\*G_i(\cdot)$ to be nonlinear without requiring it. And if they are nonlinear, the only assumption we make is that the functions should be smooth. Hence, in our paper, $g_i(\cdot)$ and $\*G_i(\cdot)$ are completely nonparametric, which means that the type of unobserved heterogeneity that can be permitted is very general. We also do not require any knowledge of the functional form of $g_i(\cdot)$ and $\*G_i(\cdot)$, which means that researchers are spared of the usual problem in practice of having to specify a parametric model for the unobserved heterogeneity.

Linear interactive effects represent the state of the art, but most empirical research is still based on additive fixed effects specifications, which makes it necessary to decide on which effects to include. Moreover, since fixed effects are known to be insufficient in many applications, there is also a need for control variables and data on those. If data are available, there is -- at least in principle -- a further need to decide on how the controls should enter the model, although linear specifications are so common that they are hardly ever questioned. Controlling for unobserved heterogeneity therefore involves a lot of choices and assumptions that need not be correct (see \citealp{LU2014194}, for an excellent, general discussion). The proposed nonparametric treatment of $g_i(\cdot)$ and $\*G_i(\cdot)$ enables a completely agnostic approach to unobserved heterogeneity, which makes it very attractive from an applied point of view. 

The new estimation procedure can be seen as a sieve extension of the CCE approach to linear interactive effects models, and is henceforth referred to as ``SCCE''. Our preference to build on CCE is first and foremost its computational simplicity. In particular, because it has a closed form, the proposed SCCE estimator is both fast and numerically stable. This is important under linearity, and it is essential in more general models where other approaches need not be feasible. Another reason for building on CCE is its excellent small-sample properties. Again, while important under linearity, good performance in small samples is not something that can be taken for granted in more general models. 

The idea is to use the cross-sectional averages of $y_{i,t}$ and $\*x_{i,t}$ to construct a first-pass ``mock'' proxy for $\*f_t$. On their own, these averages are unlikely to span the full factor space. SCCE addresses this by expanding the averages via the method of sieves (see \citealp{chen2007}, for a review). This method generates a sequence of nonlinear transformations of the averages that enrich their span, which makes it possible to approximate $g_i(\cdot)$ and $\*G_i(\cdot)$ without specifying their functional form. We then project out the variation explained by this sequence and estimate $\+\beta$ by applying pooled LS to the resulting projection errors. Our asymptotic results establish that the new estimator is consistent at the usual parametric rate and asymptotically normal. Extensive Monte Carlo simulations confirm good small sample properties under a variety of data generating processes.

The present paper is not the first to allow for nonlinearities in panel data. However, existing studies typically do not consider the same type of nonlinearity as we do. The closest study in terms of estimation strategy is that of \citet{su12}, who also use CCE and sieve estimation. However, they assume that the factors enter linearly and place the nonlinearity on the regressors, which is the complete opposite of what we do.\footnote{Other CCE extensions to nonlinear panel models include \citet{boneva17}, \citet{DeVos2019}, and \citet{chenzhang2025}.} For reasons already given, we believe that our model has greater appeal not only from an empirical point of view but also theoretically because the nonlinearity is placed on the factors, which -- in contrast to the regressors -- are unobserved. 

The only other papers that we are aware of to consider the same type of nonlinearity as we do are \citet{free}, and \citet{beymug2025}.\footnote{An incomplete list of other -- more distant -- studies that consider nonlinear panel data models include \citet{freefernweid2021}, \citet{athey2021}, \citet{arkathey2021}, \citet{arkhirsh2023}, \citet{frey}, \citet{athey2025}, \citet{feng2023}, and \citet{zele2026}.} However, their estimation approaches are computationally burdensome, and difficult to implement. \citet{free} consider two approaches. In the first, the nonlinearity of the interactive effects is handled through an approximation that involves an infinite number of factors, and the estimation of slope coefficients, factors and loadings is carried out jointly using QML. The second approach instead relies on  discretization \`a la \citet{bon}. The cross-sectional units and time periods are divided into groups within which the nonlinear interactive effects can be approximated by additive two-way fixed effects. A two-step estimation approach is proposed in which the first step involves estimating the unknown groups using hierarchical clustering, while the second step estimates the slope coefficients conditional on the first-step clusters. The approach considered by \citet{beymug2025} is similar although the first-step clustering is carried out using the $k$-means algorithm. Because of their complexity, the proposed procedures are difficult to study asymptotically. Both papers establish consistency and convergence rates under general conditions, but asymptotic normality requires sample-splitting (see \citealp{rob1988}, for an early discussion). The simplicity of SCCE makes asymptotic analysis possible without the need for sample-splitting. 

The rest of the paper proceeds as follows: Section \ref{model} introduces the SCCE estimator, whose asymptotic and small-sample properties are investigated in Sections \ref{sec-ass} and \ref{mc}, respectively. In Section \ref{emp}, we apply the new estimator to investigate the drivers of the increasing wage inequality between high- and low-skilled workers in U.S. manufacturing. Section \ref{conc} concludes. The proofs of the main results are reported in Appendix \ref{appA}. The proofs and derivations of the auxiliary results as well as some additional empirical and Monte Carlo results are reported in Appendix \ref{AppB}.

\smallskip

\no \textbf{Notation:} For a real vector $\*a$, let $\|\*a\| \equiv (\*a'\*a)^{\frac{1}{2}}$ denote its Euclidean norm. For a real matrix $\*A$, let $\mathrm{tr}(\*A)$, $\mathrm{rank}(\*A)$, $\|\*A\|_{F} \equiv \{\mathrm{tr}(\*A'\*A)\}^{\frac{1}{2}}$, and $\mathrm{vec}(\*A)$ denote its trace, rank, Frobenius norm, and column-stacking vectorization, respectively. The operator norm induced by the Euclidean norm is denoted as $\|\*A\|_{\mathrm{op}}$, which equals the largest singular value of $\*A$. The largest and smallest eigenvalues of $\*A$ are denoted by $\lmax(\*A)$ and $\lmin(\*A)$. For matrices $\*A_1,\dots,\*A_N$, write $\overline{\*A} \equiv N^{-1}\sum_{i=1}^N \*A_i$. For any $T$-rowed matrix $\*A$, define $\*P_A \equiv \*A(\*A'\*A)^+\*A'$ and $\*M_A \equiv \*I_T-\*P_A$, where $(\cdot)^+$ denotes the Moore--Penrose inverse. For a vector-valued function $\*g(\*f)=[g_1(\*f),\dots,g_n(\*f)]'$, with $\*f\in\mathbb{R}^m$, let $\nabla \*g(\*f) \equiv \tfrac{\partial \*g(\*f)}{\partial \*f'} \in \mathbb{R}^{n\times m}$. For a multi-index $\*a=(a_1,\dots,a_m)'$ of nonnegative integers, let $|\*a|=\sum_{j=1}^m a_j$ and write $\nabla^{\*a}g(\*f) \equiv \tfrac{\partial^{|\*a|}g(\*f)}{\partial f_1^{a_1}\cdots \partial f_m^{a_m}}$. The symbols $\sim$, $O_p(\cdot)$, $o_p(\cdot)$, $\overset{p}{\to}$, and $\overset{d}{\to}$ have their usual meanings. We use w.p.1 and w.p.a.1 for ``with probability one'' and ``with probability approaching one'', respectively. Finally, $C<\infty$ denotes a generic positive constant whose value may change from line to line.

%%%%%%%%%%%%%%%%%%%%%%%%%%%%%%%%%%%%%%%%%%%%%%%
	%%%%%%% MOTIVATION_ESTIMATION %%%%%%%
%%%%%%%%%%%%%%%%%%%%%%%%%%%%%%%%%%%%%%%%%%%%%%%
\section{Motivation and Estimation Procedure} \label{model}
It is convenient to write the model presented in Section \ref{intro} in time-stacked vector form. Let us therefore introduce $\*y_i = \left[y_{i,1}, \dots, y_{i,T}\right]' \in \mathbb{R}^{T}$, $\*X_{i}= \left[\*x_{i,1}, \dots ,\*x_{i,T}\right]' \in \mathbb{R}^{T \times d}$, $\+\varepsilon_i = \left[\varepsilon_{i, 1}, \dots, \varepsilon_{i, T}\right]' \in \mathbb{R}^{T}$, $\*V_i = [\*v_{i, 1}, \dots, \*v_{i, T}]' \in \mathbb{R}^{T \times d}$, $\*g_i(\*F) = [g_i(\*f_1), \dots, g_i(\*f_T)]' \in \mathbb{R}^{T}$ and $\*G_i(\*F) = [\*G_i(\*f_1), \dots, \*G_i(\*f_T)]' \in \mathbb{R}^{T \times d}$. In this notation, \eqref{yi} and \eqref{xi} can be written as
\begin{align}
\*y_i &= \*X_i\+\beta + \*g_i(\*F) + \+\varepsilon_i,\label{y} \\
\*X_i &= \*G_i(\*F) + \*V_i.\label{x}
\end{align}
Let us further denote by $\*z_{i,t} \equiv [y_{i,t}, \*x'_{i,t}]' \in \mathbb{R}^{d+1}$ the vector of observables and by $\*Z_i \equiv [\*y_i, \*X_i] = [\*z_{i,1}, \dots, \*z_{i,T}]' \in \mathbb{R}^{T \times (d+1)}$ its stacked version. Combining \eqref{y} and \eqref{x}, the data generating process of this last matrix can be written in the following way:
\begin{align}
\*Z_i = \mathcal{G}_i(\*F) + \*U_i, \label{z}
\end{align}
where $\mathcal{G}_i(\*F) \equiv [\*g_i(\*F)+\*G_i(\*F)\+\beta, \*G_i(\*F)] = [\cg_i(\*f_1), \dots, \cg_i(\*f_T)]' \in \mathbb{R}^{T \times (d+1)}$, $\cg_i(\cdot):\br^{m}\to\cg_i\left(\br^m\right)\subseteq\br^{d+1}$,  and $\*U_i \equiv [\+\varepsilon_i + \*V_i \+\beta , \*V_i] =  [\*u_{i, 1}, \dots, \*u_{i, T}]' \in \mathbb{R}^{T \times (d+1)}$. 

Equation \eqref{z} is a nonlinear factor model for $\*Z_i$. One of the few papers to discuss estimation of such models is \citet{Amemiya}. However, their QML-style estimation approach requires that the functional form of $\mathcal{G}_i(\cdot)$ is known. The same is true for the principal components-based approach of \citet{WANG2022180}. These approaches are therefore not of any use to us. \citet{feng2023} recently considered the unknown $\mathcal{G}_i(\cdot)$ case. In our context with time as one of the two panel data dimensions, his approach involves time period-wise sample splitting, $k$-nearest neighbors matching of cross-sectional units and finally local-linear principal component estimation. Not only do these steps make for a computationally costly procedure, but the data also have to be independent over time, which is unrealistic. We therefore need a different approach. 

One possible way to estimate \eqref{z} is to ignore the functional form of $\cg_i(\cdot)$ altogether and to treat its columns as $d+1$ distinct factors. One could then try to approximate these factors using the cross-sectional average $\widehat{\*F} \equiv\overline{\*Z}$ of $\*Z_i$, as prescribed by standard CCE. However, this is not as simple as it may sound. To appreciate the issues involved, it is useful to write out $\widehat{\*F}$ in terms of the elements of \eqref{z} as 
\begin{align}
\widehat{\*F} =   \overline{\mathcal{G}}(\*F) + \overline{\*U}. \label{fhat}
\end{align} 
Because $\overline{\*U} \overset{p}{\to}\*0_{T \times (d+1)}$ as $N \to \infty$ for a given $T$ under general conditions, $\widehat{\*F} \overset{p}{\to} \overline{\mathcal{G}}(\*F)$.\footnote{As pointed out in Section \ref{intro}, the asymptotic results of this paper are based on letting both $N$ and $T$ to infinity. The fact that $\overline{\*U} \overset{p}{\to}\*0_{T \times (d+1)}$ as $N \to \infty$ for a given $T$ is therefore not enough. It is intuitive, though, and the main purpose of this discussion is to provide intuition. Formal proofs are provided in the appendix.} However, this does not necessarily mean that $\widehat{\*F}$ is consistent for the space spanned by $\*F$. In fact, there are a number of conditions that have to be met for $\widehat{\*F}$ to be consistent. Suppose for a moment that $\*g_i(\*F)$ and $\*G_i(\*F)$ are linear such that $\*g_i(\*F) = \*F \+\gamma_i$ and $\*G_i(\*F) = \*F \+\Gamma_i$. In this case, $\overline{\mathcal{G}}(\*F) = \*F[\overline{\+\gamma}+\overline{\+\Gamma}\+\beta, \overline{\+\Gamma}] \equiv \*F\overline{\*C}$, and hence $\widehat{\*F} \overset{p}{\to} \*F\overline{\*C}$. Not even this is enough to ensure that $\widehat{\*F}$ spans the space of $\*F$ asymptotically, unless we assume that $\rank (\overline{\*C})=m = d+1$. The reason for this last condition is that we need $\overline{\*C}$ to be invertible in order to show that $\*M_{F\overline{C}} = \*M_{F}$.\footnote{If $m < d+1$, there exists a full rank matrix $\overline{\*B}$, say, such that $m$ of the columns of $\widehat{\*F}\overline{\*B}$ converge to $\*F$, while the remaining $d+1-m$ columns converge to zero (see \citealp{KARABIYIK201760}).} Hence, since $\widehat{\*F}$ is consistent for $\*F\overline{\*C}$, it spans the space of $\*F$ asymptotically. However, this is possible only under very special circumstances. In general, there is no reason to expect $\widehat{\*F}$ to converge to anything meaningful, which means that there is also no reason to expect the resulting CCE estimator of $\+\beta$ to be consistent. In fact, as \citet{feng2023} points out, the number of primitive factors required to capture the nonlinearity of $\cg_i(\cdot)$ is likely large, and much larger than the value of $d$ implied by parsimonious economic models. To take one example, \citet{HOLLY2010160} use CCE to study the effect of income ($d=1$) on housing prices in the U.S. They apply the information criteria of \citet{bai_ng} with which they find evidence of up to six factors. 

The present paper can be seen as a reaction to the problem described in the previous paragraph. The proposed solution consists of the following three-step procedure designed to deal with the potential nonliearity of $\cg_i(\cdot)$: 

\begin{step}[Initial estimation]\ \label{step1}\normalfont Compute $\widehat{\*F} = \overline{\*Z}$ as before. This is our ``mock'' proxy of the factor component.
\end{step}

\begin{step}[Sieve approximation of the factor space]\ \label{step2}\normalfont
As pointed out earlier, $\widehat{\*F}$ is likely inconsistent for the space spanned by $\cg_i(\cdot)$. To address this issue, in the second step we employ the method of sieves. The main idea is to approximate $\cg_i(\cdot)$ using a set of user-specified basis functions with $\widehat{\*F}$ as their argument. Let us therefore denote by $\*p(\cdot)=[p_1(\cdot), \dots, p_K(\cdot)]' \in \mathbb{R}^{K}$ a vector of $K$ such basis functions. The sought sieve basis matrix is given by 
$\widehat{\*P} = \left[\*p(\widehat{\*f}_1), \dots, \*p(\widehat{\*f}_T)\right]' \in \br^{T \times K}$. 
\end{step}

\begin{step}[Estimation of $\+\beta$]  \label{step3} \normalfont
Given $\widehat{\*P}$, our proposed estimator of $\boldsymbol{\beta}$, denoted by $\widehat{\boldsymbol{\beta}}_{SCCE}$, is given by
\begin{align}
		\widehat{\boldsymbol{\beta}}_{SCCE} \equiv \left(\sum_{i=1}^N  \mathbf{X}_{i}'\mathbf{M}_{\widehat{P}} \mathbf{X}_{i}\right)^{-1}\sum_{i=1}^N  \mathbf{X}_{i}' \mathbf{M}_{\widehat{P}} \mathbf{y}_{i}.\label{betahat}
	\end{align} 
\end{step}

A word on the choice of sieve in Step \ref{step2}: As \citet{chen2007} points out, the approximation properties of a sieve typically depend on the support, smoothness, and other functional characteristics of the target function. However, these features are often unknown in practice. In such cases, the choice of sieve space is less important, provided it achieves the desired rate of approximation. Commonly used sieve methods with well-understood approximation properties include finite-dimensional linear sieves such as splines, wavelets, power series and Fourier series, all of which achieve comparable approximation rates (see \citealp{chen2007}, and \citealp{bel15}).

An alternative to the method of sieves is kernel estimation. However, the method of sieves has (at least) two advantages that in our opinion makes it relatively more attractive (see \citealp{chen2007}, for a comprehensive review on sieve estimation). First, the method is computationally very simple, which means that it preserves the empirical appeal of CCE. Second, the sieve approach naturally nests the conventional linear factor specification, implying that it can be used to construct formal statistical tests of the presence of factor nonlinearities. In the empirical illustration of Section \ref{emp}, we elaborate on this point.

The pooled estimator considered in Step \ref{step3} is our version of \citeauthor{pesaran2006}'s \citeyearpar{pesaran2006} pooled CCE (CCEP) estimator. This estimator is based on within pooling; that is, we sum the data over the cross-section before taking the ratio. \citet{pesaran2006} also introduces a mean-group CCE (CCEMG) estimator, which is based on between pooling. Here, one would first compute $N$ estimators obtained by applying CCE to each cross-sectional unit, and then average those. Both estimators -- CCEP and CCEMG -- usually show good finite sample properties. However, within pooling is relatively more efficient; hence, our choice of estimator. In the empirical illustration of Section \ref{emp}, we compare SCCE to CCEP and CCEMG.

\begin{remark}
Our analysis assumes homogeneous slope coefficients. One way to allow for some heterogeneity is to assume instead that the slopes vary randomly across units with a common mean $\+\beta$. Provided that the random slopes are independent of all other random elements of the model, SCCE should be estimating this mean (see \citealp{pesaran2006}).
\end{remark}

%%%%%%%%%%%%%%%%%%%%%%%%%%%%%%%%%%%%%%%%%%%%%%%
		%%%%%%% ASSUMPTIONS %%%%%%%
%%%%%%%%%%%%%%%%%%%%%%%%%%%%%%%%%%%%%%%%%%%%%%%

\section{Assumptions and Asymptotic Results} \label{sec-ass}
%%%%%% A1 %%%%%%
\begin{assumption}[Idiosyncratic errors]\label{A1}
\leavevmode
\begin{enumerate}[label=(\alph*)]
\item\label{A1a} For each $i$, $\{(\varepsilon_{i,t}, \*v_{i,t}): t \geq 1\}$ is a strictly stationary and $\alpha$-mixing process with mixing coefficients $\alpha_{i}(j)$ satisfying $\sum_{j=1}^\infty j^2 \alpha_i(j)^{\frac{\eta}{4+\eta}} \leq C$ for some $\eta > 0$. The process is also independent across $i$ with $\mathbb{E}(\varepsilon_{i,t}) = 0$, $\mathbb{E}(\*v_{i,t}) = \*0_{d \times 1}$, $\max_{t=1,\dots,T} \max_{i=1,\dots,N} \mathbb{E}(\|\*v_{i,t}\|_{F}^{4+\eta}) \leq C$ and $\max_{t=1,\dots,T} \max_{i=1,\dots,N}\mathbb{E}(|\varepsilon_{i,t}|^{4 + \eta}) \leq C$.

\item\label{A1b} Let $\sigma^2_{\varepsilon, i} \equiv \lim_{T \to \infty} \frac{1}{T} \mathbb{E}(\+\varepsilon_i'\+\varepsilon_i) \in \mathbb{R}$ and $\+\Sigma_{v,i} \equiv \lim_{T \to \infty} \frac{1}{T} \mathbb{E}(\*V_i' \*V_i) \in \mathbb{R}^{d \times d}$. The following applies to these variances: $\sup_{i=1, \dots, N} \sigma_{\varepsilon, i}^2 \leq C$, $\sigma^2_{\varepsilon} = \lim_{N \to \infty} \frac{1}{N} \sum_{i=1}^N \sigma_{\varepsilon, i}^2 \leq C$, $\sup_{i=1, \dots, N} \|\+\Sigma_{v,i} \|_{F} \leq C$ and $\+\Sigma_v \equiv \lim_{N \to \infty} \frac{1}{N} \sum_{i=1}^N \+\Sigma_{v,i} \in \mathbb{R}^{d \times d}$ is positive definite. Moreover, the covariance matrices $\+\Omega_{\varepsilon, i} \equiv \mathbb{E}(\+\varepsilon_i \+\varepsilon_i')\in \mathbb{R}^{T \times T}$ and $\+\Omega_{v, i} \equiv \mathbb{E}(\*V_{i}\*V'_{i})\in \mathbb{R}^{T \times T}$ have eigenvalues bounded away from zero and infinity.
\end{enumerate}
\end{assumption}

%%%%%% A2 %%%%%%
\begin{assumption}[Common factors]\label{A2}
$\{\*f_t: t \geq 1\}$ is a strictly stationary and $\alpha$-mixing process with mixing coefficient $\alpha_0(j)$ satisfying $\sum_{j=1}^{\infty} j^2 \alpha_0(j)^{\frac{\eta}{4+ \eta}} \leq C$ with $\eta > 0$.  
\end{assumption}

%%%%%% A3 %%%%%%
\begin{assumption}[Independence] \label{A3} $\*f_t$, $\varepsilon_{i, s}$ and $\*v_{j, l}$ are mutually independent for all $t$, $i$, $j$, $s$ and $l$.
\end{assumption}

%%%%%% DISCUSSION A1--A4 %%%%%%
Assumptions \ref{A1}--\ref{A3} are similar to Assumptions 1--2 in \citet{pesaran2006} and are standard in the CCE literature. Hence, we will only discuss main differences and restrictions here. Assumptions \ref{A1}\ref{A1a} and \ref{A2} impose strict stationarity on $\varepsilon_{i,t}$, $\*v_{i,t}$ and $\*f_t$, and specify $\alpha$-mixing conditions as in \citet{su12} with mixing rates vanishing to zero sufficiently fast. This implies that $\alpha_i(j) = o(j^{-(3+ \frac{12}{\eta})})$, assuming that $\alpha_i(j)$ is approximately a $p$-series. The decay of the mixing rates depends inversely on the parameter $\eta$; the smaller is $\eta$, the faster the mixing rates vanish. As pointed out by \citet{su12}, $\alpha$-mixing is not particularly restrictive. The Monte Carlo results reported in the Appendix \ref{AppB} suggest that Assumptions \ref{A1}\ref{A1a} and \ref{A2} can be relaxed to allow for weak cross-sectional correlation in $\varepsilon_{i,t}$ and $\*v_{i,t}$, and non-stationarity in $\*f_t$. Assumption \ref{A3} is a standard independence condition in the CCE literature.

To ensure accurate approximation of $\cg_i(\*f_t)$ using our proposed method, we need to impose smoothness conditions with respect to $\*f_t$. Let $\crc_f \subseteq \mathbb{R}^m$ denote the support of $\*f_t$. Classical series methods often require $\mathcal{R}_f$ to be bounded, such that, for instance, $\mathcal{R}_f=[0,1]^m$, which can be restrictive. We therefore follow \citet{su12}, and \citet{chen2005}, and allow the support to be unbounded, taking $\mathcal{R}_f = \mathbb{R}^m$ and using the following weighted sup-norm metric:
\begin{align}
\|\cg_i\left(\*f\right)\|_{\infty,\omega} & \equiv \sup_{\*f\in\mathbb R^{m}} \left\|\cg_i(\*f)\right\|_{F}\, [1+\|\*f\|_{F}^{2}]^{-\frac{\omega}{2}} = \sup_{\*f\in\mathbb R^{m}} \left\|\cg_i(\*f)\right\|_{F}\, z(\*f)^{\frac{1}{2}}, \label{normg}
\end{align} where $\omega \geq 0$ and $z(\*f) \equiv [1+\|\*f\|_{F}^{2}]^{-\omega}$ is a weight function. When $\omega = 0$, \eqref{normg} reduces to the usual sup-norm, which would suffice if $\mathcal{R}_f$ would indeed be a bounded subset of $\mathbb{R}^m$ (see, for example, \citealp{lee16}, \citealp{su12}, \citealp{chen2005}, and \citealp{chen2007}).

A typical smoothness class of functions is the H\"older-smooth -- or ``$p$-smooth'' -- class of functions. This class is particularly popular in econometrics as functions belonging to this class can be well approximated by the method of sieves (see \citealp{chen2007}, for a discussion). Let $\lceil \lambda \rceil$ be the largest integer which is strictly smaller than $\lambda$. The H\"older space $\Lambda^\lambda(\mathcal{R}_f)$ with smoothness $\lambda  >0$ is a space of functions $g: \mathcal{R}_f \to \mathbb{R}$ in which the first $\lceil \lambda \rceil $ derivatives are bounded and the $\lceil\lambda \rceil$-th highest derivatives are H\"older continuous with exponent $\lambda - \lceil \lambda \rceil \in (0,1]$. The H\"older norm is given by
\begin{align}
\|g\|_{\Lambda^\lambda}	\equiv \sup_{\*f \in \mathcal{R}_f} \left|g(\*f)\right| + \max_{\left|\*a\right| = \lceil \lambda \rceil} \sup_{\*f \neq \*f^*} \frac{|\nabla^{\*a} g(\*f) - \nabla^{\*a} g(\*f^*)|}{(\|\*f - \*f^*\|_{F})^{\lambda- \lceil \lambda \rceil}} < \infty.
\end{align}
We borrow the following definition of H\"older-smooth functions from \citet{chen2007}, and \citet{chen2005}.

\begin{definition}[H\"older-smooth functions] \label{D31}
Let $\Lambda^\lambda(\mathcal{R}_f, \omega)$ denote a weighted H\"older space of functions $g: \mathcal{R}_f \to \mathbb{R}$ such that $g(\cdot)[1 + \|\cdot\|_{F}^2]^{-\frac{\omega}{2}}$ is in $\Lambda^\lambda(\mathcal{R}_f)$. A weighted H\"older ball with radius $r$ is given by $\Lambda_r^\lambda(\mathcal{R}_f, \omega) \equiv \{g \in \Lambda^\lambda(\mathcal{R}_f, \omega): \|g(\cdot)[1 + \|\cdot\|_{F}^2]^{-\frac{\omega}{2}}\|_{\Lambda^\lambda} \leq r $\}. A function $g(\cdot)$ is said to be $H(\lambda, \omega)$-smooth on $\mathcal{R}_f$ if it belongs to the weighted H\"older ball $\Lambda_r^\lambda(\mathcal{R}_f, \omega)$ for some $\lambda>0$, $r \in (0,\infty)$ and $\omega \geq0$.
\end{definition}

If $\omega=0$, the weighted H\"older ball reduces to the standard H\"older ball $\Lambda_r^\lambda(\mathcal{R}_f) \equiv \{g \in \Lambda^\lambda(\mathcal{R}_f): \|g(\cdot)\|_{\Lambda^\lambda} \leq r \}$. As before, if $\mathcal{R}_f$ would be a bounded subset of $\mathbb{R}^m$ this would suffice. Here, however, $\mathcal{R}_f = \mathbb{R}^m$, and a standard H\"older ball would exclude any function whose magnitude grows without bound. Hence, even simple linear interactive effects with deterministic factors, such as $g_i(\*f_t) = \+\gamma_i'\*f_t=\*1_{m}'\*f_t$, would be ruled out, which is clearly not desirable in our set-up.

%%%%%% A4 %%%%%%
\begin{assumption}[Function class] \label{A4}
Define the set $\mathcal{S}\equiv \{g_i(\cdot),G_{j,i}(\cdot): i=1, \dots, N, j=1, \dots, d\}$, where the mappings $g_i(\cdot)$ and $G_{j,i}(\cdot)$ are defined as before. Further, let $\overline{\mathcal G}(\cdot):\br^m \to \overline{\mathcal G}(\br^m) \subseteq \br^{d+1}$, where $\overline{\mathcal G}(\cdot)$ denotes the cross-sectional average of $\mathcal{G}_i(\cdot)$.
\begin{enumerate}[label=(\alph*)] 
\item\label{A4a} For any $\widetilde{g}_i(\cdot)\in \mathcal{S}$, $\widetilde{g}_i(\cdot) \in \Lambda_r^{\lambda_i}(\crc_f, \omega_i)$ for some $\lambda_i > 0$, $\omega_i \geq 0$.
\item\label{A4b} $\sup_{t=1,\hdots,T}\int_{\mathbb{R}^m} [1+\|\*f_t\|_{F}^{2}]^{\overline{\omega}} d\mu(\*f_t) < \infty$ for some $\overline{\omega} > \sup_{i=1,\dots, N}\left(\omega_i + \lambda_i\right)$, where $d\mu(\*f_t) = w(\*f_t)d\*f_t$ and $w(\*f_t)$ is the probability density function of $\*f_t$.
\item\label{A4c} Let $\mathcal{R}_{\overline{\cg}}\equiv\overline{\cg}(\br^m)$ and $h_i(\*u)\equiv\cg_i(\overline{\cg}^{-1}(\*u))$ for $\*u\in\mathcal{R}_{\overline{\cg}}$, where the inverse is well defined under Assumption \ref{A5}. For each $i$, there are sieve coefficients $\+\alpha_{g_i}\in\br^K$ and $\+\alpha_{G_i}\in\br^{K\times d}$, with $\+\alpha_{h_i}\equiv[\+\alpha_{g_i}+\+\alpha_{G_i}\+\beta,\,\+\alpha_{G_i}]$, such that
\begin{align}
\sup_{\*u\in\mathcal{R}_{\overline{\cg}}}
\left\|h_i(\*u)-\+\alpha_{h_i}'\*p(\*u)\right\|_{F}
\left(1+\|\*u\|_{F}^2\right)^{-\frac{\widetilde{\omega}}{2}}
=O\left(K^{-\frac{\lambda_i}{m}}\right), \label{eq:proxyweightedapprox}
\end{align}
where $\widetilde{\omega}\geq0$ is a weight satisfying $\omega\widetilde{\omega}\leq\overline{\omega}$ with $\omega$ being from Assumption \ref{A5} and $\overline{\omega}$ is from part \ref{A4b}. 
\item\label{A4d} $\mathbb{E}\left[\widetilde{g}_i(\*f_t)\right] = 0$.
\end{enumerate}
\end{assumption}

%%%%%% A5 %%%%%%
\begin{assumption}[Rank condition] \label{A5}
$\overline{\cg}(\cdot): \br^m \to \overline{\cg}\left(\br^m\right) \subseteq \br^{d+1}$ is $H(\lambda, \omega)$-smooth, continuously differentiable, and injective. Its Jacobian $\nabla\overline{\cg}(\*f)\in\br^{(d+1)\times m}$ has full column rank $m\leq d+1$ for every $\*f\in\br^m$.
\end{assumption}

%%%%%% DISCUSSION A4--A5 %%%%%%
Assumption \ref{A4}\ref{A4a} specifies a weighted smoothness condition for each function in the set $\mathcal{S}$. Recall that $\cg_i(\*f_t)\equiv[g_i(\*f_t) + \+\beta'\*G_i(\*f_t), \, \*G_i(\*f_t)']'\in \br^{d+1}$, which we can write equivalently as 
\begin{align}
\cg_i(\*f_t)= \left[\begin{array}{c}
    g_i(\*f_t)  \\
    \*G_i(\*f_t) 
\end{array}\right]  + \left[\begin{array}{c}
    \+\beta'\*G_i(\*f_t)  \\
    \*0_{d\times 1} 
\end{array}\right] . \label{cgide}
\end{align} This decomposition shows that $\cg_i(\*f_t)$ consists of a nonlinear component $[g_i(\*f_t), \*G_i(\*f_t)']'$ and a linear transformation of $\*G_i(\*f_t)$, $[\+\beta'\*G_i(\*f_t), \*0_{d\times 1}']'$. Because linear transformations and constant shifts preserve H\"older smoothness and approximation rates, both terms in \eqref{cgide} belong to the same smoothness class. Consider the function $\widetilde{g}_i(\cdot) \in \mathcal{S}$. Equation \eqref{cgide} implies that any assumption placed on $\widetilde{g}_i(\cdot)$ applies also to $\cg_i(\cdot)$, and vice versa. We will make use of this feature throughout the paper. Assumption \ref{A4}\ref{A4b} restricts the tails of the marginal density of $\*f_t$. This assumption is necessary since we allow $\mathcal{R}_f$ to be unbounded. Assumption \ref{A4}\ref{A4c} quantifies the weighted error from approximating $\cg_i(\cdot)$ by $K$-dimensional linear sieves evaluated at the noise-free factor proxies. The condition is imposed on $h_i(\cdot)$ because the sieve uses these proxies as its arguments. Smoothness of the original factor functions alone does not ensure the same approximation rate when they are expressed as functions of the proxies. Since $h_i(\overline{\cg}(\*f))=\cg_i(\*f)$, the growth bound in Assumption \ref{A5} and $\omega\widetilde{\omega}\leq\overline{\omega}$ imply
\begin{align*}
\sup_{\*f\in\br^m}
\left\|\cg_i(\*f)-\+\alpha_{h_i}'\*p(\overline{\cg}(\*f))\right\|_{F}
\left(1+\|\*f\|_{F}^2\right)^{-\frac{\overline{\omega}}{2}}
=O\left(K^{-\frac{\lambda_i}{m}}\right).
\end{align*}
The same rate therefore holds for $G_{j,i}$ and $g_i$ as discussed above. Assumption \ref{A4}\ref{A4d} is a standard identification condition that holds whenever the model in \eqref{yi} and \eqref{xi} includes an intercept.

Assumption \ref{A5} can be viewed as the nonlinear counterpart of the rank condition in the CCE literature.\footnote{See condition (21) in the seminal paper of \citet{pesaran2006}.} Several points are in order. First, provided that $\nabla \overline{\cg}(\*f_t)$ has full column rank $m \leq d+1$, the inverse function theorem ensures that $\overline{\cg}(\cdot)$ is locally invertible into its image, which is sufficient for our analyses. Second, the restriction concerns the number of factors, $m$, and not the number of nonlinear transformations, $K$. Even when $m$ is small, these transformations can generate a much larger linear function space. Nevertheless, Assumption \ref{A5} rules out the case when $m>d+1$, which can be restrictive. Alternative factor estimation approaches such as principal components are not inherently subject to this restriction. Third, the injectivity condition targets only the cross-sectional average $\overline{\cg}(\cdot)$, and does not impose additional restrictions on $g_i(\cdot)$ and $G_{j,i}(\cdot)$ beyond those already stated in Assumption \ref{A4}. It rules out cases in which two distinct factor realizations $\*f_t \neq \*f_s$ produce the same average interactive effects, so that $\overline{\cg}(\*f_t) = \overline{\cg}(\*f_s)$, which is a standard condition in CCE even in the linear case. To see this, suppose that $g_i(\*f_t) \equiv \+\gamma_i'\*f_t$  and $\*G_i(\*f_t) \equiv \+\Gamma_i'\*f_t$, so that $\overline{\cg}(\*f_t) = [\overline{\+\gamma} + \overline{\+\Gamma}\+\beta,\, \overline{\+\Gamma}]'\*f_t \equiv \overline{\*C}'\*f_t$. Since $\overline{\cg}(\cdot)$ is linear in this case, the usual rank condition $\rank(\overline{\*C}')=m\leq d+1$ directly implies injectivity. Thus, in the linear case, Assumption \ref{A5} reduces to the conventional CCE rank condition. Conversely, if $\rank(\overline{\*C}')<m$, then different factor realizations can generate the same cross-sectional average, so that the relevant factor variation cannot be recovered from the cross-sectional averages. An extreme example is when $\mathbb{E}(\+\gamma_i)=\*0_{m\times 1}$ and $\mathbb{E}(\+\Gamma_i)=\*0_{m\times d}$, in which case $\overline{\*C}$ converges to zero. Cases like this are already ruled out by the standard CCE rank condition. Assumption \ref{A5} therefore extends the usual rank requirement to the nonlinear setting by requiring the map from factors to cross-sectional averages to be informative. A similar injectivity condition is imposed by \citet{beymug2025} in a related setting.\footnote{\citet{beymug2025} suggest a way to circumvent this condition, although their simulations show that doing so leads to substantially wider confidence intervals and poorer finite-sample performance.}

\begin{remark} \label{R1}
By Assumption \ref{A5}, $\nabla\overline{\cg}(\*f_t)$ has full column rank $m$. Hence, locally, the image $\mathcal{R}_{\overline{\cg}} \equiv \overline{\cg}(\mathbb{R}^m) \subseteq \mathbb{R}^{d+1}$ varies along only $m$ independent dimensions. Thus, although the factor proxy $\widehat{\*f}_t = \overline{\cg}(\*f_t) + o_p(1)$ is a $(d+1)$-dimensional vector, asymptotically its variation is determined by the $m$ latent factors in $\*f_t$. This explains why $d$ does not enter the approximation rates in Assumption \ref{A4}\ref{A4c}. Since $\widehat{\*f}_t$ approximately lies on the image of the $m$-dimensional factor space, functions of $\widehat{\*f}_t$ can be treated as functions of the $m$ underlying factors. The sieve approximation error is therefore governed by $m$, which leads to rates of the $K^{-\frac{\lambda_i}{m}}$ type given in Assumption \ref{A4}\ref{A4c}. 
\end{remark}

%%%%%% A6 %%%%%%
\begin{assumption}[Basis functions and sieve coefficients] \label{A6}
Let $\*u_t\equiv\overline{\cg}(\*f_t)$. The population Gram matrix of the sieve basis $\*p(\cdot)$ is such that $\be[\*p(\*u_t)\*p(\*u_t)']=\*I_K$. Further define $\widetilde{\*p}(\*f)\equiv\*p(\overline{\cg}(\*f))$ and the set of sieve matrices $\mathcal{P} \equiv \{\widetilde{\*P}, \widehat{\*P}\}$, where $\widehat{\*P}$ is as before.
\begin{enumerate}[label=(\alph*)]
\item\label{A6a} For any $\*P \in \mathcal{P}$, uniformly in $K$ and $T$, there exist constants $0 < \underline{C} < \overline{C} < \infty$, such that $\underline{C} \leq \lmin\left(\frac{1}{T}\*P' \*P\right)\leq \lmax\left(\frac{1}{T}\*P' \*P\right) \leq \overline{C}$ w.p.a.1.
\item\label{A6b} $\left\|\frac1T\TP'\TP-\*I_K\right\|_{\mathrm{op}}=o_p(1)$.
\item\label{A6c} For large enough $K$, there exist sequences $\zeta_0(K),\zeta_1(K)\geq1$ satisfying $\sup_{\*u\in\br^{d+1}}\|\*p(\*u)\|_{F}\leq\zeta_0(K)$ and $\sup_{\*u\in\br^{d+1}}\|\nabla\*p(\*u)\|_{F}\leq\zeta_1(K)$. Further, $\tfrac KT\to0$, $\sqrt T K^{-\frac{\lambda_i}{m}}\to0$, $\sqrt{NT}K^{-\frac{2\lambda_i}{m}}\to0$, $\sqrt{\frac TN}K\zeta_1(K)^2\to0$, $\frac{T}{N} \to 0$.
\item\label{A6d} The sieve approximation coefficients satisfy $\sup_{i=1,\dots,N}\|\+\alpha_{g_i}\|_{F}\leq C$ and $\sup_{i=1,\dots,N}\|\+\alpha_{G_i}\|_{F}\leq C$, uniformly in $K$.
\end{enumerate}
\end{assumption}

%%%%%% DISCUSSION A6%%%%%%
Assumption \ref{A6}\ref{A6a} is standard in the sieve estimation literature (see, for example, \citealp{newey97}, or \citealp{chen2007}). Similarly, Assumption \ref{A6}\ref{A6b} is a mild condition commonly imposed in the literature (see, for example, \citealp{chenchris2015}). Assumption \ref{A6}\ref{A6c} restricts how fast $K$ may grow and how large or steep the normalized basis can be. As usual, increasing $K$ reduces approximation bias while increasing estimation variance (see, for example, \citealp{bel15}, \citealp{su12}, or \citealp{chen2007}). The sequences $\zeta_{0}(K)$ and $\zeta_{1}(K)$ bound the size of the basis vector and its first derivative, respectively. Bounding $\zeta_0(K)$ avoids overly large series terms, while bounding $\zeta_1(K)$ ensures that small input errors do not explode when $\widehat{\*f}_t$ is inserted into $\*p(\cdot)$. In practice, many common bases such as splines and Fourier series satisfy $\zeta_0(K) = O(\sqrt{K})$ and $\zeta_1(K)=O\left(K^{\frac{3}{2}}\right)$, while others such as Legendre polynomials or power series can have $\zeta_0(K) = O(K)$ and $\zeta_1(K)=O\left(K^3\right)$ (see, for example, \citealp{bel15}, and \citealp{newey97}).

The conditions $\sqrt{T}K^{-\frac{\lambda_i}{m}}\to 0$ and $\sqrt{NT}K^{-\frac{2\lambda_i}{m}}\to 0$ ensure that the sieve approximation bias is asymptotically negligible. Restrictions of this type are standard in the sieve literature (see \citealp{su12}). The condition $\sqrt{\frac{T}{N}}K\zeta_{1}(K)^{2}\to 0$ controls the error coming from the evaluation of the sieve basis at the proxy $\widehat{\*f}_t$ rather than at its probability limit $\overline{\cg}(\*f_t)$. It is analogous to the $\frac{T}{N^2}\to0$ condition employed by \citet{pesaran2006} in the linear factor case. The restriction is stronger here because of the way the factor approximation error enters the nonlinear basis. In particular, the proxy error, which is of order $O_p(N^{-\frac{1}{2}})$, is propagated through the derivatives of the basis functions, which are bounded by $\zeta_1(K)$.\footnote{This differs from \citet{su12}, who apply the sieve to the observed regressors and use the CCE terms only as linear controls. In our setting, the basis $\*p(\widehat{\*f}_t)$ is evaluated directly at an estimated proxy. This generates derivative terms that lead to stronger growth conditions than in \citet{su12}.}

The requirement that $\frac{T}{N} \to 0$ is not uncommon in the panel data literature and is natural in so-called ``short'' panels in which $N\gg T$, as is common in microeconomic applications (see \citealp{arkim2024}, for a discussion). \citet{su12} implicitly impose a similar requirement in their Assumption 4(ii), but with a weaker condition on $\lambda_i$ because their sieve basis is evaluated at observed regressors rather than at estimated factor proxies.

Assumption \ref{A6}\ref{A6d} requires the norms of the sieve coefficients to remain bounded across units as $K$ grows. Since the basis is normalized to have identity Gram matrix, $\be \left(|\+\alpha_{g_i}'\*p(\*u_t)|^2\right) = \|\+\alpha_{g_i}\|_{F}^2$ and $\be \left(\|\+\alpha_{G_i}'\*p(\*u_t)\|_{F}^2\right) = \|\+\alpha_{G_i}\|_{F}^2$.
Hence, the assumption implies that the sieve approximations have bounded second moments. Together with the rates in part \ref{A6c}, these bounds ensure that the error coming from the estimation of the factors is sufficiently small.

\begin{remark} \label{R2}
Conditional on the Gram-matrix condition in Assumption \ref{A6}\ref{A6b}, part \ref{A6c} determines an admissible corridor for $K$. For splines, where $\zeta_j(K)=O(K^{\frac{1}{2}+j})$ for $j\in \{0,1\}$, $K$ should satisfy\footnote{For derivations of the inequalities stated in this remark, see the Appendix \ref{AppB}.}
\begin{align}
    \max \left\{ T^{\frac{m}{2\lambda_i}}, \, \left(NT\right)^{\frac{m}{4\lambda_i}}\right\} \ll K \ll 
    \min\left\{ T, \, \left(\frac{N}{T}\right)^{\frac{1}{8}}\right\}. 
    \label{eq:Kcorridor}
\end{align}
Let $T=N^{\rho}$ for $\rho\in(0,1)$. Then the corridor is non-empty if $\frac{m}{\lambda_i} < \frac{4\rho}{1+\rho}$ for $\rho\leq \frac{1}{9}$ and $\frac{m}{\lambda_i} < \frac{1-\rho}{2(1+\rho)}$ for $\rho> \frac{1}{9}$. The weakest requirement is obtained at $\rho=\frac{1}{9}$, where it reduces to $\lambda_i>\frac{5}{2}m$. The corridor collapses as $\rho\to1$, so our conditions for asymptotic normality exclude proportional growth of $N$ and
$T$. This does not rule out consistency, which requires
weaker rate restrictions. The Monte Carlo results in Section \ref{mc} also show good
finite-sample performance when $N=T$. The condition that $\lambda_i>\frac{5}{2}m$ is a smoothness condition that is satisfied by many transformations, such as linear, polynomial, trigonometric, logistic, and suitably restricted exponential transformations. It is restrictive mainly for functions with limited differentiability.
\end{remark}

%%%%%% LEMMA 1 %%%%%%

We start our asymptotic analysis by justifying Step \ref{step2} of the SCCE estimation procedure. As already pointed out, the number of factors required to capture the nonlinearity of $\cg_i(\cdot)$ is likely large. We also know from Section \ref{model} that $\widehat{\*f}_t \overset{p}{\to} \overline{\mathcal{G}}(\*f_t)$, where $\overline{\cg}(\cdot): \br^m \to \overline{\cg}\left(\br^m\right) \subseteq \br^{d+1}$. Hence, $\widehat{\*f}_t$ effectively recovers a low-dimensional projection of the space spanned by $\cg_i(\*f_t)$. Applying the sieve basis to $\widehat{\*f}_t$ re-expands this projection into a high-dimensional function space, which can be made rich enough to approximate the relevant factor transformations. Had $\overline{\cg}(\*f_t)$ been observed, we could have formed $\*p(\overline{\cg}(\*f_t))$, and under suitable approximation conditions $\cg_i(\*f_t)$ would be well approximated by $\+\alpha_{\cg_i}'\*p(\overline{\cg}(\*f_t))$ for some $\+\alpha_{\cg_i}\in \br^{K \times (d+1)}$. The coefficient matrix $\+\alpha_{\cg_i}$ need not be known but will be absorbed in the estimation of the model. The key point is that the same type of approximation remains valid when the basis is evaluated at $\widehat{\*f}_t$. That is, $\+\alpha_{\cg_i}'\*p(\widehat{\*f}_t)$ can be used as a proxy for $\cg_i(\*f_t)$. Lemma \ref{L1}, whose proof is given in the appendix, formalizes this argument.

\begin{lemma} \label{L1}
Suppose that Assumptions \ref{A1}--\ref{A6} hold. Then, the following holds as $N,T \to \infty$:
\begin{enumerate}[label=(\roman*)]
\item \label{L11} $\left\|\*p (\widehat{\*f}_t) - \*p\left(\overline{\cg}\left(\*f_t\right)\right)\right\|_{F} =o_p(1)$.
\item \label{L12} There exists a $\+\alpha_{\cg_i}\in\br^{K\times(d+1)}$ such that $\left\|\cg_i\left(\*f_t\right) - \+\alpha_{\cg_i}^\prime \*p(\widehat{\*f}_t)\right\|_{F} = o_p(1)$.
\end{enumerate}
\end{lemma}

Hence, $\+\alpha_{\cg_i}^\prime \*p(\widehat{\*f}_t)$ can provide a consistent approximation to $\cg_i(\*f_t)$ even though $\widehat{\*f}_t$ does not consistently estimate (the space spanned by) $\*f_t$ itself. To the best of our knowledge, this result is new to the literature and is therefore interesting on its own. However, the main reason for including it here is that it is needed to establish Theorem \ref{T1}, which is our main asymptotic result.
 
%%%%%% MAIN THEOREM %%%%%%
\begin{theorem}\label{T1}
Under Assumptions \ref{A1}--\ref{A6}, as $N,\,T\to\infty$, 
\begin{eqnarray}
\sqrt{NT}(\widehat{\+\beta}_{SCCE} - \+\beta) \overset{d}{\to} \mathcal{N}(\*0_{d\times 1},\+\Sigma_v^{-1}\+\Theta \+\Sigma_v^{-1}),
\end{eqnarray} where
\begin{align}
\+\Theta \equiv \lim_{N,T\to\infty}\frac{1}{NT}\sum_{i=1}^{N}\mathbb{E}\left(\mathbf{V}_{i}'\boldsymbol{\Omega}_{\varepsilon,i}\mathbf{V}_{i}\right),
\end{align} with $\boldsymbol{\Omega}_{\varepsilon,i} \equiv \mathbb{E}\left(\+\varepsilon_i\+\varepsilon_i'\right)$.
\end{theorem}

Theorem \ref{T1} establishes that our proposed SCCE estimator is consistent, and that the rate of convergence is the usual parametric one. The estimator is also asymptotically normally distributed, which means that it supports standard inference based on Student-$t$ and Wald tests. However, this requires a consistent estimator of the covariance matrix $\+\Sigma_v^{-1}\+\Theta \+\Sigma_v^{-1}$. A naturally consistent estimator of $\+\Sigma_v$ is given by  $\widehat{\+\Sigma}_{v}\equiv\frac{1}{NT}\sum_{i=1}^N \widehat{\*V}_i'\widehat{\*V}_i$, where $\widehat{\*V}_i = [\widehat{\*v}_{i,1}, \dots, \widehat{\*v}_{i,T}]' \equiv \*M_{\widehat{P}}\*X_i$. To estimate $\+\Theta$, analogously to \citet{pesaran2006}, one can use the following heteroskedasticity and autocorrelation consistent (HAC) estimator: 
\begin{align}
\widehat{\+\Theta}\equiv \widehat{\+\Theta}_0 + \sum_{l=1}^L\left(1- \frac{l}{L+1}\right)\left(\widehat{\+\Theta}_l + \widehat{\+\Theta}_l'\right),
\end{align} where $\widehat{\+\Theta}_l\equiv \frac{1}{NT} \sum_{i=1}^N \sum_{t=l+1}^T \widehat{\varepsilon}_{i,t}\widehat{\varepsilon}_{i,t-l}\widehat{\*v}_{i,t}\widehat{\*v}_{i,t-l}'$ with $\widehat{\+\varepsilon}_i = [\widehat{\varepsilon}_{i,1}, \dots, \widehat{\varepsilon}_{i,T}]' \equiv \*M_{\widehat{P}}(\*y_i - \*X_i \widehat{\+\beta}_{SCCE})$ and $L$ being the window size. Consistency of this last estimator requires $L \to \infty$ such that $\frac{L}{T}\to 0$. 

As an alternative to the above HAC estimator, one could use bootstrapping, which tends to work better in small samples. We resample $\*Z_i^* \equiv [\*y_i^*,\*X_i^*]$ from $\{\*Z_1,\ldots, \*Z_N\}$ with replacement, as proposed by \citet{staus_devos}. For each bootstrap sample, we compute $\widehat{\+\beta}_{SCCE}^*$, which is $\widehat{\+\beta}_{SCCE}$ with $[\*y_i,\*X_i]$ replaced by $[\*y_i^*,\*X_i^*]$. This procedure is repeated many times. Confidence intervals can then be constructed from the resulting bootstrapped distribution of $\widehat{\+\beta}_{SCCE}$. In the empirical illustration of Section \ref{emp}, we focus on the bootstrapped confidence intervals, although we also report HAC-based results.

%%%%%%%%%%%%%%%%%%%%%%%%%%%%%%%%%%%%%%%%%%%%%%%
	%%%%%%% MONTE CARLO SIMULATIONS %%%%%%%
%%%%%%%%%%%%%%%%%%%%%%%%%%%%%%%%%%%%%%%%%%%%%%%
\section{Monte Carlo Simulations} \label{mc}

A large-scale Monte Carlo study was conducted to evaluate the small-sample properties of the new estimator. The full set of results is too numerous to report here in full. Consequently, in this section, we focus on a representative subset. The full set of results is provided in the Appendix \ref{AppB}. The data generating process is given by a restricted version of \eqref{yi} and \eqref{xi} with $m=d=2$ and $\+\beta = [1, 1]'$. All elements of the idiosynchratic errors $\varepsilon_{i,t}$ and $\*v_{i,t}$ are drawn independently from $\mathcal{N}(0,1)$.

Let us again denote by $G_{j,i}\left(\*f_t\right)\in \br$ row $j \in \{1,2\}$ of $\*G_i\left(\*f_t\right)$. We consider the following two experiments:

\smallskip

\no \textbf{E1} (Nonlinear factor structure)\textbf{.} We generate \begin{align}
    g_i\left(\*f_t\right) &= \gamma_{1,i}f_{1,t} + \gamma_{2,i}f_{1,t}f_{2,t}+\frac{1}{2}\left(f_{1,t}-\gamma_{3,i}\right)^2, \notag\\
    G_{j,i}\left(\*f_t\right)&=\frac{3}{5}\left[\exp\left(\Gamma_{1j,i}\right)f_{1,t}f_{2,t}^2+\Gamma_{2j,i}\exp\left(f_{2,t}\right)\right]+\frac{2}{5}\sin \left(\Gamma_{3j,i}f_{1,t} + \exp\left(\Gamma_{4j,i}\right)f_{1,t}f_{2,t}\right), \notag
\end{align} where the elements of $\*f_t = [f_{1,t}\, f_{2,t}]'\in \br^2$ and $\gamma_{1,i}, \gamma_{2,i}, \gamma_{3,i}, \Gamma_{1s,i}, \Gamma_{2s,i}$ are drawn independently from $\mathcal{N}(0,1)$, whereas $\Gamma_{3j,i}, \Gamma_{4j,i}$ are drawn from $\mathcal{N}(1,1)$.
\smallskip 

\no \textbf{E2} (Linear factor structure)\textbf{.} In this experiment, 
\begin{align}
    g_i\left(\*f_t\right) &= \+\gamma_i'\*f_t, \notag \\
    G_{j,i}\left(\*f_t\right) &= \+\Gamma_{j,i}'\*f_t, \notag
\end{align} where $\+\gamma_i = [\gamma_{1,i}, \gamma_{2,i}]' \in \br^2$ and $\+\Gamma_{j,i} = [\Gamma_{1j,i}, \Gamma_{2j,i}]' \in \br^2$, where $\*f_t$, $\gamma_{1,i}$, $\gamma_{2,i}$, $\Gamma_{j1,i}$ and $\Gamma_{j2,i}$ are as in E1.

A word on implementation: We use univariate cubic spline polynomials as the sieve basis, with a single common intercept. Denote by $\widehat{f}_{r,t} \in \br$ the $r$-th element of $\widehat{\*f}_t$, for $r=1,\dots,d+1$. Define the coordinate-wise terms
\begin{align}
\*q_r(\widehat{f}_{r,t}) \equiv [\widehat{f}_{r,t}, \widehat{f}_{r,t}^{2}, \widehat{f}_{r,t}^{3}, (\widehat{f}_{r,t} - \theta_{r,1})_+^{3}, \dots, (\widehat{f}_{r,t} - \theta_{r,J})_+^{3}]'\in \br^{3+J},
\end{align}
where $(\widehat{f}_{r,t} - \theta_{r,j})_+^{3} = \max\{(\widehat{f}_{r,t} - \theta_{r,j})^3, 0\}$ and $\theta_{r,1},\dots, \theta_{r,J}$ are ``knot'' values computed as the $\frac{j}{J+1}$-th empirical quantile of $\widehat{f}_{r,1},\dots, \widehat{f}_{r,T}$. In terms of the notation of Section \ref{model}, the combined basis is $\*p(\widehat{\*f}_t)=[1,\*q_1(\widehat{f}_{1,t})',\dots,\*q_{d+1}(\widehat{f}_{d+1,t})']'\in\br^K$. Since $d=2$ in this section, the basis contains $K=1+(d+1)(3+J)=10+3J$ terms in total. The number of knots is set to $J = C \lfloor T^{\frac{1}{4}} \rfloor$, so that $K$ increases with $T$. Tables \ref{tab:tableA2}--\ref{tab:tableA4} in the Appendix \ref{AppB} show that the results are relatively insensitive to the knot choice
for the larger values of $T$ considered, but larger knot
specifications produce instability when $T=20$. We focus
on $C=1$ and report the results for alternative choices in
the Appendix \ref{AppB}.\footnote{The Appendix \ref{AppB} also
reports the simulation results depicted in all four figures
in table form as well as for different specifications of $J$.}

%%%%%%%% MC figure %%%%%%%%
\begin{figure}
\centering
\caption{Simulation results for experiments E1 and E2.}
\begin{subfigure}{0.49\textwidth}
\centering
\includegraphics[width=0.8\hsize]{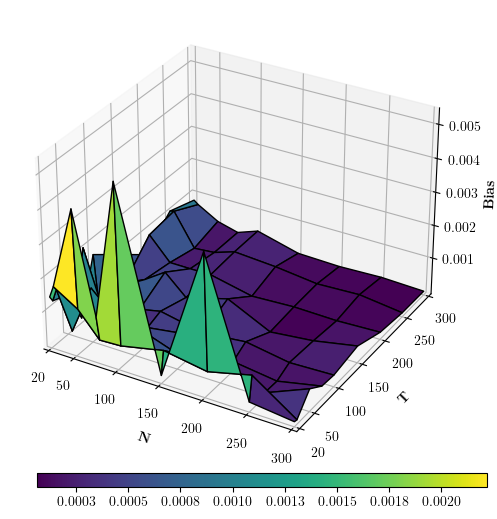}
\caption{Absolute bias E1.}
\end{subfigure}
\hfill
\begin{subfigure}{0.49\textwidth}
\centering
\includegraphics[width=0.8\hsize]{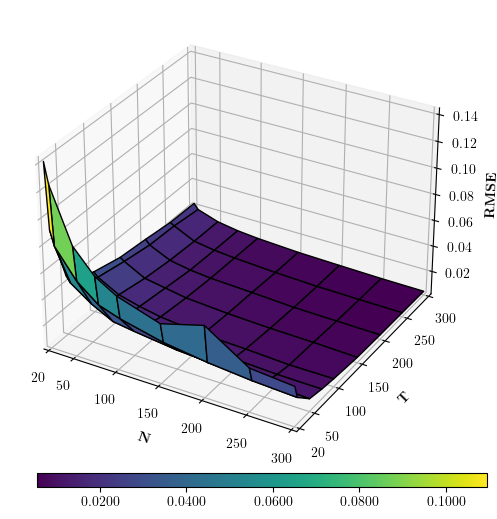}
\caption{RMSE E1.}
\end{subfigure}
\hfill
\begin{subfigure}{0.49\textwidth}
\centering
\includegraphics[width=0.8\hsize]{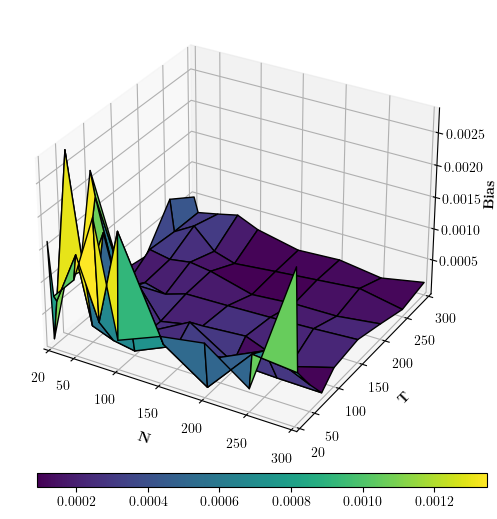}
\caption{Absolute bias E2.}
\end{subfigure}
\hfill
\begin{subfigure}{0.49\textwidth}
\centering
\includegraphics[width=0.8\hsize]{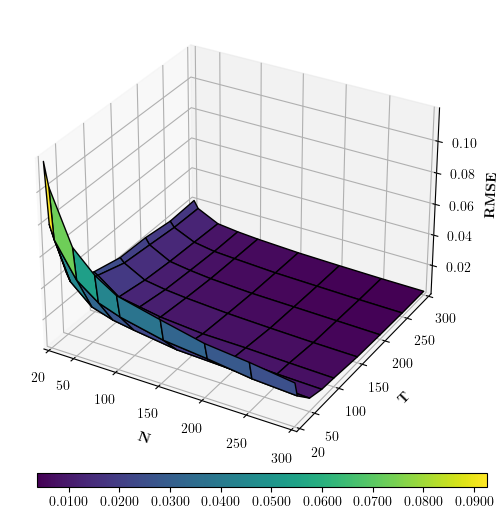}
\caption{RMSE E2.}
\end{subfigure}
\scriptsize
\begin{minipage}{\textwidth}
\vspace{0.5cm}
\textit{Notes:} The absolute bias and RMSE are computed as
$\left|\frac{1}{S} \sum_{s=1}^S (\widehat{\beta}_{1,s} - \beta_1)\right|$ and $\sqrt{\frac{1}{S}\sum_{s=1}^S (\widehat{\beta}_{1,s} - \beta_1)^2}$, respectively, where $S =1,000$ is the number of simulations, and $\widehat{\beta}_{1,s}$ is the estimate of $\beta_1$ from replication $s$. E1 (E2) refers to the experiment with a nonlinear (linear) factor structure. While we focus here on the first coefficient, $\beta_1$, the results for the second coefficient, $\beta_2$, were similar and can be made are available upon request.
\end{minipage}
\label{fig:resultmc}
\end{figure}

Figure \ref{fig:resultmc} presents absolute bias and root mean squared error (RMSE) based on 1,000 replications. According to the results, the SCCE estimator performs well even in samples as small as $N=T=20$, which is desirable given that such sample sizes are not uncommon in practice. Performance improves as both $N$ and $T$ increase, providing evidence of good finite-sample performance across the
designs considered. Performance is good in both experiments, although it is generally better in E2 than in E1, which is to be expected given the relatively simple, linear factor structure in this experiment. The fact that SCCE works well regardless of the factor structure being considered is reassuring because it means that it can be applied without prior knowledge in this regard.

%%%%%%%%%%%%%%%%%%%%%%%%%%%%%%%%%%%%%%%%%%%%%%%
	%%%%%%% EMPIRICAL ILLUSTRATION %%%%%%%
%%%%%%%%%%%%%%%%%%%%%%%%%%%%%%%%%%%%%%%%%%%%%%%
\section{Empirical Illustration} \label{emp}
We apply our method to study the rise in wage inequality between high-skilled and low-skilled workers in U.S. manufacturing over recent decades.\footnote{See Figure \ref{fig:wagedevelop} in the Appendix \ref{AppB} for the development of average wage inequality in the U.S. manufacturing sector.} Following \citet{voig}, we investigate whether persistent differences in the skill premium can be attributed to ``intersectoral technology skill complementarity (ITSC)''. ITSC refers to the idea that skilled labor in intermediate production stages complements skilled labor in downstream sectors through supply-chain linkages. Skill demand in one sector may therefore increase skill demand in related sectors, creating a multiplier effect that raises the wage premium for skilled workers and widens the wage gap.
	
\citet{voig} uses unbalanced data on 358 U.S. manufacturing sectors for 1958--2005. More recently, \citet{yin21}, and \citet{juo22} study the impact of ITSC on wage inequality using CCEP and CCEMG estimators applied to a balanced subset of 313 sectors. Our interest in these studies stems from their assumption that the unobserved factors enter linearly. This restriction may be inappropriate here, since technological change can affect skill demand indirectly through intersectoral linkages and nonlinear adjustment mechanisms. For example, automation may initially raise demand for high-skilled workers but later accelerate the decline of mid-skilled jobs, contributing to wage polarization. Relatedly, \citet{ace22} conjecture that automation-driven task displacement is a central driver of U.S. wage inequality, operating in ways that standard linear or factor-augmented models may fail to capture. Ignoring such nonlinearities can lead to omitted variables bias and misleading conclusions.
	
Our SCCE approach is designed to capture such nonlinearities in the unobserved factors, making it well-suited for studying the effect of ITSC on wage inequality. We use the same balanced subsample of \citeauthor{voig}'s \citeyearpar{voig} dataset as \citet{yin21}, and \citet{juo22}.\footnote{The dataset is available from the website of Journal of Business \& Economic Statistics at: \texttt{https://www.tandfonline.com/doi/full/10.1080/07350015.2019.1623044\#d1e18402}.} Thus, $N=313$ and $T=48$, which according to the Monte Carlo results in Section \ref{mc} should ensure good performance. The model is the same as in these papers, except that we do not impose a linear factor structure:
\begin{align}
	\ln\left(\frac{w_{L,i,t}}{w_{H,i,t}}\right) = \beta_{1} \ln(\sigma_{i,t}) + \beta_{2} \ln\left(\frac{H_{i,t}}{L_{i,t}}\right) + \+\beta'_{3} \*z_{i,t} + g_i(\*f_t) + \varepsilon_{i,t}. \label{wage}
\end{align}
Here, $\tfrac{w_{L,i,t}}{w_{H,i,t}}$ is the relative wage of low-skilled to high-skilled workers. The two main regressors are $\ln(\sigma_{i,t})$, the input skill intensity and proxy for ITSC constructed by \citet{voig}, and $\ln\left(\tfrac{H_{i,t}}{L_{i,t}}\right)$, the relative demand for high- versus low-skilled labor. The corresponding coefficients of interest are $\beta_1$ and $\beta_2$. The vector $\*z_{i,t}\in\br^6$ contains the control variables used in the previous literature:
\begin{align}
	\*z_{i,t} = \left[\begin{array}{c}
                     z_{1,i,t} \\
                     z_{2,i,t} \\
                     z_{3,i,t} \\
                     z_{4,i,t} \\
                     z_{5,i,t} \\
                     z_{6,i,t}
                   \end{array}\right]= \left[\begin{array}{c}
                     k_{i,t}^{\mathrm{equip}} \\
                     (\frac{OCAM}{K})_{i,t} \\
                     (\frac{HT}{K}-\frac{OCAM}{K})_{i,t} \\
                     R\&D_{\mathrm{lag},i,t} \\
                     OS^{\mathrm{narr}}_{i,t} \\
                     OS^{\mathrm{broad}}_{i,t}-OS^{\mathrm{narr}}_{i,t}
                   \end{array}\right],
\end{align} where $k_{i,t}^{\text{equip}}$ is real capital equipment per worker, $(\tfrac{OCAM}{K})_{i,t}$ is the sectoral share of office, computing and accounting equipment, $(\tfrac{HT}{K})_{i,t}$ is the sectoral share of high-technology capital, $R\&D_{\text{lag},i,t}$ is lagged R\&D intensity, and $OS_{i,t}^{\text{broad}}$ and $OS_{i,t}^{\text{narr}}$ are broad and narrow measures of outsourcing. Finally, $g_i(\*f_t)$ captures unobserved heterogeneity with essentially unrestricted functional form. This is the key difference relative to \citet{yin21}, and \citet{juo22}, who assume $g_i(\*f_t)=\+\gamma_i'\*f_t$ despite reasons to suspect that this restriction may be violated.
			
\begin{figure}
\centering
\caption{Components of the factor proxy over time.}
\begin{subfigure}{0.31\textwidth}
\centering
\includegraphics[width=1.0\hsize]{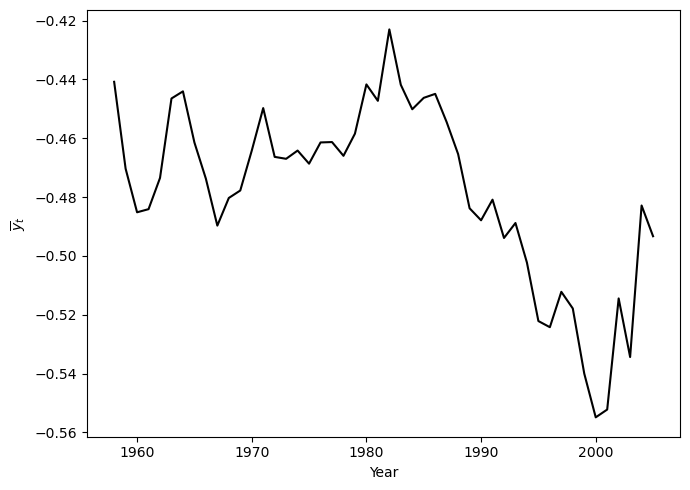}
\caption{$\overline{y}_t$.}
\end{subfigure}
\hfill
\begin{subfigure}{0.31\textwidth}
\centering
\includegraphics[width=1.0\hsize]{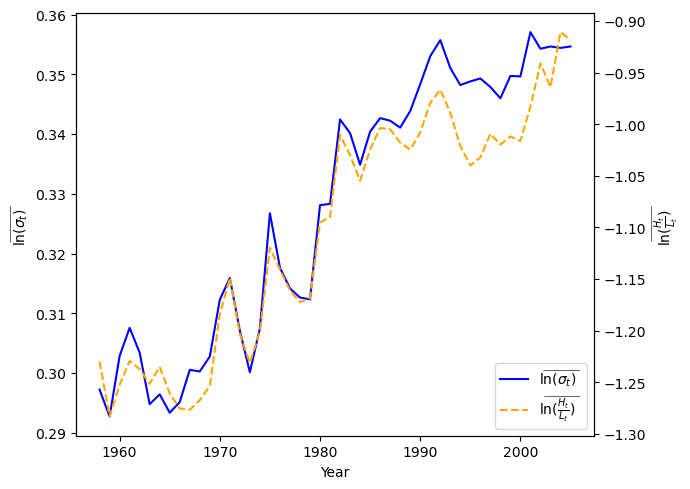}
\caption{$\overline{\ln\left(\sigma_t\right)}$ and $\overline{\ln \left(\frac{H_t}{L_t}\right)}$.}
\end{subfigure}
\hfill
\begin{subfigure}{0.31\textwidth}
\centering
\includegraphics[width=1.0\hsize]{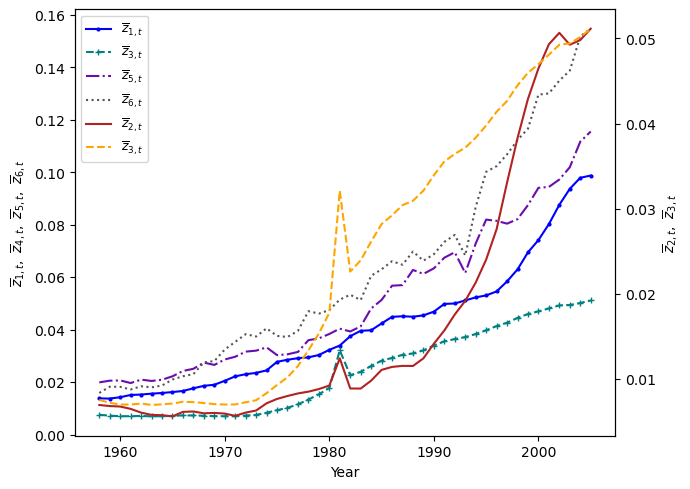}
\caption{The elements of $\overline{\mathbf{z}}_t$.}
\end{subfigure}
\hfill
\vspace{0.5cm} % Adds space between caption and notes
\scriptsize
\begin{minipage}[t]{\textwidth}
\textit{Notes:} The figure plots the components of the factor proxy over the 1958-2005 period. Panel (a) plots the cross-sectional averages of the outcome variable $y_{i,t} = \ln \left(\frac{w_{L, i, t}}{w_{H, i, t}}\right)$, panel (b) plots the cross-sectional averages of the two key regressors, $\ln\left(\sigma_{i,t}\right)$ and $\ln\left(\frac{H_{i,t}}{L_{i,t}}\right)$, while panel (c) plots the cross-sectional averages of the controls included in $\mathbf{z}_{i,t}$. 
\end{minipage}
\label{factors}
\end{figure}

As for the number of knots, $J$, we use the same specification as in Section \ref{mc} with $C=1,\dots,5$, focusing on $C=1$ and treating the remaining values as robustness checks. Consistent with Step \ref{step1}, SCCE estimation uses cross-sectional averages of both the outcome and the regressors. Writing $y_{i,t}=\ln\left(\tfrac{w_{L,i,t}}{w_{H,i,t}}\right)$, the full-model proxy in levels is
\begin{align*}
\widehat{\*f}_t=\left[\overline{y}_t,\ \overline{\ln(\sigma_t)},\ \overline{\ln\left(\frac{H_t}{L_t}\right)},\ \overline{\*z}_t'\right]'.
\end{align*}
The narrow-model proxy omits the control-variable averages $\overline{\*z}_t$. Figure \ref{factors} displays the components of the full-model proxy. The displayed series appear to be trending, and the augmented Dickey-Fuller tests reported in Table \ref{tab:adftest} of the Appendix \ref{AppB} do not reject the unit-root null for these series. Since unit-root non-stationarity is not permitted under our assumptions, we first-difference the data before applying our estimation procedure. The proxies used in estimation are therefore averages of the differenced outcome and regressors.\footnote{Figure \ref{fig:factorsd} in the Appendix \ref{AppB} shows that, after this transformation, the displayed regressor averages appear mean-reverting and no longer exhibit a trend.}

Similarly to \citet{yin21}, we distinguish between a ``narrow'' and a ``full'' model. The full model is given by \eqref{wage}, while the narrow model excludes the controls $\*z_{i,t}$. We estimate both models using SCCE, as well as the CCEP and CCEMG estimators of \citet{pesaran2006}. The latter two are included to assess whether we can replicate the results of \citet{yin21}, and \citet{juo22}, and because their linear factor specification is nested within our more general nonlinear model.\footnote{\citet{yin21}, and \citet{juo22} do not transform their data by taking first differences. We use the same approach for implementing CCEP and CCEMG to ensure comparability with their results. Unreported results suggest that differencing has little effect, which is partly expected given that CCE has been shown to be robust to non-stationarity (see \citealp{kap2011}).}

\begin{sidewaystable}
\vspace*{15.0cm}
\centering
\caption{Estimation results.}
\label{tab:appli}
\scriptsize
\setlength{\tabcolsep}{3pt}
\renewcommand{\arraystretch}{0.9}
\begin{adjustbox}{width=0.70\textwidth, center}
\begin{tabular}{lc c c c c c c c c}
    \toprule
    & \multicolumn{4}{c}{Narrow model} & \multicolumn{4}{c}{Full model} \\
    \cmidrule(lr){2-5} \cmidrule(lr){6-9}
    Regressor & SCCE & CCEP & CCEMG & & SCCE & CCEP & CCEMG & \\
    \midrule
    $\ln(\sigma_{i,t})$ &  -0.34 & -0.61 &  -0.59 &  & -0.33 & -0.52 & -0.71 & \\
                      & (-0.67; -0.03) & (-0.97; -0.25) & (-1.01; -0.15) &  & (-0.70; 0.00) & (-0.88; -0.17) & (-1.17; -0.24) & \\

    $\ln\left(\frac{H_{i,t}}{L_{i,t}}\right)$ & 0.56  & 0.40 & 0.36 &  & 0.61 & 0.48 & 0.45 & \\
                        & (0.50; 0.61) & (0.33; 0.46) & (0.33; 0.38) &  & (0.55; 0.66) & (0.42; 0.55) & (0.43; 0.48) & \\

    $z_{1,i,t}$ & &  &  & & -0.03 & -0.19 & -1.64 &\\
                        & & &  &  & (-0.47; 0.51) & (-0.44; 0.06) & (-2.74; -0.63) &\\

    $z_{2,i,t}$ & &  &  & & 3.15 & 1.23 & -3.29 & \\
                        & &  &  & & (0.40; 6.02) & (-0.37; 2.78) & (-6.66; 0.10) & \\

    $z_{3,i,t}$ & &  &  & & -1.03 & 0.25 & 2.28 & \\
                        & &  &  & & (-2.57; 0.42) & (-0.43; 1.01) & (0.19; 4.28) & \\

    $z_{4,i,t}$ & &  &  & & -0.20 & 0.20 & 1.44 & \\
                        & &  &  & & (-0.74; 0.35) & (-0.29; 0.64) & (-0.23; 3.04) & \\

    $z_{5,i,t}$ & &  &  & & -0.17 & -0.08 & -1.20 & \\
                        & &  &  & & (-0.46; 0.08) & (-0.19; 0.06) & (-2.28; -0.22) & \\

    $z_{6,i,t}$ & &  &  & & 0.07 & -0.09 & 0.20 & \\
                        & &  &  & & (-0.23; 0.38) & (-0.26; 0.08) & (-0.50; 0.98) & \\
    \bottomrule
\end{tabular}
\end{adjustbox}
\begin{minipage}{0.95\textheight}
\scriptsize
\singlespacing
\textit{Notes}: The table reports point estimates of the coefficients of the model in \eqref{wage} with the associated 95\% bootstrap confidence intervals appearing within parentheses. Results are reported for three estimators, SCCE, CCEP and CCEMG. Results with 95\% confidence intervals based on HAC standard errors for our SCCE estimator can be found in the Appendix \ref{AppB}, Table \ref{tab:apphac}. The regressors of interest are $\ln\left(\sigma_{i,t}\right)$, which is a measure of the input skill intensity, and $\ln\left(\frac{H_{i,t}}{L_{i,t}}\right)$, which is the ratio of high-skilled to low-skilled labor. The six control variables are: real capital equipment per worker ($z_{1,i,t}$), the sectoral share of office, computing and accounting equipment ($z_{2,i,t}$), the difference between high-tech and computer capital share ($z_{3,i,t}$), R\&D intensity ($z_{4,i,t}$), narrow outsourcing ($z_{5,i,t}$), and the difference between broad and narrow outsourcing ($z_{6,i,t}$). The dependent variable is the log of the relative wage of low-skilled to high-skilled workers.
\end{minipage}
\end{sidewaystable}

The estimation results are presented in Table \ref{tab:appli}. The CCEP and CCEMG results replicate those reported by \citet{yin21}, and \citet{juo22}, which is reassuring. Comparing these linear-factor estimates with SCCE suggests that ignoring nonlinearities leads to an overestimation of the effect of ITSC on wage inequality in absolute terms. For $\ln(\sigma_{i,t})$, the narrow-model estimate decreases in absolute value from -0.59 for CCEMG and -0.61 for CCEP to -0.34 for SCCE, implying that the linear estimates are about 40\% larger. The full model leads to the same conclusion. Since the sign is always negative, the results support the view that ITSC increases wage inequality. For $\ln \left(\frac{H_{i,t}}{L_{i,t}}\right)$, the sign is positive throughout, but CCEP and CCEMG underestimate the effect relative to SCCE. This difference is statistically more pronounced, since the CCEP and CCEMG estimates are no longer covered by the SCCE confidence intervals. The positive sign is consistent with the view that high-skilled and low-skilled labor are substitutes.\footnote{To ensure that the results are not driven by the chosen number of knots, we re-estimate the model for $C=2,\dots,5$, corresponding to $J\in\{2,4,6,8,10\}$. Figure \ref{fig:figknots} in the Appendix \ref{AppB} shows that the estimated effects of $\ln(\sigma_{i,t})$ and $\ln \left(\frac{H_{i,t}}{L_{i,t}}\right)$ are very stable across values of $J$, suggesting that the results are robust in this regard. We also re-estimate \eqref{wage} using HAC standard errors. Table \ref{tab:apphac} and Figure \ref{fig:figknotsapp} show that the HAC-based confidence intervals are slightly tighter than the bootstrapped ones reported in Figure \ref{fig:figknots}.}

The SCCE estimates differ quite markedly from those of CCEP and CCEMG. To assess whether these differences are due to nonlinearities in the factor component, we test the null hypothesis that the factors enter linearly. Since CCEP can be viewed as a restricted version of SCCE, we apply \citeauthor{ney}'s \citeyearpar{ney} $C(\alpha)$ statistic to test the significance of the nonlinear factor transformations included in the sieve base. Under the null, these transformations should be insignificant and the statistic should be asymptotically chi-squared distributed. The observed test value is $94.6$ in the narrow model and almost identical in the full model, leading us to reject linearity in both cases. This suggests that the differences between SCCE and the CCE-based estimates are likely due to omitted nonlinearities in the factor component.

%%%%%%%%%%%%%%%%%%%%%%%%%%%%%%%%%%%%%%%%%%%%%%%
		%%%%%%% CONCLUSION %%%%%%%
%%%%%%%%%%%%%%%%%%%%%%%%%%%%%%%%%%%%%%%%%%%%%%%
\section{Concluding Remarks} \label{conc}
The existing econometric literature on interactive effects panel data models supposes that the common factors enter in a linear fashion. However, since the factors are unobserved, so is their functional form, and there is typically little or no empirical or theoretical guidance. In fact, in most scenarios of empirical relevance, it is not possible to rule out factor nonlinearities. As a response to this, the present paper proposes a new approach -- called ``SCCE'' -- in which the factors are permitted to enter the model through an unknown functional form. The factors can enter linearly but they are not required to, and if they enter nonlinearly the new approach does not require any knowledge thereof. It is therefore very general in this regard. Interestingly, despite this generality, SCCE is not only computationally simple, easy to implement, and fast, but it also has excellent small-sample and asymptotic properties.

\section*{AI Statement}
Generative AI tools were used for editorial support, including language refinement and \LaTeX{} formatting. The authors reviewed and verified all AI-assisted output and remain fully responsible for the paper’s content, results, and any remaining errors.

\bibliographystyle{apalike} 

\bibliography{sample}

\clearpage
\setcounter{page}{1}

% Appendix title, names, summary, and contents
\begin{appendix}\label{app}
\begin{bibunit}[apalike]
% Keep appendix citation hyperlinks distinct from the manuscript bibliography.
\makeatletter
\let\appendixnatlinkstart\hyper@natlinkstart
\let\appendixnatanchorstart\hyper@natanchorstart
\def\hyper@natlinkstart#1{\appendixnatlinkstart{appendix-#1}}
\def\hyper@natanchorstart#1{\appendixnatanchorstart{appendix-#1}}
\makeatother
% Center portrait floats vertically on appendix float pages.
\makeatletter
\setlength{\@fptop}{0pt plus 1fil}
\setlength{\@fpbot}{0pt plus 1fil}
\makeatother

%%%%%%%%%%%%%%%%%%%%%%%%%%%%%%%%%%%%%%%%%%%%%%%
		%%%%%%% APPENDIX %%%%%%%
%%%%%%%%%%%%%%%%%%%%%%%%%%%%%%%%%%%%%%%%%%%%%%%

\setcounter{lemma}{0}
\renewcommand{\thelemma}{A\arabic{lemma}}
\providecommand{\theHlemma}{appendix.\arabic{lemma}}
\setcounter{theorem}{0}
\renewcommand{\thetheorem}{A\arabic{theorem}}
\renewcommand{\theHtheorem}{appendix.\arabic{theorem}}
\setcounter{corollary}{0}
\renewcommand{\thecorollary}{A\arabic{corollary}}
\providecommand{\theHcorollary}{appendix.\arabic{corollary}}
\setcounter{table}{0}
\renewcommand{\thetable}{\Alph{section}\arabic{table}}
\setcounter{figure}{0}
\renewcommand{\thefigure}{\Alph{section}\arabic{figure}}
% Appendix opening page (within the combined manuscript).
\begin{center}
  \begin{minipage}{0.9\textwidth}
    \centering
    \setstretch{1.0}
    {\normalfont\Large\bfseries Appendix to\\[0.5em]
    ``Estimation of Panel Data Models with Nonlinear Factor Structure''\par}
    \vspace{1.5em}
    {\normalfont\normalsize Christina Maschmann$^\blacktriangle$ and Joakim Westerlund$^\vartriangle$\par}
  \end{minipage}
\end{center}
% Match the author markers and contact details on the manuscript title page.
\begingroup
  \renewcommand{\thefootnote}{}
  \footnotetext[0]{$^\blacktriangle$Maschmann: Department of Economics, Lund University. Email: \texttt{christina.maschmann@nek.lu.se}. \\ $^\vartriangle$Westerlund: Department of Economics, Lund University, and Department of Economics, Deakin University. Email: \texttt{joakim.westerlund@nek.lu.se}.}
\endgroup
\bigskip

\no \normalfont\small \textbf{Summary.} This appendix provides the proofs and additional results accompanying the main manuscript. Appendix \ref{appA} proves our main results, Lemma \ref{L1} and Theorem \ref{T1}. Appendix \ref{AppB} provides proofs of auxiliary Lemmas \ref{appL1}--\ref{appL4} and Corollaries \ref{appC1}--\ref{appC2}, It also provides the derivations of the admissible corridor discussed in Remark \ref{R2} of Section \ref{sec-ass} in the main paper, as well as additional results for the empirical application and the Monte Carlo simulations, which are omitted from the main text.
\par\bigskip
\normalsize
\appendixtableofcontents
\clearpage
\appendixsection{Proofs of Main Results}\label{appA}
This section is organized as follows. The first subsection proves Lemma \ref{L1} from Section \ref{sec-ass}. The second subsection states auxiliary lemmas and corollaries used in the proof of Theorem \ref{T1}, and then proves the theorem. The proofs of the auxiliary results used here are given in Appendix \ref{AppB}.

\appendixsubsection{Proof of Lemma \ref{L1}} \label{sec:proofL1}
%%%%%%%%%%%% PART (i) %%%%%%%%%%%%
Consider \ref{L11}. By the mean-value inequality, there exists a vector $\*f_t^* \in \br^{d+1}$ which lies element-wise between $\overline{\cg}(\*f_t)$ and $\widehat{\*f}_t$, such that
\begin{align}
\left\|\*p(\widehat{\*f}_t)-\*p\left(\overline{\cg}\left(\*f_t\right)\right)\right\|_{F}
&\leq \sup_{\*f_t^*}\left\|\nabla \*p(\*f_t^*)\right\|_{F}\left\|\widehat{\*f}_t-\overline{\cg}\left(\*f_t\right)\right\|_{F} \notag \\
&\leq \zeta_1(K)\left\|\overline{\*u}_t\right\|_{F} = O_p\left(\frac{\zeta_1(K)}{\sqrt{N}}\right)=o_p(1),
\end{align}
which holds since $\left\|\overline{\*u}_t\right\|_{F}=O_p\left(\frac{1}{\sqrt{N}}\right)$ and by Assumption \ref{A6}\ref{A6c}. This establishes part \ref{L11}.

%%%%%%%%%%%% PART (ii) %%%%%%%%%%%%
Consider \ref{L12}. First note that
\begin{align}
  \left\|\cg_i\left(\*f_t\right) - \+\alpha_{\cg_i}^\prime \*p(\widehat{\*f}_t)\right\|_{F} &= \left\|\cg_i\left(\*f_t\right) - \+\alpha_{\cg_i}^\prime \*p\left(\overline{\cg}\left(\*f_t\right)\right) - \+\alpha_{\cg_i}^\prime \left[\*p(\widehat{\*f}_t) - \*p\left(\overline{\cg}\left(\*f_t\right)\right) \right]\right\|_{F} \notag\\
  & \leq \left\|\cg_i\left(\*f_t\right) - \+\alpha_{\cg_i}^\prime \*p\left(\overline{\cg}\left(\*f_t\right)\right)\right\|_{F} + \left\|\+\alpha_{\cg_i}^\prime \left[\*p(\widehat{\*f}_t) - \*p\left(\overline{\cg}\left(\*f_t\right)\right) \right]\right\|_{F} \notag\\
  & \leq \underbrace{\left\|\cg_i\left(\*f_t\right) - \+\alpha_{\cg_i}^\prime \*p\left(\overline{\cg}\left(\*f_t\right)\right)\right\|_{F}}_{\equiv \*I} + \underbrace{\left\|\+\alpha_{\cg_i}\right\|_{F}}_{\equiv \*{II}} \underbrace{\left\|\*p(\widehat{\*f}_t) - \*p\left(\overline{\cg}\left(\*f_t\right)\right)\right\|_{F}}_{\equiv \*{III}} \notag \\
  & \equiv \*I + \*{II} \cdot \*{III},
\end{align}
where $\+\alpha_{\cg_i}=[\+\alpha_{g_i}+\+\alpha_{G_i}\+\beta,\,\+\alpha_{G_i}]$ and $\+\beta$ is fixed. Hence, $\*{II}=O(1)$ by Assumption \ref{A6}\ref{A6d} and $\*{III} = O_p\left(\frac{\zeta_1(K)}{\sqrt{N}}\right)$ by Lemma \ref{L1}\ref{L11}. Therefore, $\*{II}\cdot \*{III} = O_p\left(\frac{\zeta_1(K)}{\sqrt{N}}\right)$.

We are left to analyze $\*I$. Recall the mapping $h_i:\mathcal{R}_{\overline{\cg}}\to\br^{d+1}$ defined in Assumption \ref{A4}\ref{A4c}. By injectivity in Assumption \ref{A5}, $h_i(\overline{\cg}(\*f_t))=\cg_i(\*f_t)$. Assumption \ref{A4}\ref{A4c} directly supplies a sieve approximation $\Pi_{\infty K}h_i\equiv\+\alpha_{h_i}'\*p(\cdot)$ on $\mathcal{R}_{\overline{\cg}}$ satisfying
\begin{align}
\left\|h_i-\Pi_{\infty K}h_i\right\|_{\infty,\widetilde{\omega}}
=\sup_{\*u\in\mathcal{R}_{\overline{\cg}}}\left\|h_i(\*u)-\+\alpha_{h_i}'\*p(\*u)\right\|_{F}\left(1+\|\*u\|_{F}^2\right)^{-\frac{\widetilde{\omega}}{2}}
=O\left(K^{-\frac{\lambda_i}{m}}\right).
\end{align}
In what follows, $\+\alpha_{\cg_i}=\+\alpha_{h_i}=[\+\alpha_{g_i}+\+\alpha_{G_i}\+\beta,\,\+\alpha_{G_i}]$, as in Assumption \ref{A4}\ref{A4c}. We use the following area formula (see, for example, \citealp{evans2015measure}, Theorem 3.9), with its extensions to locally Lipschitz maps by localization and to nonnegative measurable integrands by monotone convergence:
\begin{theorem} \label{changeofvariables}
Let $f:\br^n\to\br^m$ be locally Lipschitz continuous, $n\leq m$. Then, for each nonnegative Lebesgue measurable function $g:\br^n\to[0,\infty]$,
\begin{align}
\int_{\br^n} g(x) J(f(x))dx = \int_{\br^m} \left[\sum_{x \in f^{-1}\{y\}} g(x)\right]	d \mathcal{H}^n(y), \label{change}
\end{align}
Here, $J(f(x))$ is the Jacobian defined almost everywhere by $J(f(x))\equiv\sqrt{\det[\nabla f(x)'\nabla f(x)]}$, and $\mathcal{H}^n(y)$ is the Hausdorff measure.\footnote{The Hausdorff measure generalizes length/area/volume to possibly lower-dimensional sets, which allows measurement on lower-dimensional subsets of $\br^m$. In contrast, Lebesgue measure can only measure $m$-dimensional volume in this case. For a formal definition of the Hausdorff measure, see \citet[Definition 2.1 on page 81]{evans2015measure}.}
\end{theorem}

We would like to point out two things regarding Theorem \ref{changeofvariables}. First, note that if the function $f$ is injective, then every $y$ in $f\left(\br^n\right)$ can be mapped to at most one point in $\br^n$, such that
\begin{align}
\sum_{x \in f^{-1}\{y\}}g(x) = \begin{cases} g\left(f^{-1}\left(y\right)\right) &\text{if} \ y \in f\left(\br^n\right) \\ 0 \ &\text{otherwise}\end{cases}.	
\end{align}
Hence, equation \eqref{change} reduces to
\begin{align}
\int_{\br^n} g(x) J(f(x))dx = \int_{f\left(\br^n\right)}	g\left(f^{-1}\left(y\right)\right) d\mathcal{H}^n(y).
\end{align}
Second, by Assumption \ref{A5}, $\overline{\cg}$ is continuously differentiable. Its derivative is therefore bounded on every closed ball in $\br^m$, so the mean-value inequality makes $\overline{\cg}$ Lipschitz on each such ball. Applying Theorem \ref{changeofvariables} on these balls and then letting their radii tend to infinity gives the formula for $\overline{\cg}$ on $\br^m$, by monotone convergence for nonnegative integrands. Injectivity removes the multiplicity term, and full column rank makes $J(\overline{\cg}(\*f))$ positive.

Now consider
\begin{align}
\be\left(\left\|h_i (\*u) - \+\alpha_{h_i}'\*p(\*u)\right\|_{F}^2\right) = \int_{\mathcal{R}_{\overline{\cg}}} \left\|h_i (\*u) - \+\alpha_{h_i}'\*p(\*u)\right\|_{F}^2 w_u(\*u) d \mathcal{H}^m(\*u), \label{Ei}
\end{align}
where $h_i(\*u) = \left[h_{i,1}(\*u), \dots, h_{i,d+1}(\*u)\right]' \in \br^{d+1}$, $\+\alpha_{h_i} \in \br^{K \times (d+1)}$ is a sieve coefficient matrix, and $\*p(\cdot) \in \br^{K}$ is a vector of basis functions. By injectivity, local Lipschitz continuity, and full column rank in Assumption \ref{A5}, the area formula gives a density $w_u$ of $\*u$ with respect to $\mathcal{H}^m$ on $\mathcal{R}_{\overline{\cg}}$, satisfying $w_u(\overline{\cg}(\*f))=\tfrac{w_f(\*f)}{J(\overline{\cg}(\*f))}$ almost everywhere. Here $w_f$ is the factor density in Assumption \ref{A4}\ref{A4b}. By multiplying and dividing \eqref{Ei} by $\left(1 + \left\|\*u\right\|_{F}^2\right)^{\widetilde{\omega}}$, we obtain
\begin{align}
&\be\left(\left\|h_i (\*u) - \+\alpha_{h_i}'\*p(\*u)\right\|_{F}^2\right) = \int_{\mathcal{R}_{\overline{\cg}}} \left\|h_i (\*u) - \+\alpha_{h_i}'\*p(\*u)\right\|_{F}^2 w_u(\*u) d \mathcal{H}^m(\*u)\notag \\
&= \int_{\mathcal{R}_{\overline{\cg}}} \left[ \left\|h_i (\*u) - \+\alpha_{h_i}'\*p(\*u)\right\|_{F}\left(1 + \left\|\*u\right\|_{F}^2\right)^{-\frac{\widetilde{\omega}}{2}}\right]^2 \left(1 + \left\|\*u\right\|_{F}^2\right)^{\widetilde{\omega}}w_u(\*u) d \mathcal{H}^m(\*u) \notag\\
& \leq  \left\|h_i (\*u) - \+\alpha_{h_i}'\*p(\*u)\right\|^2_{\infty, \widetilde{\omega}} \int_{\mathcal{R}_{\overline{\cg}}}\left(1 + \left\|\*u\right\|_{F}^2\right)^{\widetilde{\omega}} w_u(\*u) d\mathcal{H}^m(\*u) \notag\\
& = O\left(K^{-\frac{2\lambda_i}{m}}\right)\int_{\mathcal{R}_{\overline{\cg}}}\left(1 + \left\|\*u\right\|_{F}^2\right)^{\widetilde{\omega}} w_u(\*u) d\mathcal{H}^m(\*u),\label{euh}
\end{align}
where the approximation bound uses Assumption \ref{A4}\ref{A4c} with its specified weight $\widetilde{\omega}$. By Assumption \ref{A5}, the weighted H\"older bound on $\overline{\cg}$ implies
\begin{align}
\sup_{\*f_t\in\br^m}\left\|\overline{\cg}(\*f_t)\right\|_{F}\left(1+\|\*f_t\|_{F}^2\right)^{-\frac{\omega}{2}}&\leq C, \notag\\
\Rightarrow \left\|\overline{\cg}(\*f_t)\right\|_{F} &\leq C \left(1+ \left\|\*f_t\right\|_{F}^2\right)^{{\frac{\omega}{2}}}, \notag\\
\Rightarrow \left\|\overline{\cg}(\*f_t)\right\|_{F}^2 &\leq C \left(1+ \left\|\*f_t\right\|_{F}^2\right)^{\omega}, \notag\\
\Rightarrow \left(1 + \left\|\overline{\cg}(\*f_t)\right\|_{F}^2\right)^{\widetilde{\omega}} &\leq C' \left(1+ \left\|\*f_t\right\|_{F}^2\right)^{\omega\widetilde{\omega}} \leq C' \left(1+ \left\|\*f_t\right\|_{F}^2\right)^{\overline{\omega}},
\end{align}
for constants $C,C'\in(0,\infty)$, where the last inequality uses $\omega\widetilde{\omega}\leq\overline{\omega}$ from Assumption \ref{A4}\ref{A4c}. Here $\overline{\omega}$ is the moment exponent in Assumption \ref{A4}\ref{A4b}. Thus,
\begin{align}
\left(1 + \left\|\*u\right\|_{F}^2\right)^{\widetilde{\omega}} \equiv \left(1 + \left\|\overline{\cg}\left(\*f_t\right)\right\|_{F}^2\right)^{\widetilde{\omega}} \leq C' \left(1+ \left\|\*f_t\right\|_{F}^2\right)^{\overline{\omega}}. \label{eq:u}
\end{align}
As discussed before, the inverse $\overline{\cg}^{-1}(\cdot)$ exists and is well defined under Assumption \ref{A5}, such that
\begin{align}
\left(1+\left\|\overline{\cg}^{-1}\left(\*u\right)\right\|_{F}^2\right)^{\overline{\omega}} &= \left(1+\left\|\overline{\cg}^{-1}\left(\overline{\cg}\left(\*f_t\right)\right)\right\|_{F}^2\right)^{\overline{\omega}} \notag \\
&=\left(1+\left\|\left(\overline{\cg}^{-1}\circ\overline{\cg}\right)\left(\*f_t\right)\right\|_{F}^2\right)^{\overline{\omega}} = \left(1+\left\|\*f_t\right\|_{F}^2\right)^{\overline{\omega}}. \label{eq:1g}
\end{align}
We can make use of the results in \eqref{eq:u} and \eqref{eq:1g} to arrive at the following expression for \eqref{euh}:
\begin{align}
&\be\left(\left\|h_i (\*u) - \+\alpha_{h_i}'\*p(\*u)\right\|_{F}^2\right) = O\left(K^{-\frac{2\lambda_i}{m}}\right)\int_{\mathcal{R}_{\overline{\cg}}}\left(1 + \left\|\*u\right\|_{F}^2\right)^{\widetilde{\omega}} w_u(\*u) d\mathcal{H}^m(\*u) \notag  \\
&\leq O\left(K^{-\frac{2\lambda_i}{m}}\right)\int_{\mathcal{R}_{\overline{\cg}}}\left(1 + \left\|\overline{\cg}^{-1}\left(\*u\right)\right\|_{F}^2\right)^{\overline{\omega}} w_u(\*u) d\mathcal{H}^m(\*u) \notag \\
&= O\left(K^{-\frac{2\lambda_i}{m}}\right) \int_{\br^m} \left(1 + \left\|\*f_t\right\|_{F}^2\right)^{\overline{\omega}}w_u(\overline{\cg}(\*f_t)) J\left( \overline{\cg}(\*f_t)\right) d\*f_t \label{apply31} \\
&= O\left(K^{-\frac{2\lambda_i}{m}}\right) \int_{\br^m} \left(1 + \left\|\*f_t\right\|_{F}^2\right)^{\overline{\omega}} w_f(\*f_t)d\*f_t,
\end{align}
where we use Theorem \ref{changeofvariables} for \eqref{apply31} and the density identity $w_f(\*f_t)=w_u(\overline{\cg}(\*f_t))J(\overline{\cg}(\*f_t))$ almost everywhere, with $J\left(\overline{\cg}(\*f_t)\right) \equiv \sqrt{\det \left[\nabla \overline{\cg}(\*f_t)'\nabla \overline{\cg}(\*f_t)\right]}$ being the Jacobian of $\overline{\cg}(\*f_t)$. By Assumption \ref{A4}\ref{A4b}, $\sup_{t=1, \dots, T}\int_{\mathcal{R}_f}\left(1+\left\|\*f_t\right\|_{F}^2\right)^{\overline{\omega}}w_f(\*f_t)d\*f_t < \infty$, which allows us to conclude that
\begin{align}
\be\left(\left\|h_i (\*u) - \+\alpha_{h_i}'\*p(\*u)\right\|_{F}^2\right)= O\left(K^{-\frac{2\lambda_i}{m}}\right) O(1) = O\left(K^{-\frac{2\lambda_i}{m}}\right). \label{lemma1Klambda}
\end{align}
The same bound holds for $g_i(\*f_t)-\+\alpha_{g_i}'\*p(\*u_t)$ and $\*G_i(\*f_t)-\+\alpha_{G_i}'\*p(\*u_t)$, since these are fixed linear combinations or subvectors of the error for $h_i$. Averaging their squared errors over $t$ therefore gives, for either $\widetilde{\*g}_i(\cdot)\in\{\*g_i(\cdot),\*G_i(\cdot)\}$,
\begin{align}
\be\left[\left\|\frac{1}{\sqrt T}\left(\TP\+\alpha_{\widetilde{\*g}_i}-\widetilde{\*g}_i(\*F)\right)\right\|_{F}^2\right]
=O\left(K^{-\frac{2\lambda_i}{m}}\right). \label{eq:proxymeansquare}
\end{align}
By Markov's inequality, \eqref{lemma1Klambda} implies
\begin{align}
\left\|h_i (\*u) - \+\alpha_{h_i}'\*p(\*u)\right\|_{F} &= \left\|h_i\left(\overline{\cg}\left(\*f_t\right)\right) - \+\alpha_{h_i}'\*p\left(\overline{\cg}\left(\*f_t\right)\right)\right\|_{F} \notag\\
&= \left\|\cg_i\left(\*f_t\right) - \+\alpha_{\cg_i}'\*p\left(\overline{\cg}\left(\*f_t\right)\right)\right\|_{F} = O_p\left(K^{-\frac{\lambda_i}{m}}\right)=o_p(1),
\end{align}
where the last equality follows from $K^{-\frac{\lambda_i}{m}}=o(1)$. Consequently, by Assumption \ref{A6}\ref{A6c}
\begin{align}
\left\|\cg_i\left(\*f_t\right) - \+\alpha_{\cg_i}^\prime \*p(\widehat{\*f}_t)\right\|_{F} &= O_p\left(K^{-\frac{\lambda_i}{m}}\right) + O(1) O_p\left(\frac{\zeta_1(K)}{\sqrt{N}}\right) \notag \\
&= O_p\left(K^{-\frac{\lambda_i}{m}}\right) + O_p\left(\frac{\zeta_1(K)}{\sqrt{N}}\right) = o_p(1).
\end{align}
This establishes part \ref{L12} of the lemma, and hence the proof of Lemma \ref{L1} is complete. \hfill $\blacksquare$

%%%%%%%%%%%%%%%%%%%%%%%%%%%%%%%%%%
%%%% PROOF OF MAIN THEOREM %%%%
%%%%%%%%%%%%%%%%%%%%%%%%%%%%%%%%%%
\appendixsubsection{Proof of Theorem \ref{T1}}
Recall from Assumption \ref{A6} that $\TP = \left[\widetilde{\*p}\left(\*f_1\right), \dots, \widetilde{\*p}\left(\*f_T\right)\right]' = \left[\*p\left(\overline{\cg}\left(\*f_1\right)\right), \dots, \*p\left(\overline{\cg}\left(\*f_T\right)\right) \right]'$, where the mapping $\overline{\cg}(\cdot)$ is defined in Assumption \ref{A5}. As a convenient way to partly deal with the growing dimension of $\TP$, similarly to \citet{newey97}, we replace $\WP$ and $\TP$ by $\WP\left(\be\left[\frac{1}{T}\TP' \TP\right]\right)^{-\frac{1}{2}}$ and $\TP\left(\be\left[\frac{1}{T}\TP' \TP\right]\right)^{-\frac{1}{2}}$, respectively, where $\*A^{\frac{1}{2}}$ is such that $\left(\*A^{\frac{1}{2}}\right)'\*A^{\frac{1}{2}}=\*A$ for any positive semidefinite matrix $\*A$. This is a normalization of the sieve basis, and Assumption \ref{A6}\ref{A6a} is understood to hold for the resulting normalized matrices. Moreover, this normalization preserves projection operators such that $\*M_{\widehat{P}} = \*M_{\widehat{P}\left(\be[\frac{1}{T} \TP'\TP]\right)^{-\frac{1}{2}}}$. Because of this normalization, we can assume without loss of generality that $\be\left(\frac{1}{T}\TP'\TP\right) = \*I_K$. Before proving our main result, Theorem \ref{T1}, we state several auxiliary lemmas and corollaries. Their proofs are provided in the Appendix \ref{AppB}.

\begin{lemma} \label{appL1}
Suppose Assumptions \ref{A1} and \ref{A3} hold. Then $\left\|\frac{1}{\sqrt{N}T} \sum_{i=1}^N \+\varepsilon_i \otimes \*V_i\right\|_{F} = O_p(1)$.
\end{lemma}

\begin{lemma}\label{appL2}
Suppose Assumptions \ref{A1}, \ref{A3}, \ref{A4}, and \ref{A5} hold and let $\widetilde{\*g}_i(\cdot) \in \widetilde{\mathcal{S}}$, where $\widetilde{\mathcal{S}}\equiv \{\*g_i(\cdot),\*G_i(\cdot)\}$. Then $\left\|\frac{1}{\sqrt{N}T} \sum_{i=1}^N \left(\TP\+\alpha_{\widetilde{\*g}_i} - \widetilde{\*g}_i(\*F)\right) \otimes \*V_i\right\|_{F} = O_p\left(K^{-\frac{\lambda_i}{m}}\right)$.
\end{lemma}

\begin{lemma}\label{appL3}
Suppose Assumptions \ref{A1}, \ref{A3}, and \ref{A6}\ref{A6d} hold and let $\widetilde{\*g}_i(\cdot)\in\widetilde{\mathcal S}$, where $\widetilde{\mathcal S}\equiv\{\*g_i(\cdot),\*G_i(\cdot)\}$. Then $\left\|\frac{1}{\sqrt{NT}} \sum_{i=1}^N \+\alpha_{\widetilde{\*g}_i} \otimes \*V_i\right\|_{F} = O_p(1)$. The same bound holds with $\+\varepsilon_i$ in place of $\*V_i$.
\end{lemma}

\begin{lemma}\label{appL4}
Suppose Assumptions \ref{A1}--\ref{A3}, and \ref{A6}\ref{A6a}-\ref{A6c} hold. Then 
\begin{enumerate}[label=(\roman*)]
    \item \label{appL42} $\left\|\frac{1}{\sqrt{T}}\left(\WP - \TP\right)\right\|_{F} = O_p\left(\frac{\zeta_1(K)}{\sqrt{N}}\right)$,
    \item \label{appL43} $\left\|\frac{1}{T}\WP^\prime \WP - \*I_K\right\|_{\mathrm{op}} = o_p(1)$.
\end{enumerate}
\end{lemma}

\begin{corollary} \label{appC1}
Suppose Assumptions \ref{A1}--\ref{A3}, and \ref{A6}\ref{A6a}-\ref{A6c} hold. Then
\begin{enumerate}[label=(\roman*)]
    \item\label{appC11} $\left\|\left(\WP - \TP\right)\left(\WP'\WP\right)^+\left(\WP - \TP\right)'\right\|_{F} = O_p\left(\frac{\zeta_1(K)^2}{N}\right)$,
    \item \label{appC12} $\left\|\left(\WP - \TP\right)\left(\WP'\WP\right)^+\TP'\right\|_{F} = o_p\left(\frac{\sqrt K\zeta_1(K)}{\sqrt N}\right)$,
    \item \label{appC13} $\left\|\TP\left[\left(\WP'\WP\right)^+ - \left(\TP'\TP\right)^+\right]\TP'\right\|_{F} = O_p\left(\frac{\zeta_1(K)}{\sqrt{N}}\right)$,
    \item \label{appC14} $\left\|(\WP - \TP)(\WP^\prime \WP)^+ \TP^\prime (\WP - \TP)\right\|_{F} = O_p\left(\frac{\sqrt T}{N}\zeta_1(K)^2\right)$,
    \item \label{appC15}$\left\|(\WP - \TP)(\WP^\prime \WP)^+(\WP - \TP)^\prime (\WP - \TP)\right\|_{F} = O_p\left(\frac{\sqrt{T}}{N^{\frac{3}{2}}}\zeta_1(K)^3\right)$.
\end{enumerate}
\end{corollary}
 
\begin{corollary}\label{appC2}
Suppose Assumptions \ref{A1}--\ref{A6} hold and let $\widetilde{\*g}_i(\cdot)\in\widetilde{\mathcal S}$, where $\widetilde{\mathcal S}\equiv\{\*g_i(\cdot),\*G_i(\cdot)\}$. Then $
\left\|\frac{1}{\sqrt{T}}\left(\WP \+\alpha_{\widetilde{\*g}_i} - \widetilde{\*g}_i(\*F)\right)\right\|_{F} = O_p\left(\frac{\zeta_1(K)}{\sqrt{N}}\right) + O_p\left(K^{-\frac{\lambda_i}{m}}\right).$
\end{corollary}

We continue the proof by re-writing equation \eqref{y} for $\*y_{i}$ in the following way:
\begin{align}
\*y_i &= \*X_i\+\beta + \*g_i\left(\*F\right) +  \+\varepsilon_i = \*X_i \+\beta + \widehat{\*P}\+\alpha_{g_i} + \+\varepsilon_i - \left(\widehat{\*P}\+\alpha_{g_i} - \*g_i\left(\*F\right)\right),
\end{align} where
\begin{align}
\widehat{\*P}\+\alpha_{g_i} - \*g_i\left(\*F\right) &= \left(\WP - \TP\right) \+\alpha_{g_i} + \left(\TP \+\alpha_{g_i} - \*g_i\left(\*F\right)\right). \label{pag}
\end{align}

Insertion into $\widehat{\+\beta}_{SCCE}$ yields
\begin{align}
\sqrt{NT}\left(\widehat{\+\beta}_{SCCE} - \+\beta\right) = \left(\frac{1}{NT}\sum_{i=1}^N \*X_i'\*M_{\widehat{P}}\*X_i \right)^{-1} \frac{1}{\sqrt{NT}}\sum_{i=1}^N\*X_i'\*M_{\widehat{P}}\left[\+\varepsilon_i - \left(\WP\+\alpha_{g_i} - \*g_i\left(\*F\right)\right)\right].\label{SCCE}
\end{align}
	
A key first step in establishing the asymptotic distribution of $\sqrt{NT}\left(\widehat{\+\beta}_{SCCE} - \+\beta\right)$ is to derive a cleaned up asymptotic representation that is free of all errors coming from the functional approximation. We start by considering $\*M_{\widehat P}  \*X_i$. Equation \eqref{x} for $\*X_i$ can be manipulated in the same way as \eqref{y} for $\*y_i$ to obtain
\begin{align}
\*X_i = \*G_i\left(\*F\right) + \*V_i = \WP \+\alpha_{G_i}  + \*V_i - \left(\WP \+\alpha_{G_i} - \*G_i\left(\*F\right)\right),
\end{align} where analogously to $\WP \+\alpha_{g_i} - \*g_i\left(\*F\right)$ we can write
\begin{align}
\WP \+\alpha_{G_i} - \*G_i\left(\*F\right) &= \left(\WP - \TP\right) \+\alpha_{G_i} + \left(\TP \+\alpha_{G_i} - \*G_i\left(\*F\right)\right). \label{eq:PalphaG}
\end{align}

\no It follows that
\begin{align}
\*M_{\widehat{P}}\*X_i = \*M_{\widehat{P}}\left[\*V_i - \left(\WP \+\alpha_{G_i} - \*G_i\left(\*F\right)\right)\right].
\end{align}

We now make use of this last expression for $\*M_{\widehat{P}}\*X_i$ in order to evaluate the numerator of \eqref{SCCE}. The following expansion will be used:
\begin{align}
&\frac{1}{\sqrt{NT}}\sum_{i=1}^N\*X_i'\*M_{\widehat{P}}\left[\+\varepsilon_i - \left(\WP\+\alpha_{g_i} - \*g_i\left(\*F\right)\right) \right] \notag \\
 &=\frac{1}{\sqrt{NT}}\sum_{i=1}^N \left[\*V_i - \left(\WP \+\alpha_{G_i} - \*G_i\left(\*F\right)\right)\right]'\*M_{\widehat{P}}\left[\+\varepsilon_i - \left(\WP\+\alpha_{g_i} - \*g_i\left(\*F\right)\right) \right] \notag \\
 &= \frac{1}{\sqrt{NT}}\sum_{i=1}^N \left[\*V_i - \left(\WP \+\alpha_{G_i} - \*G_i\left(\*F\right)\right)\right]'\*M_{\widetilde{P}}\left[\+\varepsilon_i - \left(\WP\+\alpha_{g_i} - \*g_i\left(\*F\right)\right) \right] \notag \\
 &- \frac{1}{\sqrt{NT}}\sum_{i=1}^N \left[\*V_i - \left(\WP \+\alpha_{G_i} - \*G_i\left(\*F\right)\right)\right]'\left(\*M_{\widetilde{P}} - \*M_{\widehat{P}}\right)\left[\+\varepsilon_i - \left(\WP\+\alpha_{g_i} - \*g_i\left(\*F\right)\right) \right] \notag \\
 &= \sum_{j=1}^8 \*D_j, \label{Dj}
\end{align} where
\begin{align}
\*D_1 &= \frac{1}{\sqrt{NT}}\sum_{i=1}^N \*V_i' \*M_{\widetilde{P}} \+\varepsilon_i, \notag \\
\*D_2 &= - \frac{1}{\sqrt{NT}}\sum_{i=1}^N \*V_i' \*M_{\widetilde{P}}\left(\WP \+\alpha_{g_i} - \*g_i\left(\*F\right)\right),\notag \\
\*D_3 &= - \frac{1}{\sqrt{NT}}\sum_{i=1}^N \left(\WP \+\alpha_{G_i} - \*G_i\left(\*F\right)\right)'\*M_{\widetilde{P}}\+\varepsilon_i, \notag\\
\*D_4 &= \frac{1}{\sqrt{NT}}\sum_{i=1}^N \left(\WP \+\alpha_{G_i} - \*G_i\left(\*F\right)\right)'\*M_{\widetilde{P}}\left(\WP \+\alpha_{g_i} - \*g_i\left(\*F\right)\right),\notag \\
\*D_5 &= - \frac{1}{\sqrt{NT}}\sum_{i=1}^N \*V_i'\left(\*M_{\widetilde{P}} - \*M_{\widehat{P}}\right)\+\varepsilon_i, \notag\\
\*D_6 &= \frac{1}{\sqrt{NT}}\sum_{i=1}^N \*V_i' \left(\*M_{\widetilde{P}} - \*M_{\widehat{P}}\right)\left(\WP \+\alpha_{g_i} - \*g_i\left(\*F\right)\right),\notag \\
\*D_7 &= \frac{1}{\sqrt{NT}}\sum_{i=1}^N \left(\WP\+\alpha_{G_i} - \*G_i\left(\*F\right)\right)'\left(\*M_{\widetilde{P}} - \*M_{\widehat{P}}\right)\+\varepsilon_i,\notag \\
\*D_8 &= - \frac{1}{\sqrt{NT}}\sum_{i=1}^N \left(\WP \+\alpha_{G_i} - \*G_i\left(\*F\right)\right)'\left(\*M_{\widetilde{P}} - \*M_{\widehat{P}}\right)\left(\WP \+\alpha_{g_i} - \*g_i\left(\*F\right)\right). 
\end{align}

%%%%%%%%%%%%%%%%%%%%%%%%%%%%%%%%%%
%%%%%%%%%%%%%%%% D_5 %%%%%%%%%%%%%
%%%%%%%%%%%%%%%%%%%%%%%%%%%%%%%%%%
We now proceed to evaluate each of the terms appearing above, starting with $\*D_5$. Analogous to (A.12) in \citet{westerlundpetrova2019}, we have
\begin{align}
\*M_{\widetilde P} - \*M_{\widehat P} &= \left(\WP - \TP\right)\left(\WP'\WP\right)^+\left(\WP - \TP\right)' + \left(\WP - \TP\right)\left(\WP'\WP\right)^+ \TP' \notag \\
& \quad +  \TP\left(\WP'\WP\right)^+\left(\WP - \TP\right)' + \TP\left[\left(\WP'\WP\right)^+ - \left(\TP'\TP\right)^+\right]\TP', \label{mpdiff}
\end{align}
from which it follows that
\begin{align}
-\*D_5 & \equiv \frac{1}{\sqrt{NT}}\sum_{i=1}^N \*V_i'\left(\*M_{\widetilde P} - \*M_{\widehat{P}}\right)\+\varepsilon_i \notag \\
	  &= \frac{1}{\sqrt{NT}}\sum_{i=1}^N \*V_i'\left(\WP - \TP\right)\left(\WP' \WP\right)^+\left(\WP - \TP\right)' \+\varepsilon_i +  \frac{1}{\sqrt{NT}}\sum_{i=1}^N \*V_i'\left(\WP - \TP\right)\left(\WP' \WP\right)^+ \TP'\+\varepsilon_i  \notag \\
	  &\quad +  \frac{1}{\sqrt{NT}}\sum_{i=1}^N \*V_i'\TP\left(\WP'\WP\right)^+\left(\WP - \TP\right)'\+\varepsilon_i +  \frac{1}{\sqrt{NT}}\sum_{i=1}^N \*V_i'\TP\left[\left(\WP'\WP\right)^+ - \left(\TP'\TP\right)^+\right]\TP'\+\varepsilon_i \notag \\
&\equiv \*d_{5,1} + \*d_{5,2} + \*d_{5,3} + \*d_{5,4}. \label{d5}
\end{align}
with an implicit definition of $\*d_{5,1}$, $\*d_{5,2}$, $\*d_{5,3}$ and $\*d_{5,4}$. In what remains of this appendix, implicit definitions like these will be used repeatedly to save space.

We start by considering the term $\*d_{5,2}$ in detail; the evaluation of the other terms is similar. By Exercise 10.20 in \citet{abadir2005}, $\left\|\mathrm{vec} \,\*A\right\|_{F}^2 = (\mathrm{vec}\, \*A)^\prime \mathrm{vec} \, \*A = \mathrm{tr}(\*A^\prime \*A) = \left\|\*A\right\|_{F}^2$. By using this and $\mathrm{vec}(\*A \*B \*C) = (\*C^\prime \otimes \*A)\mathrm{vec} \, \*B$ (see \citet{abadir2005}, Exercise 10.18), we obtain 
\begin{align}
    \left\|\*d_{5,2}\right\|_{F} &\equiv \left\|\frac{1}{\sqrt{NT}}\sum_{i=1}^N \*V_i'\left(\WP - \TP\right)\left(\WP' \WP\right)^+ \TP'\+\varepsilon_i\right\|_{F} \notag \\
    & =\left\|\mathrm{vec}\left(\frac{1}{\sqrt{NT}}\sum_{i=1}^N \*V_i'\left(\WP - \TP\right)\left(\WP' \WP\right)^+ \TP'\+\varepsilon_i\right)\right\|_{F} \notag \\
    &= \left\|\frac{1}{\sqrt{NT}}\sum_{i=1}^N (\+\varepsilon_i^\prime  \otimes \*V_i^\prime )\mathrm{vec}\left[\left(\WP - \TP\right)\left(\WP' \WP\right)^+ \TP'\right]\right\|_{F} \notag \\
    &= \left\|\frac{1}{\sqrt{NT}}\sum_{i=1}^N (\+\varepsilon_i \otimes \*V_i)^\prime \mathrm{vec}\left[\left(\WP - \TP\right)\left(\WP' \WP\right)^+ \TP'\right]\right\|_{F} \notag \\
    & \leq \left\|\frac{1}{\sqrt{N}T}\sum_{i=1}^N (\+\varepsilon_i \otimes \*V_i)\right\|_{F} \sqrt{T} \left\|\left(\WP - \TP\right)\left(\WP' \WP\right)^+ \TP'\right\|_{F}. \label{eq:d52}
\end{align} 
By Lemma \ref{appL1} and Corollary \ref{appC1}\ref{appC12}, it follows that
\begin{align}
\left\|\*d_{5,2}\right\|_{F}
= O_p(1)\sqrt{T}\,o_p\left(\frac{\sqrt K\zeta_1(K)}{\sqrt N}\right) = o_p\left(\sqrt{\frac{TK}{N}}\zeta_1(K)\right). \label{eq:d52rate}
\end{align}
The same argument applies to $\*d_{5,3}$. Indeed,
\begin{align}
\left\|\TP\left(\WP'\WP\right)^+\left(\WP - \TP\right)'\right\|_{F}
= \left\|\left[\left(\WP - \TP\right)\left(\WP'\WP\right)^+\TP'\right]'\right\|_{F}= \left\|\left(\WP - \TP\right)\left(\WP'\WP\right)^+\TP'\right\|_{F}.
\end{align}
Hence, by Lemma \ref{appL1} and Corollary \ref{appC1}\ref{appC12},
\begin{align}
\left\|\*d_{5,3}\right\|_{F}
&= o_p\left(\sqrt{\frac{TK}{N}}\zeta_1(K)\right). \label{eq:d53rate}
\end{align}

For $\*d_{5,4}$, we can again use exactly the same vectorization argument as for $\*d_{5,2}$ to obtain
\begin{align}
\left\|\*d_{5,4}\right\|_{F} &= \left\|\frac{1}{\sqrt{NT}}\sum_{i=1}^N \*V_i'\TP\left[\left(\WP'\WP\right)^+ - \left(\TP'\TP\right)^+\right]\TP'\+\varepsilon_i\right\|_{F} \notag \\
&\leq \left\|\frac{1}{\sqrt{N}T}\sum_{i=1}^N (\+\varepsilon_i \otimes \*V_i)\right\|_{F}\sqrt{T}\left\|\TP\left[\left(\WP'\WP\right)^+ - \left(\TP'\TP\right)^+\right]\TP'\right\|_{F}.
\end{align}
By Lemma \ref{appL1} and Corollary \ref{appC1}\ref{appC13}, it follows that
\begin{align}
\left\|\*d_{5,4}\right\|_{F}
= O_p(1)\sqrt{T}O_p\left(\frac{\zeta_1(K)}{\sqrt{N}}\right)=O_p\left(\sqrt{\frac{T}{N}}\zeta_1(K)\right). \label{eq:d54rate}
\end{align}

Conducting a similar exercise for $\*d_{5,1}$ yields
\begin{align}
\left\|\*d_{5,1}\right\|_{F} &\equiv \left\|\frac{1}{\sqrt{NT}}\sum_{i=1}^N \*V_i'\left(\WP - \TP\right)\left(\WP' \WP\right)^+\left(\WP - \TP\right)'\+\varepsilon_i\right\|_{F} \notag \\
&\leq \left\|\frac{1}{\sqrt{N}T}\sum_{i=1}^N (\+\varepsilon_i \otimes \*V_i)\right\|_{F}\sqrt{T}\left\|\left(\WP - \TP\right)\left(\WP' \WP\right)^+\left(\WP - \TP\right)'\right\|_{F}.
\end{align}
By Lemma \ref{appL1} and Corollary \ref{appC1}\ref{appC11}, we obtain
\begin{align}
\left\|\*d_{5,1}\right\|_{F}
= O_p(1)\sqrt{T}O_p\left(\frac{\zeta_1(K)^2}{N}\right)= O_p\left(\frac{\sqrt{T}}{N}\zeta_1(K)^2\right). \label{eq:d51rate}
\end{align}

The final rate for $\*D_5$ is now given by
\begin{align}
\left\|\*D_5\right\|_{F} &= \left\|\frac{1}{\sqrt{NT}}\sum_{i=1}^N \*V_i'\left(\*M_{\widetilde P} - \*M_{\widehat{P}}\right)\+\varepsilon_i\right\|_{F} \notag \\
&= O_p\left(\frac{\sqrt{T}}{N}\zeta_1(K)^2\right) + O_p\left(\sqrt{\frac{T}{N}}\zeta_1(K)\right) + o_p\left(\sqrt{\frac{TK}{N}}\zeta_1(K)\right) \notag \\
&=o_p\left(\sqrt{\frac{TK}{N}}\zeta_1(K)\right)=o_p(1),
\end{align}
To verify the final equality, Assumption \ref{A6}\ref{A6c} gives
\begin{align}
\sqrt{\frac{TK}{N}}\zeta_1(K)
&=\frac{1}{\sqrt K\,\zeta_1(K)}\sqrt{\frac TN}K\zeta_1(K)^2\to0. \label{eq:proxyGramScales}
\end{align}

%%%%%%%%%%%%%%%%%%%%%%%%%%%%%%%%%%
%%%%%%%%%%%%%%%% D_6 %%%%%%%%%%%%%
%%%%%%%%%%%%%%%%%%%%%%%%%%%%%%%%%%

We now move to $\mathbf{D}_{6},$ which we
again expand using \eqref{mpdiff};
\begin{align}
\mathbf{D}_{6} & =\frac{1}{\sqrt{NT}}\sum_{i=1}^{N}\mathbf{V}_{i}'(\mathbf{M}_{\widetilde{P}}-\mathbf{M}_{\widehat{P}})\left(\widehat{\mathbf{P}}\+\alpha_{g_i}-\*g_i(\mathbf{F})\right)\nonumber \\
 & =\frac{1}{\sqrt{NT}}\sum_{i=1}^{N}\mathbf{V}_{i}'\left(\widehat{\mathbf{P}}-\widetilde{\mathbf{P}}\right)\left(\widehat{\mathbf{P}}'\widehat{\mathbf{P}}\right)^{+}\left(\widehat{\mathbf{P}}-\widetilde{\mathbf{P}}\right)'\left(\widehat{\mathbf{P}}\+\alpha_{g_i}-\*g_i(\mathbf{F})\right)\nonumber \\
 & \quad  +\frac{1}{\sqrt{NT}}\sum_{i=1}^{N}\mathbf{V}_{i}'\left(\widehat{\mathbf{P}}-\widetilde{\mathbf{P}}\right)\left(\widehat{\mathbf{P}}'\widehat{\mathbf{P}}\right)^{+}\widetilde{\mathbf{P}}'\left(\widehat{\mathbf{P}}\+\alpha_{g_i}-\*g_i(\mathbf{F})\right)\nonumber \\
 & \quad  +\frac{1}{\sqrt{NT}}\sum_{i=1}^{N}\mathbf{V}_{i}'\widetilde{\mathbf{P}}\left(\widehat{\mathbf{P}}'\widehat{\mathbf{P}}\right)^{+}\left(\widehat{\mathbf{P}}-\widetilde{\mathbf{P}}\right)'\left(\widehat{\mathbf{P}}\+\alpha_{g_i}-\*g_i(\mathbf{F})\right)\nonumber \\
 & \quad  +\frac{1}{\sqrt{NT}}\sum_{i=1}^{N}\mathbf{V}_{i}'\widetilde{\mathbf{P}}\left[\left(\widehat{\mathbf{P}}'\widehat{\mathbf{P}}\right)^{+}-\left(\widetilde{\mathbf{P}}'\widetilde{\mathbf{P}}\right)^{+}\right]\widetilde{\mathbf{P}}'\left(\widehat{\mathbf{P}}\+\alpha_{g_i}-\*g_i(\mathbf{F})\right)\nonumber \\
 & \equiv \*d_{6,1}+ \*d_{6,2} + \*d_{6,3} + \*d_{6,4}.\label{eq:D6}
\end{align}
As before, we proceed by evaluating each of the four terms on the
right-hand side of \eqref{eq:D6}. In light of the decomposition in
\eqref{pag}, it is important to note that each of these four terms actually comprises two terms, as each component
in $\left(\widehat{\mathbf{P}}\+\alpha_{g_i}-\*g_i(\mathbf{F})\right)$
must be evaluated separately. We start by considering the second term in \eqref{eq:D6}:
\begin{align}
\*d_{6,2} & \equiv\frac{1}{\sqrt{NT}}\sum_{i=1}^{N}\mathbf{V}_{i}'\left(\widehat{\mathbf{P}}-\widetilde{\mathbf{P}}\right)\left(\widehat{\mathbf{P}}'\widehat{\mathbf{P}}\right)^{+}\widetilde{\mathbf{P}}'\left(\widehat{\mathbf{P}}\+\alpha_{g_i}-\*g_i(\mathbf{F})\right)\nonumber \\
 & =\frac{1}{\sqrt{NT}}\sum_{i=1}^{N}\mathbf{V}_{i}'\left(\widehat{\mathbf{P}}-\widetilde{\mathbf{P}}\right)\left(\widehat{\mathbf{P}}'\widehat{\mathbf{P}}\right)^{+}\widetilde{\mathbf{P}}'\left(\widehat{\mathbf{P}}-\widetilde{\mathbf{P}}\right)\+\alpha_{g_i}\nonumber \\
 & \quad +\frac{1}{\sqrt{NT}}\sum_{i=1}^{N}\mathbf{V}_{i}'\left(\widehat{\mathbf{P}}-\widetilde{\mathbf{P}}\right)\left(\widehat{\mathbf{P}}'\widehat{\mathbf{P}}\right)^{+}\widetilde{\mathbf{P}}'\left(\widetilde{\mathbf{P}}\+\alpha_{g_i}-\*g_i(\mathbf{F})\right) \notag \\
 &\equiv \*d^{(1)}_{6,2} + \*d^{(2)}_{6,2}. \label{eq:D62}
\end{align}
Both terms on the right can be treated the same way as $\frac{1}{\sqrt{NT}}\sum_{i=1}^{N}\mathbf{V}_{i}'\left(\widehat{\mathbf{P}}-\widetilde{\mathbf{P}}\right)\left(\widehat{\mathbf{P}}'\widehat{\mathbf{P}}\right)^{+}\widetilde{\mathbf{P}}'\boldsymbol{\varepsilon}_{i}.$
The treatment of the second term is actually almost identical by Assumptions \ref{A1} and \ref{A3}.
Hence, using the same arguments as for $\*d_{5,2}$ we get
\begin{align}
\left\|\*d^{(2)}_{6,2}\right\|_{F} &= \left\|\frac{1}{\sqrt{NT}}\sum_{i=1}^{N}\mathbf{V}_{i}'\left(\widehat{\mathbf{P}}-\widetilde{\mathbf{P}}\right)\left(\widehat{\mathbf{P}}'\widehat{\mathbf{P}}\right)^{+}\widetilde{\mathbf{P}}'\left(\widetilde{\mathbf{P}}\+\alpha_{g_i}-\*g_i(\mathbf{F})\right)\right\|_{F} \notag \\
& \leq \left\|\frac{1}{\sqrt{N}T}\sum_{i=1}^N \left(\left(\widetilde{\mathbf{P}}\+\alpha_{g_i}-\*g_i(\mathbf{F})\right) \otimes \*V_i\right)\right\|_{F} \sqrt{T}\left\|\left(\widehat{\mathbf{P}}-\widetilde{\mathbf{P}}\right)\left(\widehat{\mathbf{P}}'\widehat{\mathbf{P}}\right)^{+}\widetilde{\mathbf{P}}'\right\|_{F} \notag \\
&= O_p\left(K^{-\frac{\lambda_i}{m}}\right)\sqrt T\,o_p\left(\frac{\sqrt K\zeta_1(K)}{\sqrt N}\right) \notag \\
&=o_p\left(\sqrt{\frac{TK}{N}}K^{-\frac{\lambda_i}{m}}\zeta_1(K)\right). \label{eq:d622}
\end{align}

The analysis of the first term in \eqref{eq:D62}, $\*d_{6,2}^{(1)}$, is similar. We can follow the same steps as before to obtain
\begin{align}
\left\|\*d_{6,2}^{(1)}\right\|_{F} &= \left\|\frac{1}{\sqrt{NT}}\sum_{i=1}^{N}\mathbf{V}_{i}'\left(\widehat{\mathbf{P}}-\widetilde{\mathbf{P}}\right)\left(\widehat{\mathbf{P}}'\widehat{\mathbf{P}}\right)^{+}\widetilde{\mathbf{P}}'\left(\widehat{\mathbf{P}}-\widetilde{\mathbf{P}}\right)\+\alpha_{g_i}\right\|_{F} \notag \\
& \leq \left\|\frac{1}{\sqrt{NT}}\sum_{i=1}^N \+\alpha_{g_i} \otimes \*V_i\right\|_{F} \left\|\left(\widehat{\mathbf{P}}-\widetilde{\mathbf{P}}\right)\left(\widehat{\mathbf{P}}'\widehat{\mathbf{P}}\right)^{+}\widetilde{\mathbf{P}}'\left(\widehat{\mathbf{P}}-\widetilde{\mathbf{P}}\right)\right\|_{F} \notag \\
&= O_p(1)O_p\left(\frac{\sqrt T}{N}\zeta_1(K)^2\right) \notag \\
&= O_p\left(\frac{\sqrt T}{N}\zeta_1(K)^2\right), \label{eq:d621}
\end{align} which holds by Lemma \ref{appL3} and Corollary \ref{appC1}\ref{appC14}. 

Consequently, the final bound for $\left\|\*d_{6,2}\right\|_{F}$ is given by combining the results in \eqref{eq:d621} and \eqref{eq:d622}, such that 
\begin{align}
&\left\|\*d_{6,2}\right\|_{F} \leq \left\|\*d^{(1)}_{6,2}\right\|_{F} + \left\|\*d^{(2)}_{6,2}\right\|_{F} \notag \\
&=O_p\left(\frac{\sqrt T}{N}\zeta_1(K)^2\right)+o_p\left(\sqrt{\frac{TK}{N}}K^{-\frac{\lambda_i}{m}}\zeta_1(K)\right). \label{eq:d62rate}
\end{align} The population approximation part of $\*d_{6,3}$ has the same bound, since the relevant projection matrix is transposed and its Frobenius norm is unchanged. For its proxy part, the trace and eigenvalue bounds give
\begin{align}
\left\|\TP(\WP'\WP)^+(\WP-\TP)'(\WP-\TP)\right\|_{F}
&\leq\sqrt T\,\lambda_{\max}\left(\frac{\TP'\TP}{T}\right)^{\frac12}
\lambda_{\min}\left(\frac{\WP'\WP}{T}\right)^{-1}\notag\\
&\quad\times\left\|\frac{\WP-\TP}{\sqrt T}\right\|_{F}^2
=O_p\left(\frac{\sqrt T}{N}\zeta_1(K)^2\right).
\end{align}
Multiplying by the same $O_p(1)$ Kronecker-sum factor as in \eqref{eq:d621} gives $O_p(\tfrac{\sqrt{T}\,\zeta_1(K)^2}{N})$. Thus the bound in \eqref{eq:d62rate} also applies to $\*d_{6,3}$.

For $\*d_{6,1}$, the same decomposition as in \eqref{eq:D62} gives
\begin{align}
\left\|\*d_{6,1}\right\|_{F}
&\leq
\left\|\frac{1}{\sqrt{NT}}\sum_{i=1}^N \+\alpha_{g_i}\otimes \*V_i\right\|_{F}
\left\|(\WP-\TP)(\WP'\WP)^+(\WP-\TP)'(\WP-\TP)\right\|_{F} \notag \\
&\quad+
\left\|\frac{1}{\sqrt{N}T}\sum_{i=1}^N \left[\left(\TP\+\alpha_{g_i}-\*g_i(\*F)\right)\otimes \*V_i\right]\right\|_{F}\sqrt{T}
\left\|(\WP-\TP)(\WP'\WP)^+(\WP-\TP)'\right\|_{F} \notag \\
&=
O_p\left(\frac{\sqrt{T}}{N^{\frac{3}{2}}}\zeta_1(K)^3\right)
+
O_p\left(\frac{\sqrt{T}}{N}K^{-\frac{\lambda_i}{m}}\zeta_1(K)^2\right),
\label{eq:d61rate}
\end{align}
where the last equality follows from Lemmas \ref{appL2}--\ref{appL3} and Corollary \ref{appC1}\ref{appC11}, \ref{appC1}\ref{appC15}.

For $\*d_{6,4}$, the same decomposition gives
\begin{align}
\left\|\*d_{6,4}\right\|_{F}
&\leq
\left\|\frac{1}{\sqrt{NT}}\sum_{i=1}^N \+\alpha_{g_i}\otimes \*V_i\right\|_{F}
\left\|\TP\left[\left(\WP'\WP\right)^+-\left(\TP'\TP\right)^+\right]\TP'(\WP-\TP)\right\|_{F} \notag \\
&\quad+
\left\|\frac{1}{\sqrt{N}T}\sum_{i=1}^N \left[\left(\TP\+\alpha_{g_i}-\*g_i(\*F)\right)\otimes \*V_i\right]\right\|_{F}\sqrt{T}
\left\|\TP\left[\left(\WP'\WP\right)^+-\left(\TP'\TP\right)^+\right]\TP'\right\|_{F} \notag \\
&\leq
O_p(1)
O_p\left(\frac{\zeta_1(K)}{\sqrt{N}}\right)
\sqrt{T}O_p\left(\frac{\zeta_1(K)}{\sqrt{N}}\right)
+
O_p\left(K^{-\frac{\lambda_i}{m}}\right)\sqrt{T}
O_p\left(\frac{\zeta_1(K)}{\sqrt{N}}\right) \notag \\
&=
O_p\left(\frac{\sqrt{T}}{N}\zeta_1(K)^2\right)
+
O_p\left(\sqrt{\frac{T}{N}}K^{-\frac{\lambda_i}{m}}\zeta_1(K)\right),
\label{eq:d64rate}
\end{align}
where we used Lemmas \ref{appL2}--\ref{appL3}, Lemma \ref{appL4}\ref{appL42}, and Corollary \ref{appC1}\ref{appC13}.

Combining \eqref{eq:d61rate}, \eqref{eq:d62rate},  and \eqref{eq:d64rate}, we obtain
\begin{align}
\left\|\*D_6\right\|_{F}
&=O_p\left(\frac{\sqrt T}{N}\zeta_1(K)^2\right)
+O_p\left(\sqrt{\frac TN}K^{-\frac{\lambda_i}{m}}\zeta_1(K)\right)
+o_p\left(\sqrt{\frac{TK}{N}}K^{-\frac{\lambda_i}{m}}\zeta_1(K)\right) \notag \\
&=O_p\left(\frac{\sqrt T}{N}\zeta_1(K)^2\right)
+o_p\left(\sqrt{\frac{TK}{N}}K^{-\frac{\lambda_i}{m}}\zeta_1(K)\right)=o_p(1),
\end{align} where the final equality follows from Assumption \ref{A6}\ref{A6c} and \eqref{eq:proxyGramScales}.

%%%%%%%%%%%%%%%%%%%%%%%%%%%%%%%%%%
%%%%%%%%%%%%%%%% D_7 %%%%%%%%%%%%%
%%%%%%%%%%%%%%%%%%%%%%%%%%%%%%%%%%
The term $\*D_7$ can be handled in the same way. Indeed, using the decomposition in \eqref{eq:PalphaG} and applying the same vectorization argument as for $\*D_6$, with $\+\varepsilon_i$ replacing $\*V_i$ and $\+\alpha_{G_i}$ replacing $\+\alpha_{g_i}$, we obtain the same rate. The moment argument in Lemma \ref{appL2} also applies with $\+\varepsilon_i$ in place of $\*V_i$, by Assumptions \ref{A1} and \ref{A3}, and the coefficient bound follows from Lemma \ref{appL3} with $\*G_i(\cdot)\in\widetilde{\mathcal S}$. Hence, $\left\|\*D_7\right\|_{F} =o_p(1)$.

%%%%%%%%%%%%%%%%%%%%%%%%%%%%%%%%%%
%%%%%%%%%%%%%%%% D_8 %%%%%%%%%%%%%
%%%%%%%%%%%%%%%%%%%%%%%%%%%%%%%%%%
We now consider $\*D_8$. Instead of expanding all terms separately, we use a direct norm bound. By Corollary \ref{appC2}, for $\widetilde{\*g}_i(\cdot)\in\widetilde{\mathcal S}$,
\begin{align}
\left\|\WP\+\alpha_{\widetilde{\*g}_i}-\widetilde{\*g}_i(\*F)\right\|_{F} = O_p\left(\sqrt{\frac{T}{N}}\zeta_1(K)\right) + O_p\left(\sqrt{T}K^{-\frac{\lambda_i}{m}}\right). \label{eq:approxfull}
\end{align}
Moreover, by \eqref{mpdiff} and Corollary \ref{appC1}\ref{appC11}--\ref{appC13},
\begin{align}
\left\|\*M_{\widetilde P}-\*M_{\widehat P}\right\|_{F} &= O_p\left(\frac{\zeta_1(K)^2}{N}\right)+o_p\left(\frac{\sqrt K\zeta_1(K)}{\sqrt N}\right)+O_p\left(\frac{\zeta_1(K)}{\sqrt N}\right) \notag\\
&=o_p\left(\frac{\sqrt K\zeta_1(K)}{\sqrt N}\right)=o_p(1). \label{eq:Mdiffbound}
\end{align}
Indeed, by Assumption \ref{A6}\ref{A6c},
\begin{align}
\frac KN\zeta_1(K)^2
&=\frac{1}{\sqrt{NT}}\sqrt{\frac TN}K\zeta_1(K)^2\to0.\notag
\end{align} and thus, $\frac{\sqrt K\zeta_1(K)}{\sqrt N}\to0$. 
Hence, combining \eqref{eq:approxfull} and \eqref{eq:Mdiffbound} and applying Cauchy-Schwarz inequality, we get 
\begin{align}
\left\|\*D_8\right\|_{F}
&\leq \frac{1}{\sqrt{NT}}\sum_{i=1}^N \left\|\WP\+\alpha_{G_i}-\*G_i(\*F)\right\|_{F}\left\|\*M_{\widetilde P}-\*M_{\widehat P}\right\|_{F}\left\|\WP\+\alpha_{g_i}-\*g_i(\*F)\right\|_{F} \notag \\
&= \frac{1}{\sqrt{NT}}\left\|\*M_{\widetilde P}-\*M_{\widehat P}\right\|_{F}\sum_{i=1}^N \left\|\WP\+\alpha_{G_i}-\*G_i(\*F)\right\|_{F}\left\|\WP\+\alpha_{g_i}-\*g_i(\*F)\right\|_{F} \notag \\
&\leq \sqrt{NT}\left\|\*M_{\widetilde P}-\*M_{\widehat P}\right\|_{F}\left(\frac{1}{N}\sum_{i=1}^N \left\|\frac{1}{\sqrt{T}} \left(\WP\+\alpha_{G_i}-\*G_i(\*F)\right)\right\|_{F}^2\right)^{\frac{1}{2}} \notag \\
& \quad \times \left(\frac{1}{N}\sum_{i=1}^N \left\|\frac{1}{\sqrt{T}} \left(\WP\+\alpha_{g_i}-\*g_i(\*F)\right)\right\|_{F}^2\right)^{\frac{1}{2}} \notag \\ 
& = \sqrt{NT}\,o_p\left(\frac{\sqrt K\zeta_1(K)}{\sqrt N}\right) \left[O_p\left(\frac{\zeta_1(K)^2}{N}\right)+O_p\left(K^{-\frac{2\lambda_i}{m}}\right)\right] = o_p(1),
\end{align} where the final equality follows from Assumption \ref{A6}\ref{A6c}.

%%%%%%%%%%%%%%%%%%%%%%%%%%%%%%%%%%
%%%%%%%%%%%%%%%% D_2 %%%%%%%%%%%%%
%%%%%%%%%%%%%%%%%%%%%%%%%%%%%%%%%%
Let us now consider $\*D_2$, which we can expand similarly to \eqref{eq:D62}. For the population approximation term, the second-moment argument underlying Lemma \ref{appL2}, together with Assumptions \ref{A1} and \ref{A3}, gives
\begin{align}
&\be\left(\left\|\frac{1}{\sqrt{NT}}\sum_{i=1}^N\*V_i'\*M_{\widetilde P}\left(\TP\+\alpha_{g_i}-\*g_i(\*F)\right)\right\|_{F}^2\right) \notag\\
&\quad\leq\frac{C}{NT}\sum_{i=1}^N\be\left(\left\|\*M_{\widetilde P}\left(\TP\+\alpha_{g_i}-\*g_i(\*F)\right)\right\|_{F}^2\right) \notag\\
&\quad\leq\frac{C}{NT}\sum_{i=1}^N\be\left(\left\|\TP\+\alpha_{g_i}-\*g_i(\*F)\right\|_{F}^2\right)
=O\left(K^{-\frac{2\lambda_i}{m}}\right). \label{eq:D2approxmoment}
\end{align}
Here the cross-sectional terms vanish by independence and zero means, the covariance eigenvalues are bounded by Assumption \ref{A1}, and the last equality uses \eqref{eq:proxymeansquare}. Markov's inequality therefore gives an $O_p(K^{-\frac{\lambda_i}{m}})$ bound for this term. Using vectorization for the remaining term, as for $\*d_{6,2}^{(1)}$, we obtain
\begin{align}
\left\|\*D_2\right\|_{F}
&\leq \left\|\frac{1}{\sqrt{NT}}\sum_{i=1}^N\*V_i'\*M_{\widetilde P}\left(\TP\+\alpha_{g_i}-\*g_i(\*F)\right)\right\|_{F} +\left\|\frac{1}{\sqrt{NT}}\sum_{i=1}^N\*V_i'\*M_{\widetilde P}(\WP-\TP)\+\alpha_{g_i}\right\|_{F} \notag\\
&\leq O_p\left(K^{-\frac{\lambda_i}{m}}\right)
+\left\|\frac{1}{\sqrt{NT}}\sum_{i=1}^N\+\alpha_{g_i}\otimes\*V_i\right\|_{F}\sqrt T\,\lmax(\*M_{\widetilde P})\left\|\frac{1}{\sqrt T}(\WP-\TP)\right\|_{F} \notag\\
&=O_p\left(K^{-\frac{\lambda_i}{m}}\right)+O_p(1)\sqrt T\,O_p\left(\frac{\zeta_1(K)}{\sqrt N}\right) \notag\\
&=O_p\left(K^{-\frac{\lambda_i}{m}}\right)+O_p\left(\sqrt{\frac{T}{N}}\zeta_1(K)\right)=o_p(1).
\end{align}
Since $\*M_{\widetilde P}$ is symmetric and idempotent, its eigenvalues are zero or one, so $\lmax(\*M_{\widetilde P})\leq1$ and $\|\*M_{\widetilde P}\*A\|_{F}\leq\lmax(\*M_{\widetilde P})\|\*A\|_{F}$ for any conformable matrix $\*A$. The last two lines follow from Lemmas \ref{appL3} and \ref{appL4}\ref{appL42}, and Assumption \ref{A6}\ref{A6c}.

%%%%%%%%%%%%%%%%%%%%%%%%%%%%%%%%%%
%%%%%%%%%%%%%%%% D_3 %%%%%%%%%%%%%
%%%%%%%%%%%%%%%%%%%%%%%%%%%%%%%%%%
$\*D_3$ has the same basic structure as $\*D_2$. Indeed,
\begin{align}
-\*D_3 &\equiv \frac{1}{\sqrt{NT}}\sum_{i=1}^N\left(\WP\+\alpha_{G_i}-\*G_i(\*F)\right)'\*M_{\widetilde P}\+\varepsilon_i \notag\\
&=\frac{1}{\sqrt{NT}}\sum_{i=1}^N\+\alpha_{G_i}'(\WP-\TP)'\*M_{\widetilde P}\+\varepsilon_i +\frac{1}{\sqrt{NT}}\sum_{i=1}^N\left(\TP\+\alpha_{G_i}-\*G_i(\*F)\right)'\*M_{\widetilde P}\+\varepsilon_i \notag\\
&\equiv\*d_{3,1}+\*d_{3,2}.
\end{align}
Using the same vectorization argument as for $\*D_2$ and Lemma \ref{appL3} with $\+\varepsilon_i$ in place of $\*V_i$, we obtain
\begin{align}
\left\|\*d_{3,1}\right\|_{F}
&\leq\left\|\frac{1}{\sqrt{NT}}\sum_{i=1}^N\+\alpha_{G_i}\otimes\+\varepsilon_i\right\|_{F}\sqrt T\,\lmax(\*M_{\widetilde P})\left\|\frac{1}{\sqrt T}(\WP-\TP)\right\|_{F} \notag\\
&=O_p(1)\sqrt T\,O_p\left(\frac{\zeta_1(K)}{\sqrt N}\right)
=O_p\left(\sqrt{\frac{T}{N}}\zeta_1(K)\right).
\end{align}
The same second-moment argument as in \eqref{eq:D2approxmoment}, with $\*G_i(\*F)$ and $\+\varepsilon_i$ in place of $\*g_i(\*F)$ and $\*V_i$, gives $\be(\|\*d_{3,2}\|_{F}^2)=O(K^{-\frac{2\lambda_i}{m}})$. Hence, $\|\*d_{3,2}\|_{F}=O_p(K^{-\frac{\lambda_i}{m}})$ by Markov's inequality. Therefore,
\begin{align}
\left\|\*D_3\right\|_{F}
&\leq\left\|\*d_{3,1}\right\|_{F}+\left\|\*d_{3,2}\right\|_{F} \notag\\
&=O_p\left(\sqrt{\frac{T}{N}}\zeta_1(K)\right)+O_p\left(K^{-\frac{\lambda_i}{m}}\right)=o_p(1),
\end{align}
where the last equality follows from Assumption \ref{A6}\ref{A6c}.

%%%%%%%%%%%%%%%%%%%%%%%%%%%%%%%%%%
%%%%%%%%%%%%%%%% D_4 %%%%%%%%%%%%%
%%%%%%%%%%%%%%%%%%%%%%%%%%%%%%%%%%
We continue with $\*D_4$. Using Cauchy--Schwarz, Corollary \ref{appC2}, and $\lambda_{\max}(\*M_{\widetilde P})\leq1$, we obtain
\begin{align}
\left\|\*D_4\right\|_{F} &\equiv \left\|\frac{1}{\sqrt{NT}}\sum_{i=1}^{N}\left(\widehat{\mathbf{P}}\+\alpha_{G_i}-\*G_i(\mathbf{F})\right)'\mathbf{M}_{\widetilde{P}}\left(\widehat{\mathbf{P}}\+\alpha_{g_i}-\*g_i(\mathbf{F})\right)\right\|_{F} \notag \\ 
&\leq \frac{1}{\sqrt{NT}}\sum_{i=1}^{N}\left\|\WP\+\alpha_{G_i}-\*G_i(\*F)\right\|_{F}\lambda_{\max}(\*M_{\widetilde P})\left\|\WP\+\alpha_{g_i}-\*g_i(\*F)\right\|_{F} \notag \\
&\leq \sqrt{NT}\left(\frac{1}{N}\sum_{i=1}^{N}\left\|\frac{1}{\sqrt{T}}\left(\WP\+\alpha_{G_i}-\*G_i(\*F)\right)\right\|_{F}^2\right)^{\frac{1}{2}}\left(\frac{1}{N}\sum_{i=1}^{N}\left\|\frac{1}{\sqrt{T}}\left(\WP\+\alpha_{g_i}-\*g_i(\*F)\right)\right\|_{F}^2\right)^{\frac{1}{2}} \notag \\
&= \sqrt{NT}\left[O_p\left(\frac{\zeta_1(K)}{\sqrt{N}}\right)+O_p\left(K^{-\frac{\lambda_i}{m}}\right)\right]^2 \notag \\
&= O_p\left(\frac{\sqrt{T}}{\sqrt{N}}\zeta_1(K)^2\right)+O_p\left(\sqrt{T}K^{-\frac{\lambda_i}{m}}\zeta_1(K)\right)+O_p\left(\sqrt{NT}K^{-\frac{2\lambda_i}{m}}\right)=o_p(1),
\end{align} where the final equality follows from Assumption \ref{A6}\ref{A6c}. 

%%%%%%%%%%%%%%%%%%%%%%%%%%%%%%%%%%
%%%%%%%%%%%%%%%% D_1 %%%%%%%%%%%%%
%%%%%%%%%%%%%%%%%%%%%%%%%%%%%%%%%%
For $\*D_1$, we have
\begin{align}
\*D_1 \equiv \frac{1}{\sqrt{NT}}\sum_{i=1}^{N}\*V_i'\*M_{\widetilde P}\+\varepsilon_i = \frac{1}{\sqrt{NT}}\sum_{i=1}^{N}\*V_i'\+\varepsilon_i-\frac{1}{\sqrt{NT}}\sum_{i=1}^{N}\*V_i'\TP(\TP'\TP)^+\TP'\+\varepsilon_i \equiv \*d_{1,1}+\*d_{1,2}.
\end{align}
Consider $\*d_{1,2}$. By Assumptions \ref{A1} and \ref{A3},
\begin{align}
\be\left(\|\*d_{1,2}\|_{F}^2\right)
&= \be\left(\left\Vert \frac{1}{\sqrt{NT}}\sum_{i=1}^{N}\*V_i'\TP(\TP'\TP)^+\TP'\+\varepsilon_i\right\Vert_{F}^2\right) \notag \\
&= \frac{1}{NT}\sum_{i=1}^{N}\be\left[\mathrm{tr}\left(\+\varepsilon_i'\TP(\TP'\TP)^+\TP'\*V_i\*V_i'\TP(\TP'\TP)^+\TP'\+\varepsilon_i\right)\right] \notag \\
&\leq \frac{1}{NT}\sum_{i=1}^{N}\lmax\left(\+\Omega_{v,i}\right)\lmax\left(\+\Omega_{\varepsilon,i}\right)\be\left[\mathrm{tr}\left((\TP'\TP)^+\TP'\TP(\TP'\TP)^+\TP'\TP\right)\right] \notag \\
&= \frac{1}{NT}\sum_{i=1}^{N}\lmax\left(\+\Omega_{v,i}\right)\lmax\left(\+\Omega_{\varepsilon,i}\right)\be\left[\rank(\TP)\right] \notag \\
&\leq \frac{K}{NT}\sum_{i=1}^{N}\lmax\left(\+\Omega_{v,i}\right)\lmax\left(\+\Omega_{\varepsilon,i}\right) = O\left(\frac{K}{T}\right). \label{eq:d12var}
\end{align}
Here $(\TP'\TP)^+\TP'\TP$ is an orthogonal projector with trace $\rank(\TP)\leq K$, whether or not $\TP$ has full column rank.
Hence, under Assumption \ref{A6}\ref{A6c}, $\left\|\*d_{1,2}\right\|_{F}=O_p\left(\sqrt{\frac{K}{T}}\right)=o_p(1)$, implying 
\begin{align}
\mathbf{D}_{1} = \frac{1}{\sqrt{NT}}\sum_{i=1}^{N}\mathbf{V}_{i}'\boldsymbol{\varepsilon}_{i}+o_p(1). 
\end{align}

We have shown that $\*D_2, \dots, \*D_8$ are negligible. Therefore, \eqref{Dj} reduces to
\begin{align}
\frac{1}{\sqrt{NT}}\sum_{i=1}^N\*X_i'\*M_{\widehat{P}}\left[\+\varepsilon_i - \left(\WP\+\alpha_{g_i} - \*g_i\left(\*F\right)\right) \right] = \sum_{j=1}^8 \*D_j = \frac{1}{\sqrt{NT}}\sum_{i=1}^{N}\*V_i'\+\varepsilon_i + o_p(1). \label{clean}
\end{align}

%%%%%%%%%%%%%%%%%%%%%%%%%%%%%%%%%%
    %%%%%% DENOMINATOR %%%%%%
%%%%%%%%%%%%%%%%%%%%%%%%%%%%%%%%%%
Let us now consider the denominator, which we expand in the usual way:
\begin{align}
\frac{1}{NT}\sum_{i=1}^{N} & \mathbf{X}_{i}'\mathbf{M}_{\widehat{P}}\mathbf{X}_{i}=\frac{1}{NT}\sum_{i=1}^{N}\left[\mathbf{V}_{i}-\left(\widehat{\mathbf{P}}\+\alpha_{G_i}-\*G_i(\mathbf{F})\right)\right]'\mathbf{M}_{\widehat{P}}\left[\mathbf{V}_{i}-\left(\widehat{\mathbf{P}}\+\alpha_{G_i}-\*G_i(\mathbf{F})\right)\right]\nonumber \\
&=\frac{1}{NT}\sum_{i=1}^{N}\left[\mathbf{V}_{i}-\left(\widehat{\mathbf{P}}\+\alpha_{G_i}-\*G_i(\mathbf{F})\right)\right]'\mathbf{M}_{\widetilde{P}}\left[\mathbf{V}_{i}-\left(\widehat{\mathbf{P}}\+\alpha_{G_i}-\*G_i(\mathbf{F})\right)\right]\nonumber \\
&\quad-\frac{1}{NT}\sum_{i=1}^{N}\left[\mathbf{V}_{i}-\left(\widehat{\mathbf{P}}\+\alpha_{G_i}-\*G_i(\mathbf{F})\right)\right]'\left(\mathbf{M}_{\widetilde{P}}-\mathbf{M}_{\widehat{P}}\right)\left[\mathbf{V}_{i}-\left(\widehat{\mathbf{P}}\+\alpha_{G_i}-\*G_i(\mathbf{F})\right)\right].\label{eq:xmx}
\end{align}

Most terms in \eqref{eq:xmx} are either already controlled above or can be bounded by the same arguments. In particular, all terms involving $\WP\+\alpha_{G_i}-\*G_i(\*F)$ are $o_p(1)$ by Cauchy--Schwarz, Corollary \ref{appC2}, $\lambda_{\max}(\*M_{\widetilde P})\leq1$, and \eqref{eq:Mdiffbound}. The only remaining terms are
\begin{align}
\*b_1 &\equiv \frac{1}{NT}\sum_{i=1}^{N}\mathbf{V}_{i}'\mathbf{M}_{\widetilde{P}}\mathbf{V}_{i}, \\
\*b_2 &\equiv \frac{1}{NT}\sum_{i=1}^{N}\mathbf{V}_{i}'\left(\mathbf{M}_{\widetilde{P}}-\mathbf{M}_{\widehat{P}}\right)\mathbf{V}_{i}.
\end{align}

Consider $\*b_2$. By \eqref{eq:Mdiffbound} and Assumptions \ref{A1}, \ref{A3}, and \ref{A6}\ref{A6c}, we get
\begin{align}
\left\|\*b_2\right\|_{F} &= \left\|\frac{1}{NT}\sum_{i=1}^N \mathbf{V}_{i}'\left(\mathbf{M}_{\widetilde{P}}-\mathbf{M}_{\widehat{P}}\right)\mathbf{V}_{i}\right\|_{F} \leq \left(\frac{1}{NT}\sum_{i=1}^N \left\|\*V_i\right\|_{F}^2\right)\left\|\mathbf{M}_{\widetilde{P}}-\mathbf{M}_{\widehat{P}}\right\|_{F} \notag \\
&= O_p(1)o_p\left(\frac{\sqrt K\zeta_1(K)}{\sqrt N}\right) \notag \\
&=o_p\left(\frac{\sqrt K\zeta_1(K)}{\sqrt N}\right)=o_p(1).
\end{align}

For $\*b_1$, note that
\begin{align}
\*b_1 = \frac{1}{NT}\sum_{i=1}^{N}\mathbf{V}_{i}'\mathbf{M}_{\widetilde{P}}\mathbf{V}_{i}= \frac{1}{NT}\sum_{i=1}^{N}\mathbf{V}_{i}'\mathbf{V}_{i}-\frac{1}{NT}\sum_{i=1}^{N}\mathbf{V}_{i}'\TP(\TP^\prime \TP)^+\TP^\prime\mathbf{V}_{i}. \label{eq:b1decomp}
\end{align}
We next show that the second term in \eqref{eq:b1decomp} is negligible. By Assumptions \ref{A1} and \ref{A3},
\begin{align}
&\be\left[\left\|\frac{1}{NT}\sum_{i=1}^{N}\mathbf{V}_{i}'\TP(\TP^\prime \TP)^+\TP^\prime\mathbf{V}_{i}\right\|_{F}\right] \leq \frac{1}{NT}\sum_{i=1}^{N}\be\left[\mathrm{tr}\left(\mathbf{V}_{i}'\TP(\TP^\prime \TP)^+\TP^\prime\mathbf{V}_{i}\right)\right] \notag \\
&= \frac{1}{NT}\sum_{i=1}^{N}\be\left[\mathrm{tr}\left(\TP(\TP^\prime \TP)^+\TP^\prime\mathbf{V}_{i}\mathbf{V}_{i}'\right)\right] \leq \frac{1}{NT}\sum_{i=1}^{N}\lmax\left(\+\Omega_{v,i}\right)\be\left[\mathrm{tr}\left((\TP^\prime \TP)^+\TP^\prime\TP\right)\right] \notag \\
&= \frac{1}{NT}\sum_{i=1}^{N}\lmax\left(\+\Omega_{v,i}\right)\be\left[\rank(\TP)\right] \notag \\
&\leq \frac{K}{NT}\sum_{i=1}^{N}\lmax\left(\+\Omega_{v,i}\right) = O\left(\frac{K}{T}\right).
\end{align}
The trace equals $\rank(\TP)\leq K$ for every realization, so the expectation bound does not require full column rank to hold with probability one.
Hence, $\left\|\frac{1}{NT}\sum_{i=1}^{N}\mathbf{V}_{i}'\TP(\TP^\prime \TP)^+\TP^\prime\mathbf{V}_{i}\right\|_{F}=O_p\left(\frac{K}{T}\right)=o_p(1)$, where the last equality follows from Assumption \ref{A6}\ref{A6c}. Therefore,
\begin{align}
\*b_1 &= \frac{1}{NT}\sum_{i=1}^{N}\mathbf{V}_{i}'\mathbf{V}_{i}+o_p(1)=\+\Sigma_v+o_p(1),
\end{align}
where the last equality follows from a law of large numbers applied to the leading term, such that $\+\Sigma_v\equiv \lim_{N,T\to\infty}\frac{1}{NT}\sum_{i=1}^{N}\be\left(\*V_i'\*V_i\right)$.

Insertion into \eqref{eq:xmx} yields
\begin{align}
\frac{1}{NT}\sum_{i=1}^{N}\mathbf{X}_{i}'\mathbf{M}_{\widehat{P}}\mathbf{X}_{i}=\+\Sigma_v+o_p(1),
\end{align} which, together with \eqref{SCCE} and \eqref{clean},
leads to the following cleaned-up asymptotic representation for $\sqrt{NT}\left(\widehat{\boldsymbol{\beta}}_{SCCE}-\boldsymbol{\beta}\right)$:
\begin{align}
\sqrt{NT}\left(\widehat{\boldsymbol{\beta}}_{SCCE}-\boldsymbol{\beta}\right) & =\left(\frac{1}{NT}\sum_{i=1}^{N}\mathbf{X}_{i}'\mathbf{M}_{\widehat{P}}\mathbf{X}_{i}\right)^{-1}\frac{1}{\sqrt{NT}}\sum_{i=1}^{N}\mathbf{X}_{i}'\mathbf{M}_{\widehat{P}}\left[\boldsymbol{\varepsilon}_{i}-\left(\widehat{\mathbf{P}}\+\alpha_{g_i}-\*g_i(\mathbf{F})\right)\right]\nonumber \\
 & =\boldsymbol{\Sigma}_{v}^{-1}\frac{1}{\sqrt{NT}}\sum_{i=1}^{N}\mathbf{V}_{i}'\boldsymbol{\varepsilon}_{i}+o_{p}(1).
\end{align} In order to arrive at the sought asymptotic distribution, we apply
a central limit law for independent processes to $\frac{1}{\sqrt{NT}}\sum_{i=1}^{N}\mathbf{V}_{i}'\boldsymbol{\varepsilon}_{i}$. This gives
\begin{equation}
\frac{1}{\sqrt{NT}}\sum_{i=1}^{N}\mathbf{V}_{i}'\boldsymbol{\varepsilon}_{i}\overset{d}{\to}N(\mathbf{0}_{d\times1},\boldsymbol{\Theta})
\end{equation}
 as $N,T\to\infty$, where
\begin{align}
\boldsymbol{\Theta} & =\lim_{N,T\to\infty}\mathbb{E}\left[\frac{1}{\sqrt{NT}}\sum_{i=1}^{N}\mathbf{V}_{i}'\boldsymbol{\varepsilon}_{i}\left(\frac{1}{\sqrt{NT}}\sum_{i=1}^{N}\mathbf{V}_{i}'\boldsymbol{\varepsilon}_{i}\right)'\right]\nonumber \\
 & =\lim_{N,T\to\infty}\frac{1}{NT}\sum_{i=1}^{N}\sum_{j=1}^{N}\mathbb{E}\left[\mathbf{V}_{i}'\mathbb{E}\left(\boldsymbol{\varepsilon}_{i}\boldsymbol{\varepsilon}_{j}^\prime \right)\mathbf{V}_{j}\right] =\lim_{N,T\to\infty}\frac{1}{NT}\sum_{i=1}^{N}\mathbb{E}\left[\mathbf{V}_{i}'\mathbb{E}\left(\boldsymbol{\varepsilon}_{i}\boldsymbol{\varepsilon}_{i}'\right)\mathbf{V}_{i}\right]\nonumber \\
 & =\lim_{N,T\to\infty}\frac{1}{NT}\sum_{i=1}^{N}\mathbb{E}\left(\mathbf{V}_{i}'\boldsymbol{\Omega}_{\varepsilon,i}\mathbf{V}_{i}\right).
\end{align}

\no We can therefore show that under Assumptions \ref{A1}--\ref{A6},
\begin{equation}
\sqrt{NT}\left(\widehat{\boldsymbol{\beta}}_{SCCE}-\boldsymbol{\beta}\right)=\boldsymbol{\Sigma}_{v}^{-1}\frac{1}{\sqrt{NT}}\sum_{i=1}^{N}\mathbf{V}_{i}'\boldsymbol{\varepsilon}_{i}+o_{p}(1)\overset{d}{\to}\mathcal{N}(\mathbf{0}_{d\times1},\boldsymbol{\Sigma}_{v}^{-1}\+\Theta\boldsymbol{\Sigma}_{v}^{-1}),
\end{equation} as $N,T\to\infty.$ \hfill $\blacksquare$

\appendixsection{Additional Results}\label{AppB}
Appendix \ref{AppB} is organized as follows. The first subsection proves auxiliary Lemmas \ref{appL1}--\ref{appL4} and Corollaries \ref{appC1}--\ref{appC2}. Subsection \ref{AppB2} derives the admissible corridor discussed in Remark \ref{R2}. Subsections \ref{empadd} and \ref{mcadd} provide additional results from the empirical application and the Monte Carlo simulations, respectively.

\appendixsubsection{Proof of Auxiliary Lemmas and Corollaries}
%%%%%%%%%%%%%%%%%%%%%%%%%%%%%%%%%%%%%%%%%%%%
            %%%%%% LEMMA A1 %%%%%%
%%%%%%%%%%%%%%%%%%%%%%%%%%%%%%%%%%%%%%%%%%%%
\begin{restatedlemma}{appL1} 
Suppose Assumptions \ref{A1} and \ref{A3} hold. Then $\left\|\frac{1}{\sqrt{N}T} \sum_{i=1}^N \+\varepsilon_i \otimes \*V_i\right\|_{F} = O_p(1)$.
\end{restatedlemma}

\paragraph{\textbf{Proof}} Note how
\begin{align}
&\be\left(\left\|\frac{1}{\sqrt{N}T} \sum_{i=1}^N \+\varepsilon_i \otimes \*V_i\right\|_{F}^2\right) \notag \\
&= \be \left[\mathrm{tr}\left(\left(\frac{1}{\sqrt{N}T} \sum_{i=1}^N \+\varepsilon_i \otimes \*V_i\right)^\prime \left(\frac{1}{\sqrt{N}T} \sum_{j=1}^N \+\varepsilon_j \otimes \*V_j\right)\right)\right] \notag \\
&= \be \left[\mathrm{tr}\left(\frac{1}{NT^2} \sum_{i=1}^N \sum_{j=1}^N \left(\+\varepsilon_i^\prime\+\varepsilon_j\right) \otimes \left(\*V_i^\prime \*V_j\right) \right)\right] \notag \\
&= \mathrm{tr}\left[\frac{1}{NT^2}\sum_{i=1}^N \sum_{j=1}^N \be \left\{\left(\+\varepsilon_i^\prime\+\varepsilon_j\right) \otimes \left(\*V_i^\prime \*V_j\right)\right\}\right] \notag\\
&= \mathrm{tr}\left[\frac{1}{NT^2}\sum_{i=1}^N \sum_{j=1}^N \be \left(\+\varepsilon_i^\prime\+\varepsilon_j\right) \otimes \be \left(\*V_i^\prime \*V_j\right)\right] \notag \\
&= \mathrm{tr}\left[\frac{1}{NT^2}\sum_{i=1}^N  \be \left(\+\varepsilon_i^\prime\+\varepsilon_i\right) \otimes \be \left(\*V_i^\prime \*V_i\right)\right] \notag \\
&= \mathrm{tr}\left[\frac{1}{N}\sum_{i=1}^N \left(\sigma^2_{\varepsilon,i} \otimes \+\Sigma_{v,i}\right)\right] + o(1) \notag \\
& \leq \mathrm{tr}\left(\frac{1}{N}\sum_{i=1}^N\+\Sigma_{v,i}\right) \sup_{i=1, \dots, N} \sigma_{\varepsilon,i}^2 + o(1) \notag \\
& = \mathrm{tr}\left(\+\Sigma_v\right)\sup_{i=1, \dots, N} \sigma_{\varepsilon,i}^2 + o(1) = O(1),
\end{align}
which holds under Assumptions \ref{A1} and \ref{A3}. Hence, since the second moment is bounded, Markov's inequality gives
\begin{align}
    \left\|\frac{1}{\sqrt{N}T} \sum_{i=1}^N \+\varepsilon_i \otimes \*V_i\right\|_{F} = O_p(1).
\end{align} 
\hfill $\blacksquare$

%%%%%%%%%%%%%%%%%%%%%%%%%%%%%%%%%%%%%%%%%%%%
            %%%%%% LEMMA A2 %%%%%%
%%%%%%%%%%%%%%%%%%%%%%%%%%%%%%%%%%%%%%%%%%%%
\begin{restatedlemma}{appL2}
Suppose Assumptions \ref{A1}, \ref{A3}, \ref{A4}, and \ref{A5} hold and let $\widetilde{\*g}_i(\cdot) \in \widetilde{\mathcal{S}}$, where $\widetilde{\mathcal{S}}\equiv \{\*g_i(\cdot),\*G_i(\cdot)\}$. Then $\left\|\frac{1}{\sqrt{N}T} \sum_{i=1}^N \left(\TP\+\alpha_{\widetilde{\*g}_i} - \widetilde{\*g}_i(\*F)\right) \otimes \*V_i\right\|_{F} = O_p\left(K^{-\frac{\lambda_i}{m}}\right)$.
\end{restatedlemma}

\paragraph{\textbf{Proof}} Using the same steps as for the proof of Lemma \ref{appL1} and the fact that $\mathrm{tr}(\*A \otimes \*B) = \mathrm{tr}(\*A)\mathrm{tr}(\*B)$ (see \citet{abadir2005}, Exercise 10.7), we get
\begin{align}
&\be\left(\left\|\frac{1}{\sqrt{N}T} \sum_{i=1}^N \left(\TP\+\alpha_{\widetilde{\*g}_i} - \widetilde{\*g}_i(\*F)\right) \otimes \*V_i\right\|_{F}^2\right) \notag \\
&= \be \left[\mathrm{tr}\left(\left(\frac{1}{\sqrt{N}T} \sum_{i=1}^N \left(\TP\+\alpha_{\widetilde{\*g}_i} - \widetilde{\*g}_i(\*F)\right) \otimes \*V_i\right)^\prime \left(\frac{1}{\sqrt{N}T} \sum_{i=1}^N \left(\TP\+\alpha_{\widetilde{\*g}_i} - \widetilde{\*g}_i(\*F)\right) \otimes \*V_i\right)\right)\right] \notag \\
&= \be \left[\mathrm{tr}\left(\frac{1}{NT^2} \sum_{i=1}^N \sum_{j=1}^N \left(\TP\+\alpha_{\widetilde{\*g}_i} - \widetilde{\*g}_i(\*F)\right)^\prime\left(\TP\+\alpha_{\widetilde{\*g}_j} - \widetilde{\*g}_j(\*F)\right) \otimes \*V_i^\prime \*V_j \right)\right] \notag \\
&= \frac{1}{NT^2}\sum_{i=1}^N \be \left[\mathrm{tr}\left(\left(\TP\+\alpha_{\widetilde{\*g}_i} - \widetilde{\*g}_i(\*F)\right)^\prime\left(\TP\+\alpha_{\widetilde{\*g}_i} - \widetilde{\*g}_i(\*F)\right) \otimes \*V_i^\prime \*V_i\right)\right] \notag\\
&= \frac{1}{NT^2}\sum_{i=1}^N \be \left[\mathrm{tr}\left(\left(\TP\+\alpha_{\widetilde{\*g}_i} - \widetilde{\*g}_i(\*F)\right)^\prime\left(\TP\+\alpha_{\widetilde{\*g}_i} - \widetilde{\*g}_i(\*F)\right)\right)\mathrm{tr}\left(\*V_i^\prime \*V_i\right)\right] \notag \\
&= \frac{1}{N}\sum_{i=1}^N \be \left[\mathrm{tr}\left(\frac{1}{T}\left(\TP\+\alpha_{\widetilde{\*g}_i} - \widetilde{\*g}_i(\*F)\right)^\prime\left(\TP\+\alpha_{\widetilde{\*g}_i} - \widetilde{\*g}_i(\*F)\right)\right)\right]\mathrm{tr}\left[\be\left(\frac{1}{T}\*V_i^\prime \*V_i\right)\right] \notag \\
&\leq \frac{1}{N}\sum_{i=1}^N \be \left[\left\|\frac{1}{\sqrt{T}}\left(\TP\+\alpha_{\widetilde{\*g}_i} - \widetilde{\*g}_i(\*F)\right)\right\|_{F}^2\right]\left[\mathrm{tr}\left(\+\Sigma_{v,i}\right) + o(1)\right] \notag \\
& =\frac{1}{N}\sum_{i=1}^N O\left(K^{-\frac{2\lambda_i}{m}}\right)\left[O(1) + o(1)\right] \notag \\
& = O\left(K^{-\frac{2\lambda_i}{m}}\right), \label{eq:appL3trace}
\end{align} which uses Assumptions \ref{A1} and \ref{A3} and the mean-square approximation bound in \eqref{eq:proxymeansquare}, established from Assumptions \ref{A4} and \ref{A5} in the proof of Lemma \ref{L1}\ref{L12}. 

Therefore, by Markov's inequality,
\begin{align}
\left\|\frac{1}{\sqrt{N}T} \sum_{i=1}^N \left(\TP\+\alpha_{\widetilde{\*g}_i} - \widetilde{\*g}_i(\*F)\right) \otimes \*V_i\right\|_{F} = O_p\left(K^{-\frac{\lambda_i}{m}}\right).
\end{align}
\hfill $\blacksquare$

%%%%%%%%%%%%%%%%%%%%%%%%%%%%%%%%%%%%%%%%%%%%
            %%%%%% LEMMA A3 %%%%%%
%%%%%%%%%%%%%%%%%%%%%%%%%%%%%%%%%%%%%%%%%%%%

\begin{restatedlemma}{appL3}
Suppose Assumptions \ref{A1}, \ref{A3}, and \ref{A6}\ref{A6d} hold and let $\widetilde{\*g}_i(\cdot)\in\widetilde{\mathcal S}$, where $\widetilde{\mathcal S}\equiv\{\*g_i(\cdot),\*G_i(\cdot)\}$. Then $\left\|\frac{1}{\sqrt{NT}} \sum_{i=1}^N \+\alpha_{\widetilde{\*g}_i} \otimes \*V_i\right\|_{F} = O_p(1)$. The same bound holds with $\+\varepsilon_i$ in place of $\*V_i$.
\end{restatedlemma}

\paragraph{\textbf{Proof}} Independence across $i$ and the zero means in Assumption \ref{A1} imply that cross terms with $i\neq j$ vanish. Taking Frobenius norms and using $\mathrm{tr}(\*A\otimes\*B)=\mathrm{tr}(\*A)\mathrm{tr}(\*B)$ gives
\begin{align}
\be\left(\left\|\frac1{\sqrt{NT}}\sum_{i=1}^N\+\alpha_{\widetilde{\*g}_i}\otimes\*V_i\right\|_{F}^2\right)
&=\frac1{NT}\sum_{i=1}^N\|\+\alpha_{\widetilde{\*g}_i}\|_{F}^2\,\be\left(\|\*V_i\|_{F}^2\right) \notag\\
&\leq\frac{C}{N}\sum_{i=1}^N\|\+\alpha_{\widetilde{\*g}_i}\|_{F}^2=O(1), \label{eq:L4generic}
\end{align}
where the inequality uses the uniform moment bounds in Assumption \ref{A1}, and the last equality follows from Assumption \ref{A6}\ref{A6d}. Thus Markov's inequality gives the asserted $O_p(1)$ bound. The same argument applies with $\+\varepsilon_i$ in place of $\*V_i$.
\hfill $\blacksquare$

%%%%%%%%%%%%%%%%%%%%%%%%%%%%%%%%%%%%%%%%%%%%
            %%%%%% LEMMA A4 %%%%%%
%%%%%%%%%%%%%%%%%%%%%%%%%%%%%%%%%%%%%%%%%%%%

\begin{restatedlemma}{appL4}
Suppose Assumptions \ref{A1}--\ref{A3}, and \ref{A6}\ref{A6a}-\ref{A6c} hold. Then 
\begin{enumerate}[label=(\roman*)]
    \item $\left\|\frac{1}{\sqrt{T}}\left(\WP - \TP\right)\right\|_{F} = O_p\left(\frac{\zeta_1(K)}{\sqrt{N}}\right)$,
    \item  $\left\|\frac{1}{T}\WP^\prime \WP - \*I_K\right\|_{\mathrm{op}} = o_p(1)$.
\end{enumerate}
\end{restatedlemma}

\paragraph{\textbf{Proof}} Consider \ref{appL42}. Taking the squared norm yields
\begin{align}
\left\|\frac{\WP-\TP}{\sqrt T}\right\|_{F}^2
&=\frac1T\sum_{t=1}^T\left\|\*p(\widehat{\*f}_t)-\widetilde{\*p}(\*f_t)\right\|_{F}^2\notag\\
&\leq\zeta_1(K)^2\frac1T\sum_{t=1}^T\|\overline{\*u}_t\|_{F}^2
=O_p\left(\frac{\zeta_1(K)^2}{N}\right),
\end{align} where the mean-value inequality gives the bound, and Assumption \ref{A1} implies $\be\left(\tfrac{1}{T}\sum_t\|\overline{\*u}_t\|_{F}^2\right)\leq \tfrac{C}{N}$. Markov's inequality therefore yields the stated rate without requiring uniform convergence over $t$. Consequently, 
\begin{align}
\left\|\frac{1}{\sqrt{T}} \left(\WP - \TP\right)\right\|_{F} = O_p\left(\frac{\zeta_1(K)}{\sqrt{N}}\right).
\end{align} 

For \ref{appL43}, consider the following expansion
\begin{align}
\WP^\prime \WP - T \*I_K = \TP^\prime \TP - T \*I_K + (\WP - \TP)^\prime \TP + \TP^\prime (\WP - \TP) + (\WP - \TP)^\prime (\WP - \TP). 
\end{align} Then 
\begin{align}
    \left\|\frac{1}{T}\WP^\prime \WP  - \*I_K\right\|_{\mathrm{op}} \leq \left\|\frac{1}{T}\TP^\prime \TP  - \*I_K\right\|_{\mathrm{op}} + 2 \left\|\frac{1}{T}(\WP - \TP)^\prime \TP\right\|_{F} + \left\|\frac{1}{\sqrt{T}}(\WP-\TP)\right\|_{F}^2. \label{eq:phatphati}
\end{align} Here the spectral norm is bounded by the Frobenius norm. By Assumption \ref{A6}\ref{A6b}, $
\left\|\frac{1}{T}\TP'\TP-\*I_K\right\|_{\mathrm{op}}	= o_p\left(1\right)$ and by Lemma,  \ref{appL4}\ref{appL42} $\left\|\frac{1}{\sqrt{T}} \left(\WP - \TP\right)\right\|_{F}^2 = O_p\left(\frac{\zeta_1(K)^2}{N}\right)$. Thus, we are only left to analyze the second term in \eqref{eq:phatphati}. For that, first note how $\mathrm{tr}(\*B \*A \*B^\prime) \leq \lmax(\*A)\mathrm{tr}(\*B\*B^\prime)$, whenever $\*A$ is positive semidefinite (see \citet{lut}, result (12) on page 44). Then, using this result, the cyclical property of the trace and Assumption \ref{A6}\ref{A6a}, we get
\begin{align}
\left\|\frac{1}{T}(\WP - \TP)^\prime \TP\right\|_{F}^2 &= \frac{1}{T^2}\mathrm{tr}\left[(\WP - \TP)^\prime\TP\TP^\prime(\WP - \TP)\right] \notag \\
&\leq \lmax \left(\frac{1}{T}\TP^\prime\TP\right) \mathrm{tr}\left[\frac{1}{T} (\WP - \TP)^\prime(\WP - \TP)\right] \notag \\
& = \lmax \left(\frac{1}{T}\TP^\prime\TP\right)\left\|\frac{1}{\sqrt{T}}(\WP - \TP)\right\|_{F}^2 \notag \\
&= O_p(1) O_p\left(\frac{\zeta_1(K)^2}{N}\right) =O_p\left(\frac{\zeta_1(K)^2}{N}\right).
\end{align} 
The inequality uses the fact that $\TP\TP'$ and $\TP'\TP$ have the same nonzero eigenvalues.

Consequently, 
\begin{align}
    \left\|\frac{1}{T}\WP^\prime \WP  - \*I_K\right\|_{\mathrm{op}} &= o_p\left(1\right) + O_p\left(\frac{\zeta_1(K)}{\sqrt{N}}\right) + O_p\left(\frac{\zeta_1(K)^2}{N}\right) \notag \\
    &=o_p(1),
\end{align} where $\tfrac{\zeta_1(K)}{\sqrt N}\to0$ follows from Assumption \ref{A6}\ref{A6c}. This establishes the result in \ref{appL43} and completes the proof of Lemma \ref{appL4}. \hfill $\blacksquare$

%%%%%%%%%%%%%%%%%%%%%%%%%%%%%%%%%%%%%%%%%%%%
        %%%%%% COROLLARY A1 %%%%%%
%%%%%%%%%%%%%%%%%%%%%%%%%%%%%%%%%%%%%%%%%%%%
\begin{restatedcorollary}{appC1} 
Suppose Assumptions \ref{A1}--\ref{A3}, and \ref{A6}\ref{A6a}-\ref{A6c} hold. Then
\begin{enumerate}[label=(\roman*)]
    \item $\left\|\left(\WP - \TP\right)\left(\WP'\WP\right)^+\left(\WP - \TP\right)'\right\|_{F} = O_p\left(\frac{\zeta_1(K)^2}{N}\right)$,
    \item  $\left\|\left(\WP - \TP\right)\left(\WP'\WP\right)^+\TP'\right\|_{F} = o_p\left(\sqrt{\frac{K}{N}}\zeta_1(K)\right)$,
    \item  $\left\|\TP\left[\left(\WP'\WP\right)^+ - \left(\TP'\TP\right)^+\right]\TP'\right\|_{F} = O_p\left(\frac{\zeta_1(K)}{\sqrt{N}}\right)$,
    \item  $\left\|(\WP - \TP)(\WP^\prime \WP)^+ \TP^\prime (\WP - \TP)\right\|_{F} = O_p\left(\frac{\sqrt T}{N}\zeta_1(K)^2\right)$,
    \item $\left\|(\WP - \TP)(\WP^\prime \WP)^+(\WP - \TP)^\prime (\WP - \TP)\right\|_{F} = O_p\left(\frac{\sqrt{T}}{N^{\frac{3}{2}}}\zeta_1(K)^3\right)$.
\end{enumerate}
\end{restatedcorollary}

\paragraph{\textbf{Proof}}

To prove \ref{appC11}, note that by Assumption \ref{A6}\ref{A6a} and Lemma \ref{appL4}\ref{appL43}, the matrix $\frac{1}{T}\WP'\WP$ is nonsingular w.p.a.1, so that the Moore--Penrose inverse reduces to the ordinary inverse. Hence,
\begin{align}
&\left\|\left(\WP-\TP\right)\left(\WP'\WP\right)^+\left(\WP-\TP\right)'\right\|_{F} = \left\|\frac{1}{T}\left(\WP-\TP\right)\left(\frac{1}{T}\WP'\WP\right)^{-1}\left(\WP-\TP\right)'\right\|_{F} \notag \\
&\leq \left\|\frac{1}{\sqrt T}\left(\WP-\TP\right)\right\|_{F}^2\lmin\left(\frac{1}{T}\WP'\WP\right)^{-1}.
\end{align}
By Lemma \ref{appL4}\ref{appL42}, $\left\|\frac{1}{\sqrt T}\left(\WP-\TP\right)\right\|_{F} = O_p\left(\frac{\zeta_1(K)}{\sqrt N}\right)$. Moreover, by Assumption \ref{A6}\ref{A6a} and Lemma \ref{appL4}\ref{appL43}, $\lmin\left(\frac{1}{T}\WP'\WP\right)$ is bounded away from zero w.p.a.1. Therefore,
\begin{align}
\left\|\left(\WP-\TP\right)\left(\WP'\WP\right)^+\left(\WP-\TP\right)'\right\|_{F} = O_p\left(\frac{\zeta_1(K)^2}{N}\right).
\end{align}

Moving on to \ref{appC12}, first note that
\begin{align}
\left(\WP-\TP\right)\left(\WP'\WP\right)^+\TP' &= \frac{1}{T}\left(\WP-\TP\right)\left(\frac{1}{T}\WP'\WP\right)^+\TP' \notag \\
&= \frac{1}{T}\left(\WP-\TP\right)\TP' + \frac{1}{T}\left(\WP-\TP\right)\left[\left(\frac{1}{T}\WP'\WP\right)^+ - \*I_K\right]\TP',
\end{align}
and hence
\begin{align}
\left\|\left(\WP-\TP\right)\left(\WP'\WP\right)^+\TP'\right\|_{F} &\leq \left\|\frac{1}{T}\left(\WP-\TP\right)\TP'\right\|_{F} + \left\|\frac{1}{T}\left(\WP-\TP\right)\left[\left(\frac{1}{T}\WP'\WP\right)^+ - \*I_K\right]\TP'\right\|_{F}. \label{eq:C12split}
\end{align}
For the first term in \eqref{eq:C12split}, using $\mathrm{tr}(\*B\*A\*B') \leq \lambda_{\max}(\*A)\mathrm{tr}(\*B\*B')$ whenever $\*A$ is positive semidefinite, Assumption \ref{A6}\ref{A6a}, and Lemma \ref{appL4}\ref{appL42}, we obtain
\begin{align}
&\left\|\frac{1}{T}\left(\WP-\TP\right)\TP'\right\|_{F}^2 = \mathrm{tr}\left[\frac{1}{T^2}\TP\left(\WP-\TP\right)'\left(\WP-\TP\right)\TP'\right] \notag \\
&= \mathrm{tr}\left[\frac{1}{T}\left(\WP-\TP\right)\frac{1}{T}\TP'\TP\left(\WP-\TP\right)'\right] \leq \lambda_{\max}\left(\frac{1}{T}\TP'\TP\right)\mathrm{tr}\left[\frac{1}{T}\left(\WP-\TP\right)\left(\WP-\TP\right)'\right] \notag \\
&= \lambda_{\max}\left(\frac{1}{T}\TP'\TP\right)\left\|\frac{1}{\sqrt T}\left(\WP-\TP\right)\right\|_{F}^2 = O_p(1)O_p\left(\frac{\zeta_1(K)^2}{N}\right) \notag \\
&= O_p\left(\frac{\zeta_1(K)^2}{N}\right).
\end{align}
Thus,
\begin{align}
\left\|\frac{1}{T}\left(\WP-\TP\right)\TP'\right\|_{F} = O_p\left(\frac{\zeta_1(K)}{\sqrt N}\right). \label{eq:C12first}
\end{align}
For the second term in \eqref{eq:C12split}, we have
\begin{align}
\left\|\frac{1}{T}\left(\WP-\TP\right)\left[\left(\frac{1}{T}\WP'\WP\right)^+ - \*I_K\right]\TP'\right\|_{F} &\leq \left\|\frac{1}{\sqrt T}\left(\WP-\TP\right)\right\|_{F}\left\|\left(\frac{1}{T}\WP'\WP\right)^+ - \*I_K\right\|_{\mathrm{op}}\left\|\frac{1}{\sqrt T}\TP\right\|_{F}. \label{eq:C12second}
\end{align}
Since $\frac{1}{T}\WP'\WP$ is nonsingular w.p.a.1,
\begin{align}
\left(\frac{1}{T}\WP'\WP\right)^+ - \*I_K = \left(\frac{1}{T}\WP'\WP\right)^{-1} - \*I_K = -\left(\frac{1}{T}\WP'\WP\right)^{-1}\left(\frac{1}{T}\WP'\WP-\*I_K\right).
\end{align}
Therefore,
\begin{align}
\left\|\left(\frac{1}{T}\WP'\WP\right)^+ - \*I_K\right\|_{\mathrm{op}} &\leq \lambda_{\min}\left(\frac{1}{T}\WP'\WP\right)^{-1}\left\|\frac{1}{T}\WP'\WP-\*I_K\right\|_{\mathrm{op}} \notag \\
&= O_p(1)o_p(1)=o_p(1), \label{eq:phatinvident}
\end{align}
where we used Assumption \ref{A6}\ref{A6a} and Lemma \ref{appL4}\ref{appL43}. Finally, by Assumption \ref{A6}\ref{A6a},
\begin{align}
\left\|\frac{1}{\sqrt T}\TP\right\|_{F}^2
&= \mathrm{tr}\left(\frac{1}{T}\TP'\TP\right)
\leq K\lambda_{\max}\left(\frac{1}{T}\TP'\TP\right)
=O_p(K), \label{eq:PtildeFrob}
\end{align}
so that $\left\|\frac{1}{\sqrt T}\TP\right\|_{F} = O_p(\sqrt K)$. Combining these bounds in \eqref{eq:C12second} gives
\begin{align}
\left\|\frac{1}{T}\left(\WP-\TP\right)\left[\left(\frac{1}{T}\WP'\WP\right)^+ - \*I_K\right]\TP'\right\|_{F} &= O_p\left(\frac{\zeta_1(K)}{\sqrt N}\right)o_p(1)O_p(\sqrt K) \notag\\
&=o_p\left(\frac{\sqrt K\zeta_1(K)}{\sqrt N}\right). \label{eq:C12secondrate}
\end{align}
Combining \eqref{eq:C12first} and \eqref{eq:C12secondrate} yields
\begin{align}
\left\|\left(\WP-\TP\right)\left(\WP'\WP\right)^+\TP'\right\|_{F} &= O_p\left(\frac{\zeta_1(K)}{\sqrt N}\right) + o_p\left(\frac{\sqrt K\zeta_1(K)}{\sqrt N}\right) \notag \\
&=o_p\left(\frac{\sqrt K\zeta_1(K)}{\sqrt N}\right).
\end{align}
The last equality uses $K\to\infty$ under Assumption \ref{A6}\ref{A6c}, so that $O_p(\frac{\zeta_1(K)}{\sqrt N})=o_p(\frac{\sqrt K\zeta_1(K)}{\sqrt N})$.

For the proof of \ref{appC13}, first note that
\begin{align}
\TP\left[\left(\WP'\WP\right)^+ - \left(\TP'\TP\right)^+\right]\TP'
&= \frac{1}{T}\TP\left[\left(\frac{1}{T}\WP'\WP\right)^+ - \left(\frac{1}{T}\TP'\TP\right)^+\right]\TP'.
\end{align}
Hence,
\begin{align}
&\left\|\TP\left[\left(\WP'\WP\right)^+ - \left(\TP'\TP\right)^+\right]\TP'\right\|_{F}\notag\\
&\quad\leq \lambda_{\max}\left(\frac{1}{T}\TP'\TP\right)\left\|\left(\frac{1}{T}\WP'\WP\right)^{-1} - \left(\frac{1}{T}\TP'\TP\right)^{-1}\right\|_{F}. \label{eq:C13sandwich}
\end{align}
where the inequality follows by applying the trace and eigenvalue bounds twice. Both $\frac{1}{T}\WP'\WP$ and $\frac{1}{T}\TP'\TP$ are nonsingular w.p.a.1, so that the Moore--Penrose inverses reduce to ordinary inverses. Their inverse difference satisfies
\begin{align}
&\left\|\left(\frac{1}{T}\WP'\WP\right)^{-1} - \left(\frac{1}{T}\TP'\TP\right)^{-1}\right\|_{F}  = \left\|-\left(\frac{1}{T}\WP'\WP\right)^{-1}\left(\frac{1}{T}\WP'\WP-\frac{1}{T}\TP'\TP\right)\left(\frac{1}{T}\TP'\TP\right)^{-1}\right\|_{F} \notag \\
&\quad \leq \lambda_{\min}\left(\frac{1}{T}\WP'\WP\right)^{-1}\lambda_{\min}\left(\frac{1}{T}\TP'\TP\right)^{-1}\left\|\frac{1}{T}\WP'\WP-\frac{1}{T}\TP'\TP\right\|_{F}. \label{eq:C13invdiff}
\end{align}
Moreover,
\begin{align}
\left\|\frac{1}{T}\WP'\WP-\frac{1}{T}\TP'\TP\right\|_{F} &= \left\|\frac{1}{T}\left(\WP-\TP\right)'\TP + \frac{1}{T}\TP'\left(\WP-\TP\right) + \frac{1}{T}\left(\WP-\TP\right)'\left(\WP-\TP\right)\right\|_{F} \notag \\
&\leq 2\left\|\frac{1}{T}\left(\WP-\TP\right)'\TP\right\|_{F} + \left\|\frac{1}{\sqrt T}\left(\WP-\TP\right)\right\|_{F}^2 \notag \\
&= O_p\left(\frac{\zeta_1(K)}{\sqrt N}\right) + O_p\left(\frac{\zeta_1(K)^2}{N}\right), \label{eq:C13gramdiff}
\end{align}
where the first term follows from the cross-Gram trace bound in Lemma \ref{appL4}'s proof and the second from Lemma \ref{appL4}\ref{appL42}. By Assumption \ref{A6}\ref{A6a} and Lemma \ref{appL4}\ref{appL43}, both $\lambda_{\min}\left(\frac{1}{T}\WP'\WP\right)$ and $\lambda_{\min}\left(\frac{1}{T}\TP'\TP\right)$ are bounded away from zero w.p.a.1. Combining \eqref{eq:C13invdiff} and \eqref{eq:C13gramdiff}, we obtain
\begin{align}
\left\|\left(\frac{1}{T}\WP'\WP\right)^{-1} - \left(\frac{1}{T}\TP'\TP\right)^{-1}\right\|_{F} &= O_p\left(\frac{\zeta_1(K)}{\sqrt N}\right) + O_p\left(\frac{\zeta_1(K)^2}{N}\right).
\end{align}
Since $\lambda_{\max}\left(\frac{1}{T}\TP'\TP\right)=O_p(1)$ under Assumption \ref{A6}\ref{A6a}, \eqref{eq:C13sandwich} gives
\begin{align}
\left\|\TP\left[\left(\WP'\WP\right)^+ - \left(\TP'\TP\right)^+\right]\TP'\right\|_{F} &= O_p\left(\frac{\zeta_1(K)}{\sqrt N}\right) + O_p\left(\frac{\zeta_1(K)^2}{N}\right).
\end{align}
Finally, since $\frac{\zeta_1(K)}{\sqrt N}=o(1)$ under Assumption \ref{A6}\ref{A6c}, the second term is dominated by the first. Hence,
\begin{align}
\left\|\TP\left[\left(\WP'\WP\right)^+ - \left(\TP'\TP\right)^+\right]\TP'\right\|_{F} = O_p\left(\frac{\zeta_1(K)}{\sqrt N}\right).
\end{align}

For \ref{appC14}, note that
\begin{align}
&\left\|(\WP - \TP)(\WP^\prime \WP)^+ \TP^\prime (\WP - \TP)\right\|_{F} =\left\|\frac{1}{T}(\WP - \TP)\left(\frac{1}{T}\WP^\prime \WP\right)^+ \TP^\prime (\WP - \TP)\right\|_{F} \notag \\
&\leq \left\|\frac{1}{T}(\WP - \TP)\TP^\prime (\WP - \TP)\right\|_{F}  + \left\|\frac{1}{T}(\WP - \TP)\left[\left(\frac{1}{T}\WP^\prime \WP\right)^+ - \*I_K\right]\TP^\prime (\WP - \TP)\right\|_{F} \notag \\
&\leq \left\|\frac{1}{\sqrt{T}}(\WP - \TP)\right\|_{F} \left\|\frac{1}{\sqrt{T}}\TP^\prime (\WP - \TP)\right\|_{F} \notag \\
&\quad+\left\|\frac{1}{\sqrt{T}}(\WP - \TP)\right\|_{F} \left\|\left(\frac{1}{T}\WP^\prime \WP\right)^+ - \*I_K \right\|_{\mathrm{op}} \left\|\frac{1}{\sqrt{T}}\TP^\prime (\WP - \TP)\right\|_{F} \notag \\
&= O_p\left(\frac{\zeta_1(K)}{\sqrt{N}}\right) O_p\left(\sqrt{\frac{T}{N}}\zeta_1(K)\right) \notag \\
&\quad + O_p\left(\frac{\zeta_1(K)}{\sqrt{N}}\right)o_p(1)O_p\left(\sqrt{\frac{T}{N}}\zeta_1(K)\right) \notag \\
&= O_p\left(\frac{\sqrt T}{N}\zeta_1(K)^2\right) + o_p\left(\frac{\sqrt T}{N}\zeta_1(K)^2\right) \notag \\
&= O_p\left(\frac{\sqrt T}{N}\zeta_1(K)^2\right),
\end{align}
where we have used Lemma \ref{appL4}\ref{appL42}, \eqref{eq:phatinvident}, and the cross-Gram trace bound in Lemma \ref{appL4}'s proof. The last equality absorbs the little-$o_p$ term into the $O_p$ bound.

For \ref{appC15}, using \ref{appC11} and Lemma \ref{appL4}\ref{appL42}, we obtain
\begin{align}
&\left\|(\WP - \TP)(\WP^\prime \WP)^+(\WP - \TP)^\prime (\WP - \TP)\right\|_{F} \leq \left\|(\WP - \TP)(\WP^\prime \WP)^+(\WP - \TP)^\prime\right\|_{F}\left\|\WP-\TP\right\|_{F} \notag \\
&= O_p\left(\frac{\zeta_1(K)^2}{N}\right)\sqrt{T}\left\|\frac{1}{\sqrt T}(\WP-\TP)\right\|_{F}= O_p\left(\frac{\zeta_1(K)^2}{N}\right)O_p\left(\sqrt{\frac{T}{N}}\zeta_1(K)\right) \notag \\
&= O_p\left(\frac{\sqrt{T}}{N^{\frac{3}{2}}}\zeta_1(K)^3\right).
\end{align}

This completes the proof of Corollary \ref{appC1}. \hfill $\blacksquare$

%%%%%%%%%%%%%%%%%%%%%%%%%%%%%%%%%%%%%%%%%%%%
        %%%%%% COROLLARY A2 %%%%%%
%%%%%%%%%%%%%%%%%%%%%%%%%%%%%%%%%%%%%%%%%%%%
\begin{restatedcorollary}{appC2}
Suppose Assumptions \ref{A1}--\ref{A6} hold and let $\widetilde{\*g}_i(\cdot)\in\widetilde{\mathcal S}$, where $\widetilde{\mathcal S}\equiv\{\*g_i(\cdot),\*G_i(\cdot)\}$. Then
\begin{align}
\left\|\frac{1}{\sqrt{T}}\left(\WP \+\alpha_{\widetilde{\*g}_i} - \widetilde{\*g}_i(\*F)\right)\right\|_{F} = O_p\left(\frac{\zeta_1(K)}{\sqrt{N}}\right) + O_p\left(K^{-\frac{\lambda_i}{m}}\right).
\end{align}
\end{restatedcorollary}

\paragraph{\textbf{Proof}} Note the following decomposition:
\begin{align}
\left\|\frac{1}{\sqrt{T}}\left(\WP \+\alpha_{\widetilde{\*g}_i} - \widetilde{\*g}_i(\*F)\right)\right\|_{F} &= \left\|\frac{1}{\sqrt{T}}(\WP-\TP)\+\alpha_{\widetilde{\*g}_i} + \frac{1}{\sqrt{T}}(\TP \+\alpha_{\widetilde{\*g}_i} - \widetilde{\*g}_i(\*F))\right\|_{F} \notag \\
&\leq \left\|\frac{1}{\sqrt{T}}(\WP-\TP)\right\|_{F} \left\|\+\alpha_{\widetilde{\*g}_i} \right\|_{F} + \left\|\frac{1}{\sqrt{T}}(\TP \+\alpha_{\widetilde{\*g}_i} - \widetilde{\*g}_i(\*F))\right\|_{F}.
\end{align}
Assumption \ref{A6}\ref{A6d} gives $\sup_i\|\+\alpha_{\widetilde{\*g}_i}\|_{F}=O(1)$. Together with Lemma \ref{appL4}\ref{appL42} and the population-proxy approximation bound used in Lemma \ref{appL2}, this yields
\begin{align}
\left\|\frac1{\sqrt T}(\WP\+\alpha_{\widetilde{\*g}_i}-\widetilde{\*g}_i(\*F))\right\|_{F}
&\leq O_p\left(\frac{\zeta_1(K)}{\sqrt N}\right)O(1)+O_p\left(K^{-\frac{\lambda_i}{m}}\right)\notag\\
&=O_p\left(\frac{\zeta_1(K)}{\sqrt N}\right)+O_p\left(K^{-\frac{\lambda_i}{m}}\right).
\end{align}
\hfill $\blacksquare$

\appendixsubsection{Derivation of the Admissible Corridor for \texorpdfstring{$K$}{K} in Section \ref{sec-ass}}\label{AppB2}
We derive the growth corridor from Assumption \ref{A6}\ref{A6c}, conditional on part \ref{A6b}. The restrictions are
\begin{align}
\frac KT&\to0,&\sqrt T K^{-\frac{\lambda_i}{m}}&\to0,&\sqrt{NT}K^{-\frac{2\lambda_i}{m}}&\to0,\notag\\
\sqrt{\frac TN}K\zeta_1(K)^2&\to0,&\frac TN&\to0.
\end{align}
For the splines, we have $\zeta_1(K)=O(K^{\frac{3}{2}})$, such that sufficient restrictions are\footnote{See, for example, \citealp{bel15}, and \citealp{newey97}}
\begin{align}
K&\ll T,&K&\gg T^{\frac{m}{2\lambda_i}},&K&\gg(NT)^{\frac{m}{4\lambda_i}},\notag\\
\sqrt{\frac TN}K^4&\to0,&\frac TN&\to0.
\end{align}
For these spline sequences it is ensured by $K\ll(\tfrac NT)^{\frac{1}{8}}$. Thus the admissible corridor from part \ref{A6c} is
\begin{align}
\max\left\{T^{\frac{m}{2\lambda_i}},(NT)^{\frac{m}{4\lambda_i}}\right\}\ll K\ll\min\left\{T,\left(\frac{N}{T}\right)^{\frac{1}{8}}\right\}. \label{eq:appKcorridor}
\end{align}

Next, let $T=N^\rho$ and $K=N^\kappa$, with $\rho\in(0,1)$. The lower bound in \eqref{eq:appKcorridor} becomes
\begin{align}
\kappa>\max\left\{\frac{\rho m}{2\lambda_i},\frac{(1+\rho)m}{4\lambda_i}\right\}=\frac{(1+\rho)m}{4\lambda_i},
\end{align}
where the equality follows because $\rho<1$. The upper bound becomes
\begin{align}
\kappa<\min\left\{\rho,\frac{1-\rho}{8}\right\}.
\end{align}
The two upper-bound terms are equal at $\rho=\frac{1}{9}$. Therefore, the corridor is non-empty if
\begin{align}
\frac{(1+\rho)m}{4\lambda_i}<\rho \qquad \text{for } \rho\leq \frac{1}{9}, \qquad \frac{(1+\rho)m}{4\lambda_i}<\frac{1-\rho}{8} \qquad \text{for } \rho>\frac{1}{9}.
\end{align}
Equivalently,
\begin{align}
\frac{m}{\lambda_i}<\frac{4\rho}{1+\rho} \qquad \text{for } \rho\leq \frac{1}{9}, \qquad \frac{m}{\lambda_i}<\frac{1-\rho}{2(1+\rho)} \qquad \text{for } \rho>\frac{1}{9}.
\end{align} Conditional on part \ref{A6b}, the weakest smoothness requirement from this corridor is obtained at the kink $\rho=\frac{1}{9}$. At this point, a sufficient condition is $\lambda_i>\frac52m$ for all $i$.

%%%%%%%%%%%%%%%%%%%%%%%%%%%%%%%%%%%%%%%%%%%%%%%%%%%%%%%%%%%%%%%%%%%%
%%%%%%%%%%%%%%%% ADDITIONAL ILLUSTRATION FOR EMPIRICS %%%%%%%%%%%%%
%%%%%%%%%%%%%%%%%%%%%%%%%%%%%%%%%%%%%%%%%%%%%%%%%%%%%%%%%%%%%%%%%%%%
\appendixsubsection{Additional Results: Empirical Application} \label{empadd}
In this section, we provide additional illustrations for our empirical application. Table \ref{tab:adftest} reports augmented Dickey-Fuller tests for the regressor components of the factor proxy, which do not reject the unit-root null for these series. Table \ref{tab:apphac} presents SCCE estimates with 95\% confidence intervals based on HAC standard errors.

Figure \ref{fig:wagedevelop} shows the development of wage inequality in U.S. manufacturing over the sample period. Figure \ref{fig:factorsd} plots the regressor components of the factor proxy after first differencing, indicating mean reversion. The outcome average is also used in estimation but is not displayed in this figure or Table \ref{tab:adftest}. Figure \ref{fig:figknots}  shows the estimated effects of our main regressors, $\ln(\sigma_{i,t})$ and $\ln\left(\frac{H_{i,t}}{L_{i,t}}\right)$, across different numbers of knots, $J$, using 95 \% bootstrap confidence intervals. Figure \ref{fig:figknotsapp} presents the corresponding estimates using confidence intervals based on HAC standard errors.

\clearpage
\begin{table}[p]
\centering
\begin{threeparttable}
\caption{Augmented Dickey-Fuller tests for the regressor components of the factor proxy.}
\renewcommand{\arraystretch}{1.0}
\small
\begin{tabular*}{0.7\textwidth}{@{\extracolsep{\fill}}lcc}
    \toprule
    Proxy component & ADF & $p$-value \\
    \midrule
    $\overline{y}_t$ & -1.97 & 0.30 \rule{0pt}{3ex}\\
    $\overline{\ln(\sigma_{t})}$ & -1.02 & 0.75 \rule{0pt}{3ex}\\
    $\overline{\ln\left(\frac{H_{t}}{L_{t}}\right)}$ & -0.57  & 0.88 \rule{0pt}{3ex}\\
    $\overline{z}_{1,t}$ &0.98 & 0.99 \rule{0pt}{3ex}\\
    $\overline{z}_{2,t}$ & 0.90 & 0.99 \rule{0pt}{3ex}\\
    $\overline{z}_{3,t}$ & 0.76 & 0.99 \rule{0pt}{3ex}\\
    $\overline{z}_{4,t}$ & -1.49 & 0.54 \rule{0pt}{3ex}\\
    $\overline{z}_{5,t}$ & 2.39 & 0.99 \rule{0pt}{3ex}\\
    $\overline{z}_{6,t}$ & 2.45 & 0.99 \rule{0pt}{3ex}\\
    \bottomrule
\end{tabular*}
\label{tab:adftest}
\begin{tablenotes}
\item \scriptsize \textit{Notes}: The lag order in the ADF test is selected using BIC. $\overline{z}_{1,t}, \dots, \overline{z}_{6,t}$ denote auxiliary regressors used as controls and are further specified in Table \ref{tab:appli} in the main text. 			
\end{tablenotes}
\end{threeparttable}
\end{table}

\clearpage

\begin{table}[p]
\centering
\begin{threeparttable}
  \caption{Estimation results for SCCE with HAC standard errors.}
  \label{tab:apphac}
  \begin{tabular*}{0.7\textwidth}{@{\extracolsep{\fill}}lcc}
\toprule
& Narrow model & Full model \\
\midrule
$\ln(\sigma_{i,t})$                &  -0.34          & -0.33 \\
                                   & (-0.62; -0.06)  & (-0.62; -0.05) \\
$\ln\!\left(\frac{H_{i,t}}{L_{i,t}}\right)$ & 0.56  & 0.61 \\
                                   & (0.52; 0.59)    & (0.58; 0.64) \\
$z_{1,i,t}$                        &                 & -0.03 \\
                                   &                 & (-0.37; 0.31) \\
$z_{2,i,t}$                        &                 & 3.15 \\
                                   &                 & (1.20; 5.11) \\
$z_{3,i,t}$                        &                 & -1.03 \\
                                   &                 & (-2.09; 0.04) \\
$z_{4,i,t}$                        &                 & -0.20 \\
                                   &                 & (-0.55; 0.16) \\
$z_{5,i,t}$                        &                 & -0.17 \\
                                   &                 & (-0.33; -0.00) \\
$z_{6,i,t}$                        &                 & 0.07 \\
                                   &                 & (-0.14; 0.28) \\
\bottomrule
\end{tabular*}
\renewcommand{\arraystretch}{1.0}
  \begin{tablenotes}
  \item \scriptsize \textit{Notes}: The table reports point estimates of the coefficients of the model in \eqref{wage} with the associated 95\% confidence intervals appearing within parentheses, based on HAC standard errors. Results are reported for our SCCE estimator for the narrow and full model. The regressors of interest are $\ln\left(\sigma_{i,t}\right)$, which is a measure of the input skill intensity, and $\ln\left(\frac{H_{i,t}}{L_{i,t}}\right)$, which is the ratio of high-skilled to low-skilled labor. The six control variables are; real capital equipment per worker ($z_{1,i,t}$), the sectoral share of office, computing and accounting equipment ($z_{2,i,t}$), the difference between high-tech and computer capital share ($z_{3,i,t}$), R\&D intensity ($z_{4,i,t}$), narrow outsourcing ($z_{5,i,t}$), and the difference between broad and narrow outsourcing ($z_{6,i,t}$). The dependent variable is the log of the relative wage of low-skilled to high-skilled workers.
  \end{tablenotes}
\end{threeparttable}
\end{table}

\clearpage

\setcounter{figure}{0}		
\begin{figure}[p]
\centering
\caption{Average wage inequality in U.S. manufacturing sector for the time period 1958 -- 2005.}
\centering
\includegraphics[width=9cm]{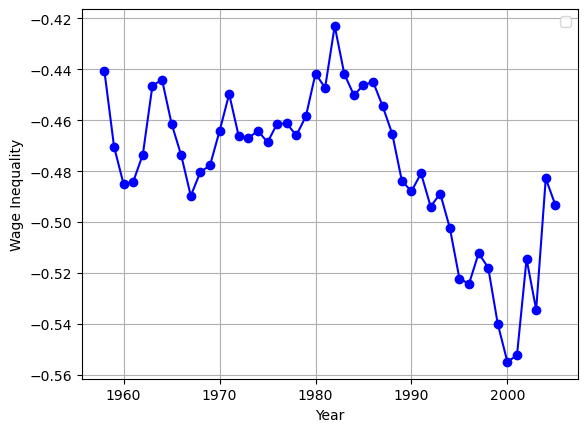}
\begin{minipage}{\textwidth}
\vspace{0.5cm}
\scriptsize \textit{Notes:} We calculate the average wage inequality as $\overline{\ln\left(\frac{w_t^{(L)}}{w_t^{(H)}}\right)}=\frac{1}{N}\sum_{i=1}^N \ln\left(\frac{w_{i,t}^{(L)}}{w_{i,t}^{(H)}}\right)$, for manufacturing sectors $i=1, \dots, N$. Here, $\frac{w_{i,t}^{(L)}}{w_{i,t}^{(H)}}$ is the relative wage of low-skilled workers to high-skilled workers in a U.S. manufacturing sector $i$ at time $t$. This illustration is based on a balanced version of the data from \citet{voig}.
\end{minipage}
\label{fig:wagedevelop}
\end{figure}

\clearpage
\begin{figure}[p]
\centering
\caption{Regressor components of the factor proxy after first differencing.}
\begin{subfigure}{0.31\textwidth}
\centering
\includegraphics[width=1.0\hsize]{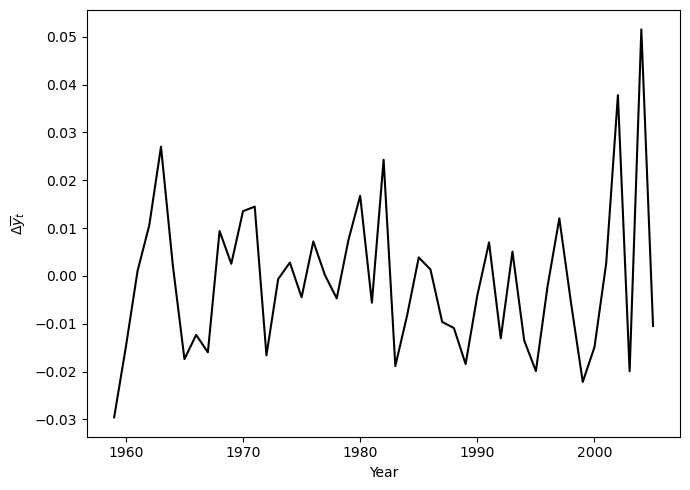}
\caption{$\overline{y}_t$.}
\end{subfigure}
\hfill
\begin{subfigure}{0.31\textwidth}
\centering
\includegraphics[width=1.0\hsize]{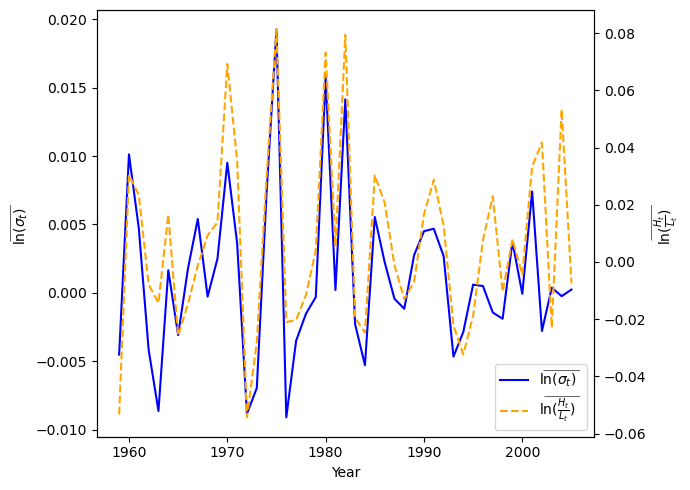}
\caption{$\overline{\ln\left(\sigma_t\right)}$ and $\overline{\ln \left(\frac{H_t}{L_t}\right)}$.}
\end{subfigure}
\hfill
\begin{subfigure}{0.31\textwidth}
\centering
\includegraphics[width=1.0\hsize]{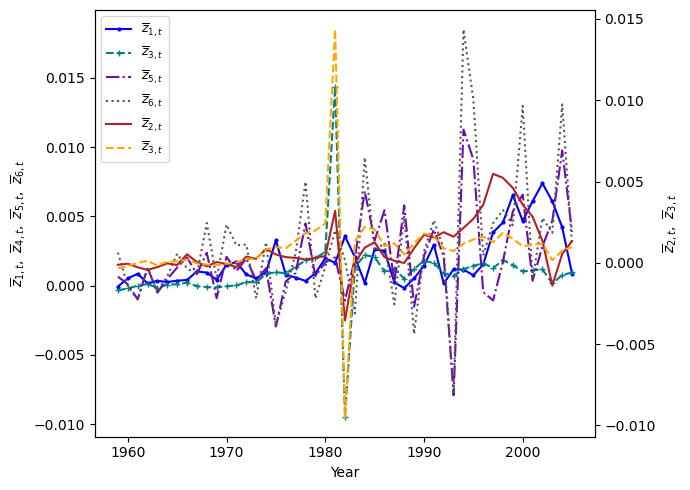}
\caption{The elements of $\overline{\mathbf{z}}_t$.}
\end{subfigure}
\hfill
\vspace{0.5cm} % Adds space between caption and notes
\scriptsize
\begin{minipage}[t]{\textwidth}
\textit{Notes:} The figure plots first differences of the regressor components of the factor proxy over the 1958-2005 period. Panel (a) plots the cross-sectional averages of the outcome variable $y_{i,t} = \ln \left(\frac{w_{L, i, t}}{w_{H, i, t}}\right)$, panel (b) plots the cross-sectional averages of the two key regressors, $\ln\left(\sigma_{i,t}\right)$ and $\ln\left(\frac{H_{i,t}}{L_{i,t}}\right)$, while panel (c) plots the cross-sectional averages of the controls included in $\mathbf{z}_{i,t}$. 
\end{minipage}
\label{fig:factorsd}
\end{figure}

\clearpage

\begin{figure}[p]
\centering
\caption{Estimated effects of $\ln \left(\sigma_{i,t}\right)$ and $\ln \left(\frac{H_{i,t}}{L_{i,t}}\right)$ for different number of knots, $J$.}
\begin{subfigure}{0.49\textwidth}
\centering
\includegraphics[width=0.92\hsize]{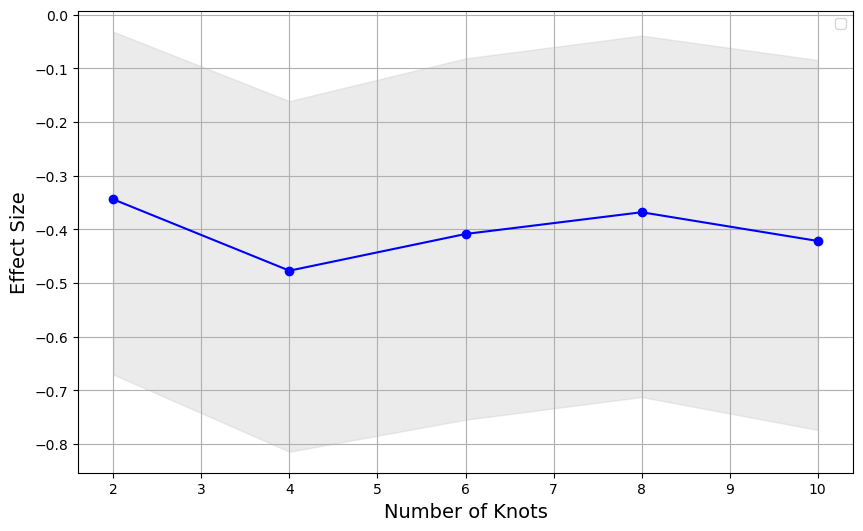}
\caption{The effect of $\ln \left(\sigma_{i,t}\right)$ in the narrow model.}
\end{subfigure}
\hfill
\begin{subfigure}{0.49\textwidth}
\centering
\includegraphics[width=0.92\hsize]{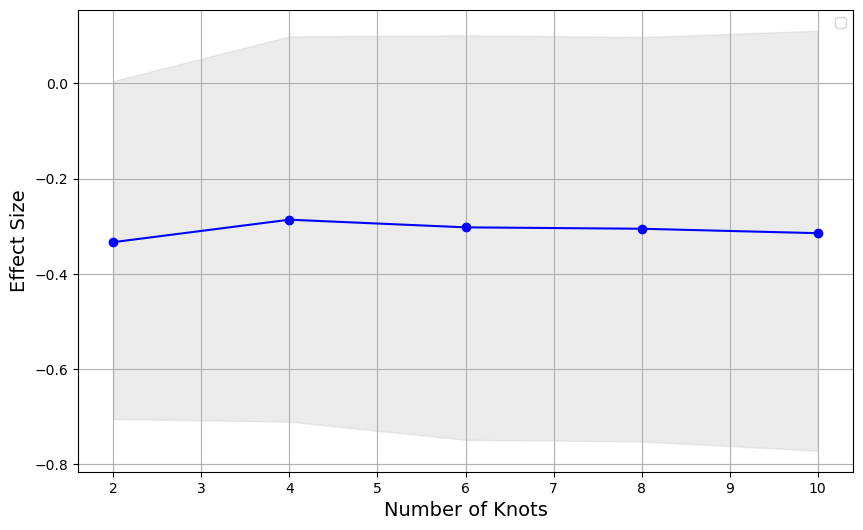}
\caption{The effect of $\ln \left(\sigma_{i,t}\right)$ in the full model.}
\end{subfigure}
\hfill
\begin{subfigure}{0.49\textwidth}
\centering
\includegraphics[width=0.92\hsize]{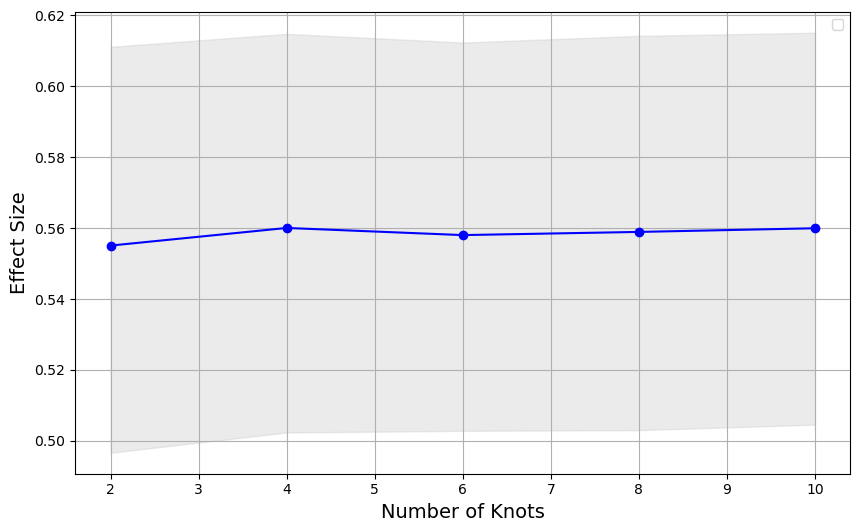}
\caption{The effect of $\ln \left(\frac{H_{i,t}}{L_{i,t}}\right)$ in the narrow model.}
\end{subfigure}
\hfill
\begin{subfigure}{0.49\textwidth}
\centering
\includegraphics[width=0.92\hsize]{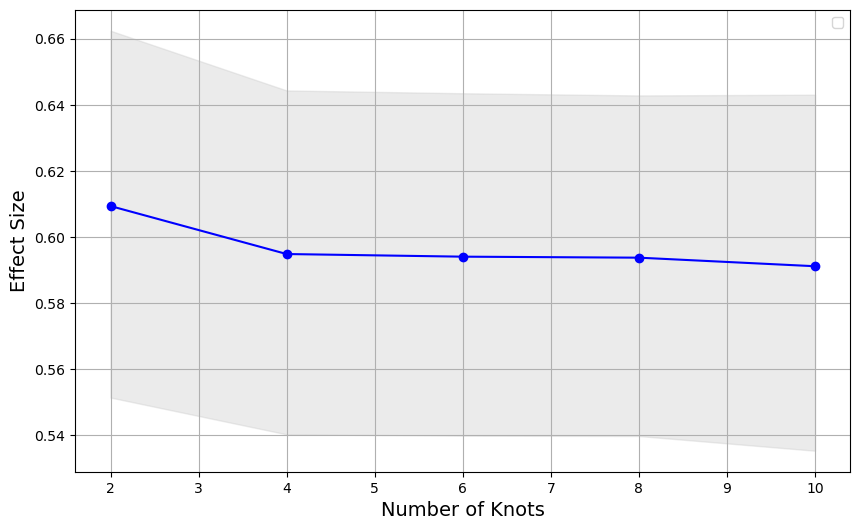}
\caption{The effect of $\ln \left(\frac{H_{i,t}}{L_{i,t}}\right)$ in the full model.}
\end{subfigure}
\scriptsize
\begin{minipage}{\textwidth}
\vspace{0.5cm}
\textit{Notes:} The figure plots point estimates obtained by re-estimating \eqref{wage} by SCCE for $J\in \{2, 4, 6, 8, 10\}$. The gray area represents 95\% bootstrap confidence intervals.
\end{minipage}
\label{fig:figknots}
\end{figure}

\clearpage
\begin{figure}[p]
\centering
\caption{Estimated effects of $\ln \left(\sigma_{i,t}\right)$ and $\ln \left(\frac{H_{i,t}}{L_{i,t}}\right)$ for different number of knots, $J$, using HAC standard errors.}
\begin{subfigure}{0.49\textwidth}
\centering
\includegraphics[width=1.0\hsize]{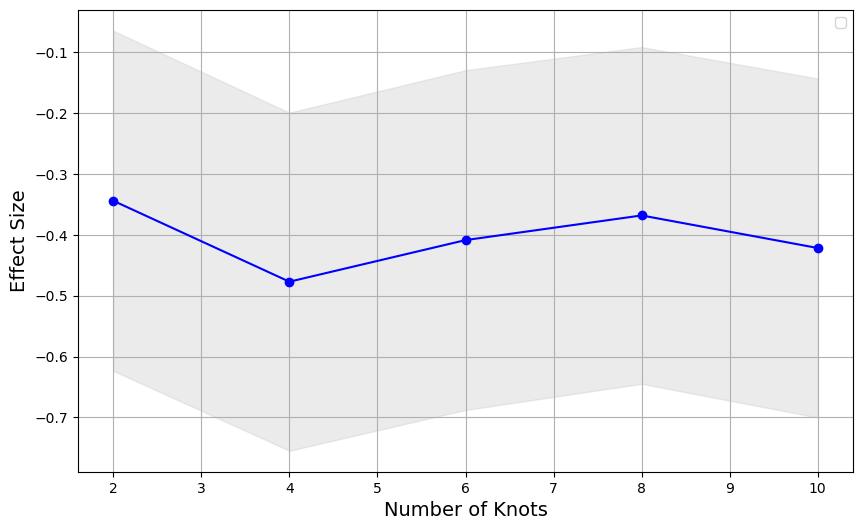}
\caption{The effect of $\ln (\sigma_{i,t})$ in the narrow model.}
\end{subfigure}
\hfill
\begin{subfigure}{0.49\textwidth}
\centering
\includegraphics[width=1.0\hsize]{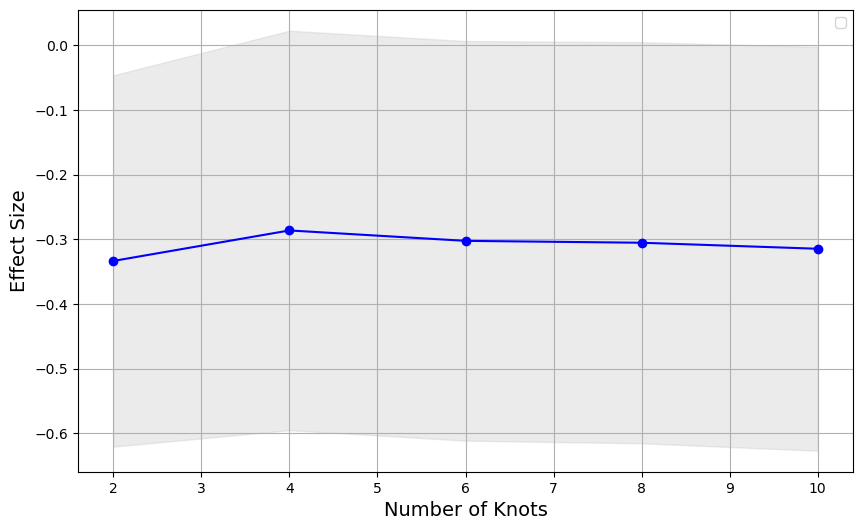}
\caption{The effect of $\ln \left(\sigma_{i,t}\right)$ in the full model.}
\end{subfigure}
\hfill
\begin{subfigure}{0.49\textwidth}
\centering
\includegraphics[width=1.0\hsize]{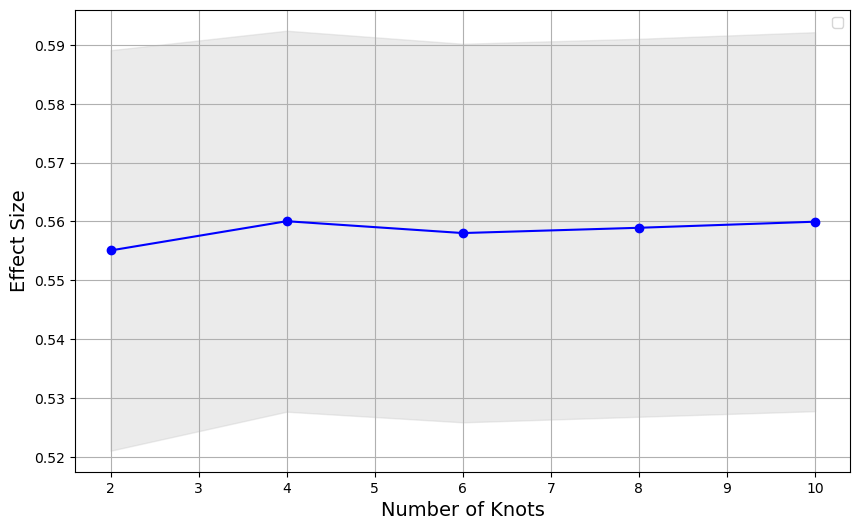}
\caption{The effect of $\ln \left(\frac{H_{i,t}}{L_{i,t}}\right)$ in the narrow model.}
\end{subfigure}
\hfill
\begin{subfigure}{0.49\textwidth}
\centering
\includegraphics[width=1.0\hsize]{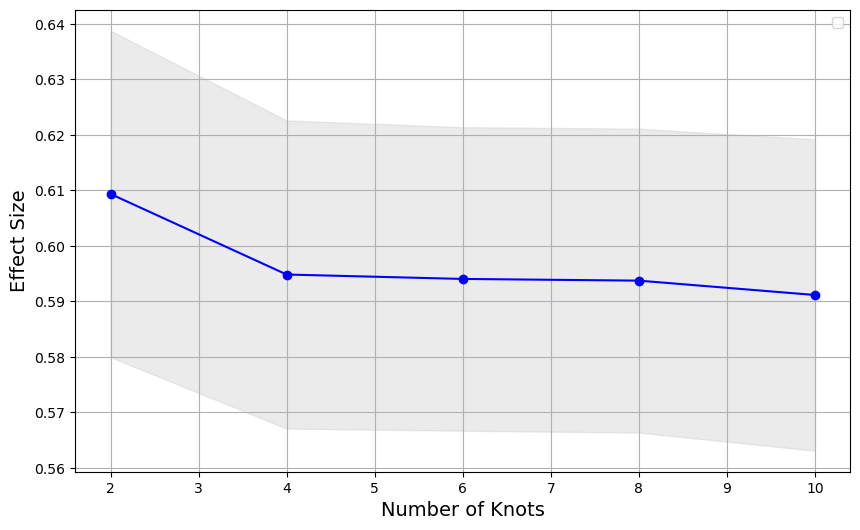}
\caption{The effect of $\ln \left(\frac{H_{i,t}}{L_{i,t}}\right)$ in the full model.}
\end{subfigure}
\scriptsize
\begin{minipage}{\textwidth}
\vspace{0.5cm}
\textit{Notes:} The figure plots point estimates obtained by re-estimating the specification in \eqref{wage} by SCCE for $J\in \{2, 4, 6, 8, 10\}$. The grey area represents 95\% confidence intervals based on HAC standard errors.
\end{minipage}
\label{fig:figknotsapp}
\end{figure}

\clearpage

%%%%%%%%%%%%%%%%%%%%%%%%%%%%%%%%%%%%%%%%%%%%%%%%%%%%%%%%%%%%%%%%%%%%
	%%%%%%%%%%%%%%%% ADDITIONAL RESULTS MC %%%%%%%%%%%%%
%%%%%%%%%%%%%%%%%%%%%%%%%%%%%%%%%%%%%%%%%%%%%%%%%%%%%%%%%%%%%%%%%%%%
\appendixsubsection{Additional Results: Monte Carlo Simulations} \label{mcadd}
This section presents additional results from a small-scale Monte Carlo simulation exercise, with a focus on various extensions that show the flexibility of our SCCE estimator. The specifications are basically Experiment \textbf{E1} in Section \ref{mc} with minor deviations. Hence, we refer the reader there for more details. Note that the simulation designs include configurations outside the sufficient conditions for asymptotic normality discussed in the main text, which allows us to assess the fininte-sample performance over a broader range of panel dimensions. The following experiments are considered:
\begin{enumerate}
\item[\textbf{E1.A}] Here, we report the numerical results of our estimator for the data-generating process as in Experiment \textbf{E1} in Section \ref{mc}.
\item[\textbf{E1.B}] Here, we report the numerical results of our estimator, assuming interactive effects as in Experiment E2 in section \ref{mc}.
\item[\textbf{E1.C}] Here, we allow the factors to be unit root non-stationary. More specifically, we generate $\*f_t = \*f_{t-1}+\*w_t$, with $\*w_t \sim \mathcal{N}(\*0_{m \times 1}, 0.05 \cdot \*I_m)$ and $\*f_0=\*0_{m \times 1}$.
\item[\textbf{E1.D}] Here, we consider an infeasible estimator, where we use the true factors as input for the sieve basis.
\item[\textbf{E1.E}] Here, we consider a DGP exactly as in E1 but use Hermite polynomials instead of splines as a sieve basis. For a definition of Hermite polynomials see, for example, \citet{chen2007}.
\item[\textbf{E1.F}] Here, relax the independence assumption on the error terms in \eqref{yi} and \eqref{xi}, and allow for weak serial and cross-sectional correlations as well as cross-sectional heteroscedasticity. We follow \citet{kad25}, and generate $\varepsilon_{i,t}=\pi \varepsilon_{i,t-1} + \theta_{i,t}+\sum_{l=1}^L \pi (\theta_{i-l,t} +\theta_{i+l,t})$, where $\varepsilon_{1,0}= \dots = \varepsilon_{N,0}=0$, $L=5$, and $\theta_{i,1} \sim \mathcal{N}(0, \sigma_i^2)$ with $\sigma_i \sim \mathcal{U}(0.5,1)$. Further, $\pi \in \{0.2, 0.5\}$, where $\pi = 0.2$ implies "low", and $\pi = 0.5 $ implies "high" dependence. $\*v_{i,t}$ is generated in the same way.
\item[\textbf{E1.G}] Here, we consider different number of knots for the spline basis in Experiment \textbf{E1}. More specifically, we will set $J = C \lfloor T^{\frac{1}{4}} \rfloor$, with $C\in \{2,3,4\}$. Further, we also consider different rates for $T$ i.e., $J=\{T^{\frac{1}{3}}, T^{\frac{1}{5}}, T^{\frac{1}{10}}\}$.
\end{enumerate}

\no We ran $1{,}000$ simulations with $N,T \in \{20,50,100,200,300\}$. Given the computational load from high-dimensional, nonlinear estimation, the Monte Carlo experiments were executed on LUNARC's COSMOS cluster at Lund University. COSMOS comprises 182 compute nodes, each with two AMD 7413 processors (Milan), resulting in 48 cores per node. All simulations were implemented in Python 3.9.18.

\newpage

\begin{landscape}
% Preserve the existing placement of landscape tables.
\makeatletter
\setlength{\@fptop}{0pt}
\makeatother
\begin{table}[p]
\centering
\begin{threeparttable}
\caption{Simulation results for Experiments E1.A, E1.B, E1.C, E1.D, and E1.E.}
\renewcommand{\arraystretch}{1.0}
\small
\begin{tabular}{rrrrrrrrrrrr}
\toprule
 &  & \multicolumn{2}{c}{E1.A} & \multicolumn{2}{c}{E1.B} & \multicolumn{2}{c}{E1.C} & \multicolumn{2}{c}{E1.D} & \multicolumn{2}{c}{E1.E} \\
\cmidrule(lr){3-4}\cmidrule(lr){5-6}\cmidrule(lr){7-8}\cmidrule(lr){9-10}\cmidrule(lr){11-12}
N & T & Bias & RMSE & Bias & RMSE & Bias & RMSE & Bias & RMSE & Bias & RMSE \\
\midrule
20 & 20 & 0.0014 & 0.1420 & 0.0016 & 0.1183 & 0.0017 & 0.1166 & 0.0001 & 0.0660 & 0.0016 & 0.0861 \\
20 & 50 & 0.0011 & 0.0478 & 0.0008 & 0.0420 & 0.0009 & 0.0406 & 0.0003 & 0.0332 & 0.0013 & 0.0449 \\
20 & 100 & 0.0016 & 0.0335 & 0.0022 & 0.0273 & 0.0001 & 0.0260 & 0.0000 & 0.0235 & 0.0017 & 0.0333 \\
20 & 200 & 0.0003 & 0.0258 & 0.0002 & 0.0220 & 0.0005 & 0.0188 & 0.0003 & 0.0177 & 0.0001 & 0.0269 \\
20 & 300 & 0.0008 & 0.0226 & 0.0005 & 0.0201 & 0.0009 & 0.0169 & 0.0003 & 0.0140 & 0.0006 & 0.0228 \\
50 & 20 & 0.0042 & 0.0859 & 0.0011 & 0.0704 & 0.0002 & 0.0714 & 0.0019 & 0.0423 & 0.0001 & 0.0534 \\
50 & 50 & 0.0015 & 0.0317 & 0.0002 & 0.0241 & 0.0008 & 0.0247 & 0.0008 & 0.0224 & 0.0009 & 0.0309 \\
50 & 100 & 0.0009 & 0.0215 & 0.0004 & 0.0159 & 0.0011 & 0.0162 & 0.0001 & 0.0149 & 0.0008 & 0.0226 \\
50 & 200 & 0.0003 & 0.0148 & 0.0002 & 0.0109 & 0.0003 & 0.0121 & 0.0000 & 0.0104 & 0.0001 & 0.0156 \\
50 & 300 & 0.0003 & 0.0131 & 0.0003 & 0.0089 & 0.0001 & 0.0100 & 0.0002 & 0.0086 & 0.0003 & 0.0138 \\
100 & 20 & 0.0054 & 0.0584 & 0.0004 & 0.0484 & 0.0005 & 0.0504 & 0.0007 & 0.0310 & 0.0028 & 0.0383 \\
100 & 50 & 0.0008 & 0.0210 & 0.0000 & 0.0175 & 0.0011 & 0.0169 & 0.0001 & 0.0157 & 0.0009 & 0.0212 \\
100 & 100 & 0.0005 & 0.0143 & 0.0001 & 0.0112 & 0.0001 & 0.0120 & 0.0007 & 0.0107 & 0.0002 & 0.0147 \\
100 & 200 & 0.0003 & 0.0102 & 0.0002 & 0.0075 & 0.0003 & 0.0080 & 0.0002 & 0.0076 & 0.0003 & 0.0108 \\
100 & 300 & 0.0004 & 0.0089 & 0.0002 & 0.0062 & 0.0001 & 0.0066 & 0.0002 & 0.0057 & 0.0005 & 0.0093 \\
200 & 20 & 0.0041 & 0.0568 & 0.0008 & 0.0356 & 0.0006 & 0.0350 & 0.0001 & 0.0206 & 0.0017 & 0.0272 \\
200 & 50 & 0.0006 & 0.0149 & 0.0004 & 0.0125 & 0.0006 & 0.0127 & 0.0002 & 0.0108 & 0.0007 & 0.0145 \\
200 & 100 & 0.0004 & 0.0100 & 0.0003 & 0.0084 & 0.0002 & 0.0081 & 0.0002 & 0.0077 & 0.0002 & 0.0102 \\
200 & 200 & 0.0001 & 0.0069 & 0.0001 & 0.0054 & 0.0002 & 0.0056 & 0.0001 & 0.0050 & 0.0003 & 0.0074 \\
200 & 300 & 0.0000 & 0.0062 & 0.0001 & 0.0043 & 0.0002 & 0.0046 & 0.0002 & 0.0042 & 0.0001 & 0.0067 \\
300 & 20 & 0.0000 & 0.0326 & 0.0024 & 0.0288 & 0.0003 & 0.0291 & 0.0008 & 0.0170 & 0.0000 & 0.0214 \\
300 & 50 & 0.0006 & 0.0120 & 0.0004 & 0.0097 & 0.0001 & 0.0100 & 0.0007 & 0.0088 & 0.0004 & 0.0118 \\
300 & 100 & 0.0002 & 0.0084 & 0.0003 & 0.0067 & 0.0003 & 0.0066 & 0.0003 & 0.0062 & 0.0003 & 0.0087 \\
300 & 200 & 0.0001 & 0.0058 & 0.0002 & 0.0042 & 0.0000 & 0.0046 & 0.0001 & 0.0043 & 0.0002 & 0.0062 \\
300 & 300 & 0.0000 & 0.0051 & 0.0002 & 0.0035 & 0.0000 & 0.0037 & 0.0001 & 0.0036 & 0.0000 & 0.0054 \\
\bottomrule
\end{tabular}
\label{tab:tableA2}
\begin{tablenotes}
\item \scriptsize \textit{Notes}: The absolute bias and RMSE are computed as
$\left|\frac{1}{S} \sum_{s=1}^S (\widehat{\beta}_{1,s} - \beta_1)\right|$ and $\sqrt{\frac{1}{S}\sum_{s=1}^S (\widehat{\beta}_{1,s} - \beta_1)^2}$, respectively, where $S =1,000$ is the number of simulations, and $\widehat{\beta}_{1,s}$ is the estimate of $\beta_1$ from replication $s$. While we focus here on the first coefficient, $\beta_1$, the results for the second coefficient, $\beta_2$, were similar and can be made are available upon request.			
\end{tablenotes}
\end{threeparttable}
\end{table}
\end{landscape}

\begin{table}[p]
\centering
\begin{threeparttable}
\caption{Simulation results for Experiment E1.F.}
\renewcommand{\arraystretch}{1.0}
\small
\begin{tabular*}{0.7\textwidth}{@{\extracolsep{\fill}}cccccc}
\toprule
&&\multicolumn{2}{c}{$\pi =0.2$} & \multicolumn{2}{c}{$\pi =0.5$}\\
\cmidrule(lr){3-4}   \cmidrule(lr){5-6}
N & T & Bias & RMSE & Bias & RMSE \\
\toprule
20 & 20 & 0.0119 & 0.1911 & 0.0057 & 0.2733 \\
20 & 50 & 0.0003 & 0.0675 & 0.0022 & 0.0971 \\
20 & 100 & 0.0010 & 0.0481 & 0.0018 & 0.0678 \\
20 & 200 & 0.0005 & 0.0341 & 0.0015 & 0.0465 \\
20 & 300 & 0.0008 & 0.0300 & 0.0001 & 0.0402 \\
50 & 20 & 0.0066 & 0.1251 & 0.0049 & 0.1804 \\
50 & 50 & 0.0006 & 0.0433 & 0.0016 & 0.0655 \\
50 & 100 & 0.0003 & 0.0282 & 0.0014 & 0.0421 \\
50 & 200 & 0.0000 & 0.0206 & 0.0003 & 0.0298 \\
50 & 300 & 0.0002 & 0.0174 & 0.0006 & 0.0253 \\
100 & 20 & 0.0018 & 0.1010 & 0.0006 & 0.1303 \\
100 & 50 & 0.0009 & 0.0296 & 0.0005 & 0.0456 \\
100 & 100 & 0.0003 & 0.0187 & 0.0001 & 0.0293 \\
100 & 200 & 0.0004 & 0.0142 & 0.0003 & 0.0201 \\
100 & 300 & 0.0005 & 0.0120 & 0.0008 & 0.0170 \\
200 & 20 & 0.0003 & 0.0856 & 0.0025 & 0.0893 \\
200 & 50 & 0.0004 & 0.0214 & 0.0002 & 0.0329 \\
200 & 100 & 0.0001 & 0.0135 & 0.0002 & 0.0208 \\
200 & 200 & 0.0004 & 0.0091 & 0.0000 & 0.0137 \\
200 & 300 & 0.0003 & 0.0082 & 0.0006 & 0.0120 \\
300 & 20 & 0.0029 & 0.0496 & 0.0047 & 0.0706 \\
300 & 50 & 0.0000 & 0.0172 & 0.0001 & 0.0256 \\
300 & 100 & 0.0003 & 0.0112 & 0.0008 & 0.0169 \\
300 & 200 & 0.0003 & 0.0083 & 0.0004 & 0.0115 \\
300 & 300 & 0.0002 & 0.0067 & 0.0005 & 0.0098 \\
\bottomrule
\end{tabular*}
\renewcommand{\arraystretch}{1.0}
\label{tab:tableA3}
\begin{tablenotes}
\item \scriptsize \textit{Notes}: See Table 1 for an explanation.	Here, $\pi$ refers to serial and cross-sectional correlation of the errors in the equations \eqref{yi} and \eqref{xi}. 		
\end{tablenotes}
\end{threeparttable}
\end{table}

\clearpage
\begin{landscape}
% Preserve the existing placement of landscape tables.
\makeatletter
\setlength{\@fptop}{0pt}
\makeatother
\begin{table}[p]
\centering
\begin{threeparttable}
  \caption{Simulation results for Experiment E1.G.}
  \label{tab:tableA4}
  \begin{tabular}{rrrrrrrrrrrrrr}
\toprule
&&\multicolumn{2}{c}{$J = 2\lfloor T^{\frac{1}{4}}\rfloor$} & \multicolumn{2}{c}{$J = 3\lfloor T^{\frac{1}{4}}\rfloor$} & \multicolumn{2}{c}{$J = 4\lfloor T^{\frac{1}{4}}\rfloor$} & \multicolumn{2}{c}{$J=\lfloor T^{\frac{1}{3}}\rfloor$} & \multicolumn{2}{c}{$J=\lfloor T^{\frac{1}{5}}\rfloor$} & \multicolumn{2}{c}{$J=\lfloor T^{\frac{1}{10}}\rfloor$}\\
\cmidrule(lr){3-4}   \cmidrule(lr){5-6} \cmidrule(lr){7-8} \cmidrule(lr){9-10} \cmidrule(lr){11-12} \cmidrule(lr){13-14}
N & T & Bias & RMSE & Bias & RMSE & Bias & RMSE & Bias & RMSE & Bias & RMSE & Bias & RMSE \\
\midrule
20 & 20 & 0.2027 & 6.5197 & 0.1158 & 3.3448 & 0.7988 & 28.9344 & 0.0014 & 0.1420 & 0.0004 & 0.1029 & 0.0004 & 0.1029 \\
20 & 50 & 0.0018 & 0.0513 & 0.0022 & 0.0573 & 0.0015 & 0.0648 & 0.0017 & 0.0499 & 0.0011 & 0.0478 & 0.0014 & 0.0459 \\
20 & 100 & 0.0010 & 0.0353 & 0.0011 & 0.0367 & 0.0008 & 0.0377 & 0.0015 & 0.0340 & 0.0018 & 0.0335 & 0.0014 & 0.0328 \\
20 & 200 & 0.0005 & 0.0261 & 0.0003 & 0.0264 & 0.0003 & 0.0266 & 0.0005 & 0.0258 & 0.0004 & 0.0267 & 0.0002 & 0.0264 \\
20 & 300 & 0.0008 & 0.0230 & 0.0007 & 0.0232 & 0.0006 & 0.0234 & 0.0009 & 0.0227 & 0.0007 & 0.0221 & 0.0003 & 0.0224 \\
50 & 20 & 0.2721 & 8.9585 & 0.0924 & 4.8175 & 0.1307 & 2.4790 & 0.0042 & 0.0859 & 0.0001 & 0.0644 & 0.0001 & 0.0644 \\
50 & 50 & 0.0019 & 0.0335 & 0.0019 & 0.0375 & 0.0028 & 0.0427 & 0.0023 & 0.0324 & 0.0015 & 0.0317 & 0.0006 & 0.0312 \\
50 & 100 & 0.0012 & 0.0221 & 0.0011 & 0.0230 & 0.0011 & 0.0241 & 0.0009 & 0.0217 & 0.0010 & 0.0219 & 0.0009 & 0.0221 \\
50 & 200 & 0.0003 & 0.0152 & 0.0003 & 0.0154 & 0.0003 & 0.0156 & 0.0003 & 0.0150 & 0.0002 & 0.0153 & 0.0003 & 0.0155 \\
50 & 300 & 0.0002 & 0.0132 & 0.0001 & 0.0133 & 0.0002 & 0.0134 & 0.0004 & 0.0131 & 0.0002 & 0.0129 & 0.0004 & 0.0135 \\
100 & 20 & 0.0034 & 3.6462 & 0.0082 & 1.2543 & 0.0746 & 3.6349 & 0.0054 & 0.0584 & 0.0031 & 0.0461 & 0.0031 & 0.0461 \\
100 & 50 & 0.0009 & 0.0230 & 0.0017 & 0.0247 & 0.0015 & 0.0269 & 0.0007 & 0.0217 & 0.0008 & 0.0210 & 0.0010 & 0.0212 \\
100 & 100 & 0.0004 & 0.0146 & 0.0004 & 0.0153 & 0.0003 & 0.0157 & 0.0006 & 0.0142 & 0.0004 & 0.0144 & 0.0003 & 0.0146 \\
100 & 200 & 0.0003 & 0.0102 & 0.0003 & 0.0102 & 0.0003 & 0.0103 & 0.0002 & 0.0101 & 0.0002 & 0.0107 & 0.0003 & 0.0106 \\
100 & 300 & 0.0004 & 0.0089 & 0.0004 & 0.0089 & 0.0004 & 0.0090 & 0.0004 & 0.0088 & 0.0004 & 0.0087 & 0.0004 & 0.0092 \\
200 & 20 & 0.0274 & 1.5439 & 0.0200 & 1.2613 & 0.0042 & 2.0946 & 0.0041 & 0.0568 & 0.0020 & 0.0309 & 0.0020 & 0.0309 \\
200 & 50 & 0.0004 & 0.0158 & 0.0009 & 0.0175 & 0.0014 & 0.0205 & 0.0004 & 0.0152 & 0.0006 & 0.0149 & 0.0006 & 0.0145 \\
200 & 100 & 0.0004 & 0.0109 & 0.0004 & 0.0104 & 0.0004 & 0.0107 & 0.0004 & 0.0101 & 0.0004 & 0.0099 & 0.0002 & 0.0099 \\
200 & 200 & 0.0001 & 0.0070 & 0.0000 & 0.0071 & 0.0000 & 0.0072 & 0.0001 & 0.0070 & 0.0001 & 0.0070 & 0.0002 & 0.0071 \\
200 & 300 & 0.0000 & 0.0062 & 0.0001 & 0.0063 & 0.0001 & 0.0064 & 0.0000 & 0.0062 & 0.0000 & 0.0062 & 0.0001 & 0.0064 \\
300 & 20 & 0.0271 & 0.4488 & 0.1172 & 5.2088 & 0.1167 & 2.3462 & 0.0000 & 0.0326 & 0.0002 & 0.0249 & 0.0002 & 0.0249 \\
300 & 50 & 0.0004 & 0.0126 & 0.0003 & 0.0139 & 0.0005 & 0.0155 & 0.0006 & 0.0122 & 0.0006 & 0.0120 & 0.0006 & 0.0117 \\
300 & 100 & 0.0001 & 0.0086 & 0.0000 & 0.0089 & 0.0000 & 0.0092 & 0.0001 & 0.0085 & 0.0002 & 0.0085 & 0.0002 & 0.0085 \\
300 & 200 & 0.0001 & 0.0058 & 0.0002 & 0.0060 & 0.0001 & 0.0060 & 0.0001 & 0.0058 & 0.0001 & 0.0060 & 0.0001 & 0.0061 \\
300 & 300 & 0.0000 & 0.0051 & 0.0000 & 0.0051 & 0.0000 & 0.0051 & 0.0000 & 0.0051 & 0.0000 & 0.0051 & 0.0000 & 0.0053 \\
\bottomrule
\end{tabular}
  \begin{tablenotes}
  \item \scriptsize \textit{Notes}: See Table 1 for an explanation. Here, $J$ is the number of knots and $T$ is the time dimension.
  \end{tablenotes}
\end{threeparttable}
\end{table}
\end{landscape}

% Only works cited within the appendix are included here.
\clearpage
\phantomsection
%\addcontentsline{atoc}{section}{References}
\putbib[sample]
\end{bibunit}
\end{appendix}

\end{document}